\documentclass[11pt]{amsart}
\usepackage{amsmath,amsfonts,amsthm,amscd,amssymb,euscript,graphics,psfrag}
\usepackage{pictex,epsfig}

\newcommand{\tmem}[1]{{\em #1\/}}
\newcommand{\tmop}[1]{\ensuremath{\operatorname{#1}}}
\newtheorem{proposition}{Proposition}
\renewenvironment{proof}{
\noindent\textbf{Proof}\ }{\hspace*{\fill}
\begin{math}\Box\end{math}\medskip}
\newtheorem{theorem}{Theorem}
\newcommand{\dueto}[1]{\textup{\textbf{(#1) }}}
\newtheorem{lemma}{Lemma}
\newcommand{\longrightarrowlim}{\mathop{\longrightarrow}\limits}
\newcommand{\tmscript}[1]{\text{\scriptsize $#1$}}
\newtheorem{varremark}{Remark}
\newenvironment{remark}{\begin{varremark}\em}{\em\end{varremark}}
\newtheorem{corollary}{Corollary}

\newcommand{\SO}{{\rm SO}}
\renewcommand{\sp}{{\rm sp}}
\newcommand{\so}{{\rm so}}
\renewcommand{\o}{{\rm o}}

\newcommand{\HS}{{\hbox{\rm LS}}}

\newcommand{\sgn}{{\hbox{\rm sgn}}}
\newcommand{\zed}{{\mathbb Z}}

\title [Averages of characteristic polynomials] {On the averages of
characteristic polynomials from classical groups }

\author{Daniel Bump}
\address{Department
of Mathematics, Stanford University, Stanford, CA 94305 }
\email{bump@math.stanford.edu}

\author{Alex Gamburd}
\address {Department of
Mathematics, University of California, Santa Cruz and  Department
of Mathematics, Stanford University, Stanford, CA 94305 }
\email{agamburd@math.stanford.edu}

\thanks{ The first author was supported in part by the NSF grant
FRG DMS-0354662. The second author was supported in part by the
NSF postdoctoral fellowship.}
\begin{document}
\begin{abstract}
We provide an elementary and self-contained derivation of formulae
for products and ratios of characteristic polynomials from
classical groups using classical results due to Weyl and
Littlewood.
\end{abstract}

\maketitle

\section{Introduction}

The study of averages of characteristic polynomials of random
matrices has attracted considerable attention in recent years. The
interest has been  motivated, in part,  by connections with number
theory, following the pioneering work of Keating and Snaith
{\cite{KS00}}; and, in part, by importance of these averages in
quantum chaos, first discussed by Andreev and Simons
{\cite{as95}}.  Over the ensuing years it has become increasingly
clear that averages of characteristic polynomials are a
fundamental characteristic of random matrix models (see, for
example, discussion in \cite[1.6]{bs04} were the authors argue
that they might be more fundamental than correlation functions).
Results in the case of Hermitian ensembles were obtained in Baik,
Deift and Strahov~{\cite{bds}}, Borodin and Strahov~{\cite{bs04}},
Br\'ezin and Hikami~{\cite{bh1}}, {\cite{bh2}}, {\cite{bh3}},
{\cite{bh4}}, Forrester and Keating~{\cite{fh04a}},
Fyodorov~{\cite{f1}}, Fyodorov and Keating~{\cite{f2}}, Fyodorov
and Strahov~{\cite{f4}}, {\cite{f3}}, {\cite{f5}}, Mehta and
Normand~{\cite{MN}} and Strahov and Fyodorov~{\cite{sf}}. Averages
of products in the case of compact classical groups were
considered by Conrey, Farmer, Keating, Rubinstein, and Snaith in
{\cite{CFKRS2}} in connection with conjectures for integral
moments of zeta and L-functions {\cite{CFKRS1}}.

Recently the averages of ratios in the case of compact classical
groups were considered by Conrey, Farmer, and Zirnbauer
{\cite{cfz}} and by Conrey, Forrester and Snaith {\cite{cfs}}. The
approach in {\cite{cfz}} is based on using supersymmetry and the
theory of dual reductive pairs. The approach in {\cite{cfs}} is
based on reducing the orthogonal and symplectic case to the case
of unitary invariant Hermitian matrices  and then invoking the
results obtained by Fyodorov and Strahov {\cite{f3}} and by Baik,
Deift and Strahov {\cite{bds}}; the case of unitary group is
treated by appealing to the formula of Day {\cite{day}} for
Toeplitz determinants, for which the authors give a self-contained
derivation using the method of Basor and Forrester {\cite{bf94}}.

The goal of this paper is to provide an elementary and
self-contained derivation of formulas for products  of
characteristic polynomials form classical groups, obtained in
{\cite{CFKRS2}}; and for their ratios, obtained in {\cite{cfz}}
and {\cite{cfs}}.   We also obtain an elementary derivation of the
formulas for integral moments of characteristic polynomials
derived by Keating and Snaith {\cite{KS00,KS00a}} using Selberg's
integral. Our proofs use classical results due to Weyl~\cite{weyl}
and Littlewood~\cite{Li} and can be  viewed as application of
symmetric function theory in random matrix theory along the lines
pioneered by Diaconis and Shahshahani {\cite{DS94}} and applied in
Rains \cite{Ra98}, Bump and Diaconis~{\cite{BD02}}, Baik and
Rains~{\cite{BR02}}, Diaconis~{\cite{di03}} and Diaconis and
Gamburd~{\cite{DG04}}. We begin with review of symmetric function
theory in Section \ref{sec:2} and consider unitary group in
Section \ref{sec:3}.

In a nutshell, our method consists of expressing the mean value of
a product or ratio of characteristic polynomials on a group $G$ in
terms of a symmetric function (such as a Schur polynomial) related
to a character of an irreducible representation whose highest
weight vector is a partition of ``rectangular shape,'' then
reducing that value as a sum over elements of $W/W_M$, where $W$
is the Weyl group of $G$, $M$ is a subgroup of~$G$ and  $W_M$ is
the Weyl group of $M$. This point is explained in
Section~\ref{comf}. Symplectic group is considered in Section
\ref{sec:sp} and Orthogonal Group is considered in section
\ref{sec:o}. Our results imply  simple derivations of formulas for
classical group characters of rectangular shape due to Okada
\cite{ok} and Krattenthaler~\cite{krat} and yield several
generalizations; this is considered in Section \ref{rect}.

\section{Review of symmetric functions theory}\label{sec:2}

\subsection{Schur functions}

A {\tmem{partition}} $\lambda$ is a sequence $\lambda_1 \geqslant
\lambda_2 \geqslant \cdots$ of nonnegative integers, eventually
zero. By abuse of notation, we write $\lambda = ( \lambda_1,
\lambda_2, \cdots, \lambda_n )$ for any $n$ such that $\lambda_{n
+ 1} = 0$. There is a unique $n$ such that $\lambda_n > 0$ but
$\lambda_{n + 1} = 0$ and this $n = l ( \lambda )$ is the
{\tmem{length}} of $\lambda$. We call $| \lambda | = \sum
\lambda_i$ the {\tmem{size}} of $\lambda$. If $i > 0$ let $m_i =
m_i ( \lambda )$ be the number of parts $\lambda_j$ of $\lambda$
equal to $i$; in this case, we write $\lambda = \left\langle
1^{m_1} 2^{m_2} 3^{m_3} \cdots \right\rangle$.

The Young diagram of a partition $\lambda$ is defined as the set
of points $(i, j) \in \zed^2$ such that $1 \leqslant i \leqslant
\lambda_j$; it is often convenient to replace the set of points
above by squares.  The conjugate partition $\lambda'$ of $\lambda$
is defined by the condition that the Young diagram of $\lambda'$
is the transpose of the Young diagram of $\lambda$; equivalently
$m_i(\lambda')=\lambda_i-\lambda_{i+1}$.

\begin{figure}[!h]
\abovedisplayskip-.5\baselineskip \belowdisplayskip-\baselineskip
\begin{equation*}
\begin{matrix} \beginpicture \setcoordinatesystem units
<0.5cm,0.5cm>         \setplotarea x from 0 to 4, y from 1 to 3
\linethickness =0.5pt                          \putrule from 0 6
to 5 6
         \putrule from 0 5 to 5 5          \putrule from 0 4 to 5 4
     \putrule from 0 3 to 3 3          \putrule from 0 2 to 2 2
\putrule from 0 2 to 0 6        \putrule from 1 2 to 1 6 \putrule
from 2 2 to 2 6        \putrule from 3 3 to 3 6 \putrule from 4 4
to 4 6        \putrule from 5 4 to 5 6
\endpicture &\qquad \qquad &

\beginpicture   \setcoordinatesystem units
<0.5cm,0.5cm>         \setplotarea x from 0 to 4, y from 1 to 3
\linethickness =0.5pt                          \putrule from 0 6
to 4 6
         \putrule from 0 5 to 4 5          \putrule from 0 4 to 4
         4
     \putrule from 0 3 to 3 3          \putrule from 0 2 to 2 2
\putrule from 0 1 to 2 1

\putrule from 0 1 to 0 6        \putrule from 1 1 to 1 6 \putrule
from 2 1 to 2 6 \putrule from 3 3 to 3 6       \putrule from 4 4
to 4 6
\endpicture\\

\hbox {Young diagram of  $\lambda$ }  &&\hbox { Young diagram of
$\lambda'$}
\end{matrix}
\end{equation*}
\end{figure}

In the figure we exhibited a partition $\lambda = (5, 5, 3, 2)
=\langle 1^0 2^1 3^1 5^2 \rangle$; $\lambda \vdash 15$ and
$l(\lambda)=4$.

Let $\lambda$ and $\mu$ be partitions. We define $\lambda + \mu$
to be the partition $( \lambda_1 + \mu_1, \lambda_2 + \mu_2,
\cdots )$. On the other hand we define $\lambda \cup \mu$ to be
the partition whose parts are the union of the parts of $\lambda$
and $\mu$, arranged in descending order. For example if $\lambda =
( 321 )$ and $\mu = ( 22 )$ then $\lambda + \mu = ( 541 )$ and
$\lambda \cup \mu = ( 32221 )$. The operations $+$ and $\cup$ are
dual to each other:
\[ ( \lambda + \mu )' = \lambda' \cup \text{} \mu' . \]
We write $\lambda \supset \mu$ if the diagram of $\lambda$
contains the diagram of $\mu$, or equivalently, if $\lambda_i
\geqslant \mu_i$ for all~$i$.

The elementary symmetric functions $e_r(x_1, \dots, x_n)$ are
defined by \begin{equation} \label{e:elsim1} e_r(x_1, \dots, x_n)
=\sum_{i_1 < \dots < i_r}x_{i_1} \dots x_{i_r}; \end{equation}
complete symmetric functions $h_r(x_1, \dots, x_N)$ are defined by
\begin{equation} \label{e:comsim1} h_r(x_1, \dots, x_n) =\sum_{i_1
\leqslant \dots \leqslant i_r}x_{i_1} \dots x_{i_r}.
\end{equation}

Now given a partition $\lambda$,  we define
\begin{equation} \label{e:elsim2} e_{\lambda}(x_1, \dots , x_n)
=\prod_{j=1}^{n} e_{\lambda_j}(x_1, \dots, x_n)\end{equation} and
similarly
\begin{equation} \label{e:comsim2} h_{\lambda}(x_1, \dots , x_n)
=\prod_{j=1}^{n} h_{\lambda_j}(x_1, \dots, x_n).\end{equation}

{\tmem{Schur functions}} are symmetric polynomials indexed by
partitions. If $\lambda$ is any partition of length $\leqslant n$
we define
\begin{equation}
\label{schurdefinition}
\begin{aligned}
& s_{\lambda} ( x_1, \cdots, x_n ) =
\frac{\det\left(x_i^{\lambda_{j}+n-j}\right)_{i, j=1}^{n}}
{\det\left(x_i^{n-j}\right)_{i, j=1}^{n}}=\\
&\\ & \frac{\left|
  \begin{array}{ccccc}
  x_1^{\lambda_1+n-1} &   x_1^{\lambda_2+n-2}&
\dots &  x_1^{\lambda_n}& \\
x_2^{\lambda_1+n-1} &   x_2^{\lambda_2+n-2}&
\dots &  x_2^{\lambda_n}& \\
    \vdots & \vdots &\ddots & \vdots & \\
x_n^{\lambda_1+n-1} &   x_n^{\lambda_2+n-2}&
\dots &  x_n^{\lambda_n}& \\
\end{array} \right|}{\left|
  \begin{array}{ccccc}
  x_1^{n-1} &   x_1^{n-2}&
\dots &  1 & \\
x_2^{n-1} &   x_2^{n-2}&
\dots &  1 & \\
    \vdots & \vdots &\ddots & \vdots & \\
x_n^{n-1} &   x_n^{n-2}&
\dots & 1 & \\
\end{array} \right|}
\end{aligned}
\end{equation}

If $l ( \lambda ) > n$ we define $s_{\lambda}(x_1,\cdots,x_n) =
0$.

The denominator in \eqref{schurdefinition} is a Vandermonde
determinant admitting the following simple evaluation:
\begin{equation} \label{vander} \det(x_{i}^{n-j}) = \prod_{1\leqslant j<k \leqslant n}(x_j-x_k).
\end{equation}

Let $\Lambda^{( n )} = \Lambda^{( n )} ( x )$ denote the ring of
symmetric polynomials with integer coefficients in $x_1, \cdots,
x_n$. Then $s_{\lambda} ( x_1, \cdots, x_n ) \in \Lambda^{( n )}$.
We have a homomorphism $\Lambda^{( n + 1 )} \longrightarrow
\Lambda^{( n )}$ in which the last variable $x_{n + 1}$ is
specialized to $0$, and under this homomorphism it is easy to see
from the definition that this specialization takes $s_{\lambda}$
to $s_{\lambda}$; that is,
\begin{equation}
\label{ratiosschurspecialize} s_{\lambda} ( x_1, \cdots, x_n, 0 )
= s_{\lambda} ( x_1, \cdots, x_n ) .
\end{equation}
This means that $s_{\lambda}$ may be regarded as an element of the
ring $\Lambda$
\[ \Lambda = \Lambda ( x ) = \lim_{\longleftarrow} \; \Lambda^{( n )} ( x ) .
\]
The ring $\Lambda$ is a polynomial ring in either $h_1, h_2,
\cdots$ or $e_1, e_2, \cdots$ where $h_n = s_{\left\langle n
\right\rangle}$ and $e_n = s_{\left\langle 1^n \right\rangle}$
(Bump~{\cite{DB}}, Proposition~36.5 or Macdonald~{\cite{Mac}},
p.~22). It is a free abelian group with basis $s_{\lambda}$, as
$\lambda$ runs through all partitions. The ring $\Lambda^{( n )}$
is a free abelian group with basis $s_{\lambda}$, as $\lambda$
runs through the partitions of length $\leqslant n$.

The ring $\Lambda$ has an involution $\iota$ which interchanges
$h_n \longleftrightarrow e_n$ (Bump~{\cite{DB}}, Theorem~36.3).
This involution takes $s_{\lambda}$ to $s_{\lambda'}$
(Bump~{\cite{DB}}, Theorem~37.2 or Macdonald~{\cite{Mac}}, (3.8)
on p.~42).

We will make crucial use of the following classical result (see
Weyl~{\cite{weyl}}, Lemma 7.6.A;  or, for modern presentations,
Bump~{\cite{DB}}, Chapter 43 or Macdonald~{\cite{Mac}}, (4.3) on
p.~63.)

\begin{proposition}
{\dueto{Cauchy identity}}Let $x_1, \cdots, x_p$ and $y_1, \cdots,
y_q$ be two sets of variables. Then
\begin{equation}
   \label{ratioscauchy} \prod_{i = 1}^p \prod_{j = 1}^q ( 1 - x_i y_j )^{- 1}
   = \sum_{\lambda} s_{\lambda} ( x_1, \ldots, x_p ) s_{\lambda} ( y_1,
   \ldots, y_q ),
\end{equation}
where the sum is over all partitions $\lambda$ of length
$\leqslant \min ( p, q )$.
\end{proposition}

Both sides are convergent provided the $|x_i |$ and $|y_j |$ are
all less than~$1$; otherwise, this may be regarded as a formal
identity.

We have also the {\tmem{dual Cauchy identity}}
\begin{equation}
\label{ratiosdualcauchy} \prod_{i = 1}^p \prod_{j = 1}^q ( 1 + x_i
y_j ) = \sum_{\lambda} s_{\lambda} ( x_1, \ldots, x_p )
s_{\lambda'} ( y_1, \ldots, y_q ) .
\end{equation}
The right hand side is a finite sum, since unless $l ( \lambda )
\leqslant p$ and $l ( \lambda' ) \leqslant q$ we have either
$s_{\lambda} = 0$ or $s_{\lambda'} = 0$, so the diagram of
$\lambda$ must fit inside a $p \times q$ rectangle.

The dual Cauchy identity can be deduced from the standard Cauchy
identity by making use of the involution on $\Lambda$.
See Bump~{\cite{DB}},
Theorem~43.5 or Macdonald~{\cite{Mac}}, (4.3$'$) on p.~65.

If $\mu$ and $\nu$ are partitions, then $s_{\mu} s_{\nu}$ can be
decomposed into Schur polynomials; we write
\begin{equation} \label{e_lr} s_{\mu}s_{\nu}=\sum_{\lambda}
c^{\lambda}_{\mu \nu} s_{\lambda}.
\end{equation}
where the sum is over partitions $\lambda$ of size $| \mu | + |
\nu |$. It follows from (\ref{ratiosschurspecialize}) that the
{\tmem{Littlewood-Richardson coefficients}} $c_{\lambda
\mu}^{\nu}$ are independent of $n$, and they are determined by
this equation provided $n \geqslant | \lambda | + | \mu |$ (so
that the $s_{\nu}$ are linearly independent).

Let $x_1, \cdots, x_p$ and $y_1, \cdots, y_q$ be two sets of
variables. The Littlewood-Richardson coefficients also appear in
\begin{equation}
   \label{explicitbranching} s_{\lambda} ( x_1, \cdots, x_p, y_1, \cdots, y_q
   ) = \sum_{\mu, \nu} c_{\mu \nu}^{\lambda} s_{\mu} ( x_1, \cdots, x_p )
   s_{\nu} ( y_1, \cdots, y_q ) .
\end{equation}
See Macdonald~\cite{Mac}, (5.9) on p.~72, who proves
\eqref{explicitbranching} from \eqref{e_lr} using the Cauchy
identity. We will also give a proof (which is closely related to
Macdonald's) in Section~\ref{sec:hd}.

Particular cases of (\ref{e_lr}) are furnished by  Pieri's formula
(see Bump~{\cite{DB}}, Theorem~42.4 or Macdonald~{\cite{Mac}},
(5.16) on p.~73.)
  If $\lambda \supseteq \mu$ are partitions, we say that
$\lambda - \mu$ is a {\tmem{horizontal strip}} if no two elements
of the set theoretic difference of their diagrams lie in the same
column; and we say that $\lambda - \mu$ is a {\tmem{vertical
strip}} no two elements of the set theoretic difference of their
diagrams lie in the same row.

The content of Pieri's formula is that
\begin{equation} \label{e:pieri}
  c^{\lambda}_{\mu \left\langle r \right\rangle} = \begin{cases}
      1 & \text{if $\lambda - \mu$ is a horizontal strip and $| \lambda | - |
      \mu | = r ;$}\\
      0 & \text{otherwise.}
    \end{cases}
\end{equation}
Equivalently, we can write
\begin{equation}
\label{e_pf1} s_{\mu} h_r = \sum_{\lambda} s_{\lambda},
\end{equation}
where the summation is over all partitions $\lambda$ such that
$\lambda - \mu$ is a horizontal $r$-strip.

Similarly we have
\begin{equation} \label{e_pf2}
s_{\mu} e_r = \sum_{\lambda} s_{\lambda},
\end{equation}
where the summation is over all partitions $\lambda$ such that
$\lambda - \mu$ is a vertical $r$-strip.

From \eqref{explicitbranching} and (\ref{e_pf1}) it follows that
\begin{equation}
\label{extvar} s_{\lambda}(x_1, \ldots, x_k, y)=
\sum_{\mu}s_{\mu}(x_1, \ldots, x_k)y^{|\lambda- \mu|},
\end{equation}
where the sum is over partitions $\mu$ such that $\lambda - \mu$
is a horizontal strip.

We denote by $k^n=\left<k^n\right>$ the partition $(k, \ldots, k)$
of length $n$; assuming that $\lambda$ is a partition of length
less than or equal to $n$ we denote
\[\lambda + k^n =(\lambda_1 +k, \ldots, \lambda_n +k).\]

We have
\begin{equation}
\label{e_pf3} s_{\lambda + k^n}(x_1,\cdots,x_n)
=e_{n}^k(x_1,\cdots,x_n)\,s_{\lambda}(x_1,\cdots,x_n).
\end{equation}
This is immediate from the definition of the Schur polynomial
together with the fact that $e_n(x_1,\cdots,x_n)=x_1\cdots x_n$.

\subsection{Connection with unitary group}
Let $g \in \tmop{GL} ( n,\mathbb{C})$ have eigenvalues $t_1,
\cdots, t_n$. Assuming that the length $l ( \lambda ) \leqslant
n$, we define $\chi_{\lambda} ( g ) = \chi_{\lambda}^{( n )} ( g )
= s_{\lambda} ( t_1, \cdots, t_n )$. (The assumption on $l (
\lambda )$ is necessary since otherwise $s_{\lambda} ( t_1,
\cdots, t_n ) = 0$.) The following result  is a special case of
the Weyl character formula \cite{weyl}, which expresses the
characters of compact Lie groups as ratios of alternating
functions; see, for example, Bump~{\cite{DB}}, Theorem~38.2. for a
modern exposition.

\begin{proposition}
If $\lambda$ is a partition of length $\leqslant n$, the function
$\chi_{\lambda}$ is the character of an irreducible  analytic
representation $\pi_{\lambda} = \pi_{\lambda}^{( n )}$ of
$\tmop{GL} ( n,\mathbb{C})$. It is irreducible; in fact, its
restriction to $U ( n )$ is irreducible.
\end{proposition}

{\noindent} As a consequence, we have Schur orthogonality for the
functions $\chi_{\lambda}$. Let $\left\langle \;, \;
\right\rangle$ denote the inner product with respect to the Haar
measure on $G = U ( n )$, normalized so that the volume of $G$
is~$1$. Then if $\lambda$ and $\mu$ are dominant weights, we have
\begin{equation}
\label{ratiosschurorthog} \left\langle \chi_{\lambda}, \chi_{\mu}
\right\rangle = \left\{\begin{array}{ll}
   1 & \text{if $\lambda = \mu$} ;\\
   0 & \text{otherwise} .
\end{array}\right.
\end{equation}

Now \eqref{e_pf3} implies that if  $\mu = \lambda + k^n$, then
$\chi_{\mu} = {\det}^k \otimes \chi_{\lambda}$. From this
observation we may extend the definition of $\chi_{\lambda}$ as
follows. If $\mu = ( \mu_1, \cdots, \mu_n )$ where $\mu_i$ are
integers (possibly negative) and $\mu_1 \geqslant \cdots \geqslant
\mu_n$, then we call $\mu$ a {\tmem{dominant weight}}. Thus the
dominant weight $\mu$ is a partition if and only if $\mu_n
\geqslant 0$. We can always write $\mu = \lambda + \left\langle
k^n \right\rangle$ where $\lambda$ is a partition for some $k$;
for example, we can take $k = \mu_n$. Then we define $\chi_{\mu} =
{\det}^k \otimes \chi_{\lambda}$. It is of course the character of
an irreducible analytic representation of $\tmop{GL} (
n,\mathbb{C})$, or its compact subgroup~$U ( n )$. The following
result is due to Weyl \cite{weyl}; see Bump~{\cite{DB}},
Theorem~38.3. for a modern presentation.
\begin{theorem}
The characters of the irreducible analytic representations of
$\tmop{GL} ( n,\mathbb{C})$ are precisely the $\chi_{\lambda}$ as
$\lambda$ runs through the dominant weights. Their restrictions to
$U ( n )$ are precisely the irreducible characters of~$U ( n )$.
\end{theorem}

If $\lambda$ is a partition, we call the representation
$\pi_{\lambda}$ of $\tmop{GL} ( n,\mathbb{C})$ with character
$\chi_{\lambda}$ a {\tmem{polynomial representation}} since its
matrix coefficients are polynomials of the coordinate functions
$g_{i j}$ of $g \in \tmop{GL} ( n,\mathbb{C})$. If $\lambda$ is a
dominant weight that is not necessarily a partition, we call
$\pi_{\lambda}$ a {\tmem{rational representation}}. Its matrix
coefficients may have denominators that are powers of the
determinant.

The value of $s_{\lambda}(1^n)$  gives the dimension of the
irreducible representation of $U(n)$ with highest weight
$\lambda$.  From \eqref{schurdefinition} using l'Hopital rule one
easily derives for $\lambda=(\lambda_1, \ldots, \lambda_n)$ the
following formula, known as the Weyl dimension formula
\cite{weyl}:
\begin{equation} \label{e_wdf} s_{\lambda}(1,
\ldots, 1)=\frac{\prod_{i<j}(\mu_i -\mu_j)}{\prod_{i<j}(i -j)},
\end{equation}
where $\mu_i=\lambda_i+n-i.$
\begin{figure}[!h]
\abovedisplayskip-.5\baselineskip \belowdisplayskip-\baselineskip
\begin{equation*}
\beginpicture \setcoordinatesystem units <0.5cm,0.5cm>
\setplotarea x from 0 to 4, y from 1 to 3    \linethickness =0.5pt
                    \put {2} at 0.5 2.5
\put {4} at 0.5 3.5 \put {7} at 0.5 4.5 \put {8}  at 0.5 5.5 \put
{1} at 1.5 2.5 \put {3} at 1.5 3.5 \put {6}  at 1.5 4.5 \put {7}
at 1.5 5.5 \put {1}  at 2.5 3.5 \put {4}  at 2.5 4.5 \put {5}  at
2.5 5.5 \put {2}  at 3.5 4.5 \put {3}  at 3.5 5.5 \put {1}  at 4.5
4.5 \put {2}  at 4.5 5.5 \putrule from 0 6 to 5 6 \putrule from 0
5 to 5 5 \putrule from 0 4 to 5 4 \putrule from 0 3 to 3 3
\putrule from 0 2 to 2 2 \putrule from 0 2 to 0 6 \putrule from 1
2 to 1 6 \putrule from 2 2 to 2 6 \putrule from 3 3 to 3 6
\putrule from 4 4 to 4 6 \putrule from 5 4 to 5 6
\endpicture
\end{equation*}
\caption{}\label{f:l1}
\end{figure}

It is easy  to deduce that the right-hand side of (\ref{e_wdf})
can be expressed in the following equivalent form (see \cite{Mac}
or \cite{stanleyv2}):
\begin{equation} \label{e_hhl} s_{\lambda}(1^n)=\prod_{u \in
\lambda}\frac{n+c(u)}{h(u)},
\end{equation}
where for a box $u$ in a diagram $\lambda$,  $h(u)$ is a hook
number of $u$ and $c(u)$ is a content number of $u$, which we now
define. Given a diagram $\lambda$ and a square $u=(i, j)\in
\lambda,$ the content of $\lambda$ at $u$ is defined by
$c(u)=j-i$. A hook with a vertex $u$ is a set of squares in
$\lambda$ directly to the right or directly below $u.$ We define
hook-length (also referred to as hook number) $h(u)$ of $\lambda$
at $u$ by
\[h(u)=\lambda_i +\lambda'_j -i -j+1.\]
Equivalently, $h(u)$ is the number of squares directly to the
right or directly below $u$, counting $u$ itself once.  For
instance, in figure \ref{f:l1} we display hook lengths for
partition $\lambda=(5, 5, 3, 2)$.

\subsection{Laplace expansion}
The following classical result from linear algebra will be used
repeatedly in what follows; its deeper meaning and significance
will be discussed in Section~\ref{comf}.

\begin{proposition}
{\dueto{Laplace expansion}} \label{prop:lexp} Fix $L$ rows of the
matrix $A$.  Then the sum of products of the minors of order $L$
that belong to these rows by their cofactors is equal to the
determinant of $A$.
\end{proposition}

Let $\Xi_{L, K}$ consist of all permutations $\sigma \in S_{K +
L}$ such that \begin{equation} \label{perms} \sigma ( 1 ) < \cdots
< \sigma ( L ), \hspace{2em} \sigma ( L + 1 ) <
  \cdots < \sigma ( L + K ) .
  \end{equation}
Given $( K + L ) \times ( K + L )$ matrix $A = ( a_{ij} )$ Laplace
expansion in the first $L$ rows can be written as follows:
\[ \det ( a_{ij} ) = \sum_{\sigma \in \Xi_{L, K}} \tmop{sgn} ( \sigma )
    \left| \begin{array}{ccc}
      a_{1, \sigma ( 1 )} & \cdots & a_{1, \sigma ( L )}\\
      \vdots &  & \vdots\\
      a_{L, \sigma ( 1 )} & \cdots & a_{L, \sigma ( L )}
    \end{array} \right| \cdot \left| \begin{array}{ccc}
      a_{L + 1, \sigma ( L + 1 )} & \cdots & a_{L + K, \sigma ( L + K )}\\
      \vdots &  & \vdots\\
      a_{L + K, \sigma ( L + 1 )} & \cdots & a_{L + K, \sigma ( L + K )}
    \end{array} \right| . \]

We now record two simple applications of the Laplace expansion.

\begin{lemma}
\label{lem:min1}Suppose $\lambda = \tau \cup \rho$ with $\lambda =
( \lambda_1, \ldots, \lambda_{L + K} )$, $\tau = ( \lambda_1,
\ldots, \lambda_L )$ and $\rho = ( \lambda_{L + 1}, \ldots,
\lambda_{L + K} )$. Then
\begin{eqnarray}
   s_{\lambda} ( \alpha_1, \cdots, \alpha_{L + K} ) = \sum_{\sigma \in
   \Xi_{L, K}} \prod_{\tmscript{\begin{array}{c}
     1 \leqslant l \leqslant L\\
     1 \leqslant k \leqslant K
   \end{array}}} ( \alpha_{\sigma ( l )} - \alpha_{\sigma ( L + k )} )^{- 1}
   &  &  \nonumber\\
   \times s_{\tau + \left\langle K^L \right\rangle} ( \alpha_{\sigma ( 1 )},
   \cdots, \alpha_{\sigma ( L )} ) \, s_{\rho} ( \alpha_{\sigma ( L + 1 )},
   \cdots, \alpha_{\sigma ( L + K )} ) & . &
\end{eqnarray}
\end{lemma}

\begin{proof}
We apply the Laplace expansion in the first $L$ rows to the
determinant in
\begin{eqnarray*}
   & s_{\lambda} ( \alpha_1, \cdots, \alpha_{L + K} ) = \Delta^{- 1} \left|
   \begin{array}{cccc}
     \alpha_1^{K + \lambda_1 + L - 1} & \cdots & \alpha_{K + L}^{K +
     \lambda_1 + L - 1} & \\
     \vdots &  & \vdots & \\
     \alpha_1^{K + \lambda_L} & \cdots & \alpha_{K + L}^{K + \lambda_L} & \\
     \alpha_1^{K + \lambda_{L + 1} - 1} & \cdots & \alpha_{K + L}^{K +
     \lambda_{L + 1} - 1} & \\
     \vdots &  & \vdots & \\
     \alpha_1^{\lambda_{K + L}} & \cdots & \alpha_{K + L}^{\lambda_{K + L}} &
   \end{array} \right| &
\end{eqnarray*}
where
\[ \Delta = \left| \begin{array}{cccc}
      \alpha_1^{K + L - 1} & \cdots & \alpha_{K + L}^{K + L - 1} & \\
      \alpha_1^{K + L - 2} & \cdots & \alpha_{K + L}^{K + L - 2} & \\
      \vdots &  & \vdots & \\
      1 & \cdots & 1 &
    \end{array} \right| = \prod_{i < j} ( \alpha_i - \alpha_j ) = \tmop{sgn}
    ( \sigma ) \prod_{i < j} ( \alpha_{\sigma ( i )} - \alpha_{\sigma ( j )}
    ) \]
and simplify.
\end{proof}


\begin{lemma} \label{unidecomp}
For $\lambda \subseteq \left\langle N^k \right\rangle$ let
$\tilde{\lambda} = ( k - \lambda_N', \cdots, k- \lambda_1' )$.
Then we have
\begin{equation} \label{e:udec}
   \prod_{i = 1}^k \prod_{n = 1}^N ( x_i  - t_n) =
   \sum_{\lambda \subseteq N^k} ( - 1 )^{| \tilde{\lambda} |}
   s_{\lambda} ( x_1, \cdots, x_k )
   s_{\tilde{\lambda}}  ( t_1, \cdots, t_N) .
\end{equation}
\end{lemma}

Using the fact that
\[s_\mu(-t_1,\cdots,-t_N)=(-1)^{|\mu|}s_\mu(t_1,\cdots,t_N)\]
the formula may be written
\begin{equation} \label{e:udecalt}
   \prod_{i = 1}^k \prod_{n = 1}^N ( x_i  + t_n) =
   \sum_{\lambda \subseteq N^k} s_{\lambda} ( x_1, \cdots, x_k )
   s_{\tilde{\lambda}}  ( t_1, \cdots, t_N) .
\end{equation}

Generalizations of this Lemma will be given below in Proposition
\ref{littlewoodsquarecase} and also in
Lemma~\ref{symplecticdecomposition}, Lemma~\ref{lem:sod}, and
Lemma~\ref{lem:soe}.

\medbreak
\begin{proof}
Using the definition of Schur functions \eqref{schurdefinition}
and the Laplace expansion we can rewrite the right-hand side of
\eqref{e:udec} as follows:
\begin{equation} \label{e:upl1}
\begin{aligned}
&\det \left|
  \begin{array}{ccccc}
  x_1^{N+k-1}&   x_1^{N+k-2} &
\dots & 1& \\
    \vdots & \vdots &\ddots & \vdots & \\
x_k^{N+k-1} &   x_k^{N+k-2}&
\dots &  1 & \\
t_1^{N+k-1} &   t_1^{N+k-2}&
\dots & 1& \\
    \vdots & \vdots &\ddots & \vdots & \\
t_N^{N+k-1} &   t_N^{N+k-2} & \dots & 1
\end{array} \right|  \\
&\times  \frac {1}{\prod_{1 \leqslant i<j \leqslant k}(x_i-x_j)}
\frac {1}{\prod_{1 \leqslant i<j \leqslant N}(t_i-t_j)}.
\end{aligned}
\end{equation}

Now the determinant in the equation \eqref{e:upl1} can be
evaluated using the formula for Vandermonde determinant formula
\eqref{vander} to be equal to
\begin{equation} \label{e:upl2}
\prod_{1 \leqslant i<j \leqslant k}(x_i-x_j)\prod_{1 \leqslant i<j
\le N}(t_i-t_j)\prod_{i = 1}^k \prod_{n = 1}^N ( x_i  - t_n).
\end{equation}
Combining \eqref{e:upl1} and \eqref{e:upl2} completes the proof.
See Remark~\ref{remzero} below.
\end{proof}

\begin{remark}\label{remzero}
The following combinatorial fact is used implicitly in the last
proof in determining which pair of Schur functions appears: if
$\lambda$ is a partition such that $\lambda_1 \leqslant N$ and
$\lambda_{1}' \leqslant k$, then the $N+k$ numbers $\lambda_i+k-i$
(for $1 \leqslant i \leqslant k$) and $k-1+j-\lambda_{j}'$ (for
$1\leqslant j\leqslant N$) are a permutation of $\{0, 1, 2, \dots,
N+k-1 \}$. See  Macdonald~\cite{Mac}, (1.7) on p.3., or
Bump~\cite{DB}, Proposition~37.2.
\end{remark}

\begin{remark} \label{remcl} Lemma \ref{unidecomp} can by proved by a simple
application of dual Cauchy identity; conversely dual Cauchy is an
immediate consequence of this Lemma.
\end{remark}

\subsection{On the role of Howe duality in symmetric function
theory} \label{sec:hd}

The Cauchy identity \eqref{ratioscauchy} and dual Cauchy
identity~\eqref{ratiosdualcauchy} play a crucial role in our
derivations.  While (as indicated in Remark \ref{remcl}) they
admit simple formal or combinatorial proofs (see for example
Macdonald \cite{Mac} or Stanley \cite{stanleyv2}), their deeper
meaning is revealed by Howe's theory of dual pairs which we
briefly review in this section.

Let $G$ and $H$ be groups, and let $\omega$ be a representation of
$G \times H$. Following Howe~{\cite{howeclassical}},
{\cite{howetheta}}, {\cite{HowePerspectives}}, we say that
$\omega$ is a {\tmem{correspondence}} if $\omega = \bigoplus_{i
\in I} \pi_i \otimes \sigma_i$ where the $\pi_i$ are irreducible
representations of $G$ and the $\sigma_i$ are irreducible
representations of $H$, and if there are no repetitions among the
(isomorphism classes of the) $\pi_i$, nor among the $\sigma_i$. In
this case $\pi_i \longleftrightarrow \sigma_i$ is the graph of a
bijection between a set of irreducible representations of $G$ and
a set of irreducible representations of~$H$. We call this
bijection the {\tmem{graph of the correspondence}}. We refer to $G
\times H$ as the {\tmem{dual pair}} of the correspondence.

A classical example is furnished by  {\tmem{Frobenius-Schur
duality}}, going back to Schur's dissertation and emphasized by
Weyl~{\cite{weyl}}. This is the fact that if $V =\mathbb{C}^n$,
then $U ( n )$ and the symmetric group $S_k$ act on
${\bigotimes}^k V$; the group $U ( n )$ acts by $g : v_1 \otimes
\cdots \otimes v_k \longrightarrow ( g v_1 ) \otimes \cdots
\otimes ( g v_k )$, while $S_k$ acts by permuting the components.
These actions commute with each other, and they therefore give
rise to a representation of $U ( n ) \times S_k$. Frobenius-Schur
duality is the fact that this representation is a correspondence.
Frobenius-Schur duality allows computations on the unitary group
sometimes to be transferred to the symmetric group, a principle
that has applications to random matrix theory. See, for example
Diaconis and Shahshahani~{\cite{DS94}} and Bump and
Diaconis~{\cite{BD02}}.

As as second example, the representation of $\tmop{GL} (
p,\mathbb{C}) \times \tmop{GL} ( q,\mathbb{C})$ on the symmetric
algebra $\bigvee \tmop{Mat}_{p \times q} (\mathbb{C})$, or
equivalently of the subgroups $U ( p ) \times U ( q )$ is a
correspondence. We now have the following interpretation of Cauchy
identity.   Let the group $\tmop{GL} ( p,\mathbb{C}) \times
\tmop{GL} ( q,\mathbb{C})$ act on $\tmop{Mat}_{p \times q}
(\mathbb{C})$ by left and right translation; that is,
\[ ( g_1, g_2 ) : X \longrightarrow g_1 \, X \, {^t g_2} . \]
Thus $\tmop{GL} ( p,\mathbb{C}) \times \tmop{GL} ( q,\mathbb{C})$
acts on the symmetric algebra $\bigvee \tmop{Mat}_{p \times q}
(\mathbb{C})$, and we consider the trace of
\[ ( g_1, g_2 ) = \left( \left(\begin{array}{ccc}
    x_1 &  & \\
    & \ddots & \\
    &  & x_p
  \end{array}\right), \left(\begin{array}{ccc}
    y_1 &  & \\
    & \ddots & \\
    &  & y_q
  \end{array}\right) \right), \hspace{2em} 0 < |x_i |, \, |y_j | < 1. \]
Since the eigenvalues of $( g_1, g_2 )$ on $\tmop{Mat}_{p \times
q} (\mathbb{C})$ are the $p q$ quantities $x_i y_j$,
\begin{eqnarray*}
\prod_{i = 1}^p \prod_{j = 1}^q ( 1 - x_i y_j )^{- 1} & = &
\sum_{k =
0}^{\infty} h_k ( x_1 y_1, \cdots, x_p y_q )\\
& = & \tmop{tr} \left( ( g_1, g_2 ) \, | \, \bigvee \tmop{Mat}_{p
\times q} (\mathbb{C}) \right) .
\end{eqnarray*}
Hence the Cauchy identity amounts to the statement that as
$\tmop{GL} ( p,\mathbb{C}) \times \tmop{GL} ( q,\mathbb{C})$
modules,
\begin{equation}
\label{ratioscauchyconcrete} \bigvee \tmop{Mat}_{p \times q}
(\mathbb{C}) \cong \bigoplus_{\lambda} \chi_{\lambda}^{( p )} (
g_1 ) \otimes \chi_{\lambda}^{( q )} ( g_2 ),
\end{equation}
where the sum is over all partitions $\lambda$ of length
$\leqslant \min ( p, q )$. This concrete interpretation is the
basis of the proofs in Bump~{\cite{DB}}, Chapter 43 and
Howe~{\cite{HowePerspectives}}.

The dual Cauchy identity describes the decomposition of the
exterior algebra over $\tmop{Mat}_{p \times q} (\mathbb{C})$,
which (unlike the symmetric algebra) is finite-dimensional; it is
discussed from the point of view of dual pairs in
Howe~{\cite{HowePerspectives}}.

We now turn to the discussion of \eqref{e_lr} and
\eqref{explicitbranching}  from the point of view of the theory of
dual pairs. Equation \eqref{e_lr} describes  the decomposition
rule for tensor products of representations of $\tmop{GL} (
n,\mathbb{C})$, or equivalently, the compact subgroup $U ( n )$ of
$\tmop{GL} ( n,\mathbb{C})$:
\[ \chi_{\lambda}^{( n )} \chi_{\mu}^{( n )} = \sum_{\nu} c_{\lambda
  \mu}^{\nu} \chi_{\nu}^{( n )}, \hspace{2em} \tmop{or} \hspace{2em}
  \pi_{\lambda} \otimes \pi_{\mu} = \bigoplus_{\nu} c_{\lambda \mu}^{\nu}
  \pi_{\nu} . \]

Formula \eqref{explicitbranching} is a reflection of the fact that
Littlewood-Richardson coefficients appear in a different context,
namely in the branching rule from $U ( p + q,\mathbb{C})$ to the
subgroup $U ( p ) \times U ( q )$.

Given completely reducible representations, $V$ and $W$ of groups
$G$ and $H$ respectively, together with an embedding $H
\hookrightarrow G$, we let $[ V, W ] = \dim \text{Hom}_H \left( W,
V \right)$, where $V$ is regarded as a representation of $H$ by
restriction. If $W$ is irreducible, then $[ V, W ]$ is the
multiplicity of $W$ in $V$. A description of the numbers $[ V, W
]$ is referred in the mathematics and physics literature as a
\textit{branching rule}. We refer to Howe, Tan and Willenbring
{\cite{HoweTanWillenbring}} for discussion of the problem of
obtaining branching rules from the point of view of dual reductive
pairs, and to King~{\cite{KingBranching}} for an extremely useful
survey of known branching rules.

Now \eqref{explicitbranching} follows from the following result by
evaluating $\chi_{\lambda}$ on a matrix of $U ( p + q )$ with
eigenvalues $x_1, \cdots, x_p$ in $U ( p )$ and $y_1, \cdots, y_q$
in $U ( q )$.
\begin{theorem}
\label{uuubranchingrule} We have
\begin{eqnarray}
   \chi_{\nu}^{( p + q )} |_{U ( p ) \times U ( q )} ( g_1, g_2 ) & = &
   \sum_{\lambda, \mu} c_{\lambda \mu}^{\nu} \chi_{\lambda} ( g_1 )
   \chi_{\mu} ( g_2 ), \nonumber\\
   \pi_{\nu}^{( p + q )} ( g_1, g_2 ) & = & \bigoplus_{\lambda, \mu}
   c_{\lambda \mu}^{\nu} \pi_{\lambda} ( g_1 ) \otimes \pi_{\mu} ( g_2 ) .
   \label{ratiosupqbranching}
\end{eqnarray}
\end{theorem}

{\noindent}Whippman~{\cite{whippman}} attributes this fact to
Coleman and (independently) Robinson. See also
King~{\cite{KingBranching}}. The following proof is essentially
the same as the one in Howe, Tan and
Willenbring~{\cite{HoweTanWillenbring}}.

\medbreak
\begin{proof}  Let $\Omega$ be a group and let $\omega$ be a representation of
$\Omega$. Let $G_1$ be a subgroup of $\Omega$, and let $H_2$ be
its centralizer. We assume that $G_1$ is the centralizer of $H_2$.
We allow the possibility that $G_1$ and $H_2$ have nontrivial
intersection. Even so, $( g, h ) \longrightarrow g h$ is a
homomorphism $G_1 \times H_2 \longrightarrow \Omega$, and so
$\omega$ gives rise to a representation of $G \times H$. We are
interested in the case where this restriction is a correspondence,
so we may write
\[ \omega |_{G_1 \times H_2} = \bigoplus_{i \in I} \pi_i^{( 1 )} \otimes
    \sigma_i^{( 2 )} \]
where $\pi_i^{( 1 )}$ and $\sigma_i^{( 2 )}$ are irreducible
representations of $G_1$ and $H_2$, respectively, and $\pi_i^{( 1
)} \longleftrightarrow \sigma_j^{( 2 )}$ is the graph of the
correspondence.

Now suppose that $H_1$ is a subgroup of $G_1$. Thus the
centralizer $G_2$ of $H_1$ contains $H_2$, and we assume that
$H_1$ is the centralizer of $G_2$. We assume that $\omega |_{H_1
\times G_2}$ is also a correspondence, so we may write
\[ \omega |_{H_1 \times G_2} = \bigoplus_{j \in J} \sigma_j^{( 1 )} \otimes
    \pi_j^{( 2 )}, \]
where $\sigma_j^{( 1 )}$ and $\pi_j^{( 2 )}$ are irreducible
representations of $H_1$ and $G_2$, respectively. We thus have a
{\tmem{seesaw}}:

{\hspace*{\fill}} \epsfig{file=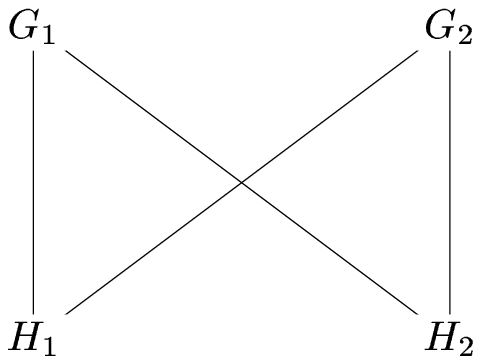} {\hspace*{\fill}}

In this diagram, the vertical lines are inclusions, and the
diagonal are the dual pairs of the two correspondences.

\begin{lemma}
   \label{ratioshoweduality}Let $G_1 \supset H_1$ and $G_2 \supset H_2$ be as
   above. If $i \in I$ then only representations from the set $\{ \sigma_j^{(
   1 )} \, | \, j \in J \}$ occur in the restriction of $\pi_i^{( 1 )}$ to
   $H_1$, so we have the ``branching rule''
   \begin{equation}
     \label{ratiosleftbranch} \pi_i^{( 1 )} = \sum_{j \in J} c_{i j}
     \sigma_j^{( 1 )}
   \end{equation}
   for some multiplicities $c_{i j}$. We have also for $j \in J$ the
   branching rule
   \[ \pi_j^{( 2 )} = \sum_{i \in I} c_{j i} \sigma_i^{( 2 )} . \]
\end{lemma}

\begin{proof}
   Any representation in the restriction of $\pi_i^{( 1 )}$ to $H_1$ occurs
   in the restriction of $\omega$ to $H_1$, so it is among the $\sigma_j^{( 1
   )}$. Thus we may write (\ref{ratiosleftbranch}). Similarly we may write
   \[ \pi_j^{( 2 )} = \sum_{i \in I} c_{j i}' \sigma_i^{( 2 )} \]
   for some multiplicities $c_{j i}'$, and is just a matter of showing that
   the $c_{i j} = c_{i j}'$. We restrict $\omega$ to $H_1 \times H_2$ and get
   isomorphisms of $H_1 \times H_2$-modules:
   \[ \bigoplus_{i, j} c_{i j} \sigma_j^{( 1 )} \otimes \sigma_i^{( 2 )}
      \cong \bigoplus_i \pi_i^{( 1 )} \otimes \sigma_i^{( 1 )} \cong \omega
      \cong \bigoplus_j \sigma_j^{( 1 )} \otimes \pi_j^{( 2 )} \cong
      \bigoplus_{i, j} c_{i j}' \sigma_j^{( 1 )} \otimes \sigma_i^{( 2 )} \]
   and the statement follows.
\end{proof}

We may now complete the proof of Theorem~\ref{uuubranchingrule}.
We will exhibit a see-saw:

{\hspace*{\fill}} \epsfig{file=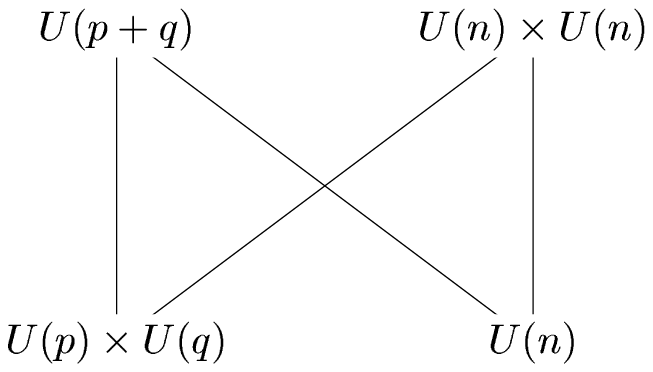} {\hspace*{\fill}}

On the right side, $U ( n )$ is embedded diagonally in $U ( n )
\times U ( n )$. The representation $\omega$ is the action of $U (
( p + q ) n )$ on the symmetric algebra on $\tmop{Mat}_{( p + q )
\times n} (\mathbb{C})$. The actions are as follows. Let
\[ X = \left(\begin{array}{c}
      X_1\\
      X_2
    \end{array}\right) \in \tmop{Mat}_{( p + q ) \times n} (\mathbb{C}),
    \hspace{2em} X_1 \in \tmop{Mat}_{p \times n} (\mathbb{C}), \; X_2 \in
    \tmop{Mat}_{q \times n} (\mathbb{C}) . \]
The action of $U ( p + q )$ is by left multiplication, and the
action of $U ( n ) \times U ( n )$  is by right multiplication on
$X_1$ and $X_2$ individually. The centralizer of $U ( p + q )$ is
the diagonal subgroup $U ( n )$ of $U ( n ) \times U ( n )$ and
the centralizer of $U ( n ) \times U ( n )$ is the subgroup $U ( p
) \times U ( q )$ of $U ( p + q )$. As we have already explained
(\ref{ratioscauchyconcrete}) shows that the restriction of
$\omega$ to either of these dual pairs is a correspondence, and
the graphs of the correspondences are $\pi_{\nu}^{( p + q )}
\longleftrightarrow \pi_{\nu}^{( n )}$ for the dual pair $( U ( p
+ q ), U ( n ) )$, and $\pi_{\lambda}^{( p )} \otimes \pi_{\mu}^{(
q )} \longleftrightarrow \pi_{\lambda}^{( n )} \otimes
\pi_{\mu}^{( n )}$ for the dual pair $( U ( p ) \times U ( q ), U
( n ) \times U ( n ) )$. The statement now follows from
Lemma~\ref{ratioshoweduality}.{\hspace*{\fill}}
\end{proof}

\section{Unitary group} \label{sec:3}
\subsection{Products}
The goal of this section is to give a simple proof of Proposition
\ref{cpaverage}, first derived by  Conrey, Farmer, Keating,
Rubinstein, and  Snaith  in \cite{CFKRS2}, and of Corollary
\ref{ratioskeatingsnaith}, first derived by Keating and
Snaith~{\cite{KS00a}}.

We always normalize the Haar measure on a compact group so the
total volume is~$1$. All integrations will be with respect to Haar
measure. We will also sometimes denote by $\mathbb{E}_G f$ the
integral of a function $f$ over the group, its expected value with
respect to the Haar probability measure.

\begin{proposition}
\label{cpaverage}Let $\alpha_1, \cdots, \alpha_{K+L}$ be complex
numbers and let $\Xi_{L, K}$ consist of all permutations $\sigma
\in S_{K + L}$ described in \eqref{perms}.

  Then
\begin{equation}\label{e:produ}
\begin{aligned}
& \int_{U ( N )}  \left\{ \prod_{l = 1}^L \det ( I + \alpha_l^{-
1} \cdot
   g^{- 1} ) \cdot \prod_{k = 1}^K \det ( I + \alpha_{L + k}g)\right\}  \hspace{0.25em} dg
   =\\
&  \frac{s_{\left\langle N^L \right\rangle}( \alpha_1, \cdots,
\alpha_{K+L})}{\prod_{l=1}^{L} (\alpha_l)^{N}}=
\\
   & \sum_{\sigma \in \Xi_{L, K}} \frac{\prod_{k = 1}^K ( \alpha_{\sigma ( L +
   k )}^{- 1} \alpha_{L + k} )^N } {\prod_{k = 1}^K \prod_{l = 1}^L ( 1 -
   \alpha_{\sigma ( l )}^{- 1} \alpha_{\sigma ( L + k )} )}  .
\end{aligned}
\end{equation}
\end{proposition}

\begin{proof}
We have
\begin{eqnarray*}
   &  & \int_{U ( N )}  \left\{ \prod_{l = 1}^L \det ( I + \alpha_l^{-
1} \cdot
   g^{- 1} ) \cdot \prod_{k = 1}^K \det ( I + \alpha_{L + k}g)\right\}  dg =\\
   &  & \prod_{l = 1}^L \alpha_l^{-N}
   \int_{U ( N )}\prod_{k = 1}^{K + L} \det ( I + \alpha_k g ) \overline{\det ( g )^L}
   \, d g.
\end{eqnarray*}
By the dual Cauchy identity, if $t_1, \cdots, t_N$ are the
eigenvalues of $g$,
\[ \prod_{k = 1}^{K + L} \det ( I + \alpha_k g ) = \sum_{\lambda}
s_{\lambda}( \alpha_1, \cdots, \alpha_{K+L}) \, s_{\lambda'} (
t_1, \cdots, t_N ), \] where $\lambda$ runs through all
partitions.

  Since
\[ \det ( g )^L = s_{\lambda'} ( t_1, \cdots, t_N) \]
where $\lambda = \left\langle N^L\right\rangle$ and $\lambda' =
\left\langle L^N \right\rangle$. Thus integrating over $g$ gives
just this term.  This proves the first line of \eqref{e:produ}.
Now the second line follows by an application of Lemma
\ref{lem:min1}.
\end{proof}

\begin{remark}
In view of discussion in Section \ref{sec:hd} , we may
characterize the preceding proof as using the dual pair $U ( N ),
U ( k )$ to transfer the computation from $U ( N )$ to $U ( k )$.
This is fully analogous to the method used (for example) in
Diaconis and Shahshahani {\cite{DS94}} of using Frobenius-Schur
duality to transfer computations from the unitary group to the
symmetric group.
\end{remark}

Keating and Snaith~{\cite{KS00a}} (see also Baker and Forrester
\cite{BF97}) proved Corollary~\ref{ratioskeatingsnaith} below
(without restriction that $k$ be an integer) using the Selberg
integral (see Forrester \cite{Fo02} for a comprehensive and
insightful discussion of the Selberg integral and its
generalizations.) This result was important because of its
relationship to conjectures of Conrey and Ghosh \cite{CG98},
Conrey and Gonek \cite{CoGo01}, and Keating and
Snaith~{\cite{KS00a}}  for the moments of the Riemann zeta
function, due to which a dramatic new aspect of predictive power
of random matrix theory for the zeta function was established .

\begin{corollary}
\label{ratioskeatingsnaith} We have
\begin{equation}
   \label{ksformula} \int_{U ( n )} | \det ( g - I ) |^{2 k} \, d g =
   \prod_{j = 0}^{n - 1} \frac{j! ( j + 2 k ) !}{( j + k ) !^2} .
\end{equation}
\end{corollary}

\begin{proof}
Propositions \ref{cpaverage} applied with $\alpha_i=1$ and $L=K=k$
implies
$$\int_{U ( n )} | \det ( g - I ) |^{2 k} \, d g =s_{\left\langle N^k
\right\rangle}(1^{2k}).$$

We now apply \eqref{e_hhl}.  It is easy to see, that for a
partition $\lambda=N^k$ the product of hook numbers is given by
\begin{equation} \label{hook123}
\prod_{j=0}^{N-1}\frac{(j+k)!}{j!}, \end{equation} whereas the
product $\prod_{u \in \lambda}(2k+c(u))$ is given by
\[\prod_{i=1}^{k}\prod_{j=1}^{N}(2k-i+j)
=\prod_{j=1}^{N-1}\frac{(j+2k)!}{(j+k)!};\] the result follows.
\end{proof}

\subsection{Littlewood-Schur symmetric functions}
While the proof of the result for ratios in the unitary case can
be given using Schur functions only, it becomes more neat and
transparent if we use the following generalization of Schur
functions. These are called {\tmem{hook Schur functions}} by
Berele and Regev {\cite{BR87}} who denoted by them
$\tmop{HS}_{\lambda} (x; y)$. Macdonald~\cite[p.27, ex. 5; p.45,
ex. 3]{Mac})  denotes them $s_{\lambda} ( x / y )$.   These
functions were considered earlier by Littlewood~{\cite{Li1}},
pp.~66-70 in 1936 (see also Littlewood~{\cite{Li}} p.~114--118,
the section entitled ``Extension to rational fractions of the
formula for the S-function as a quotient of determinants'').
Littlewood  gave a formula for the $\tmop{LS}_{\lambda}$ when
$\lambda = \left\langle p^q \right\rangle$, which is important for
us (a particular case of his formula is given in Proposition
\ref{littlewoodsquarecase} below). We have not seen reference to
this work of Littlewood in any of the post 1980-s papers. Perhaps
an appropriate name for these functions is {\tmem{Littlewood-Schur
symmetric functions}}, so we will denote
them~$\tmop{LS}_{\lambda}$.

Let $x_1, \cdots, x_k$ and $y_1, \cdots, y_l$ be two sets of
variables. Define
\[ \tmop{LS}_{\lambda} ( x_1, \cdots, x_k ; \, y_1, \cdots, y_l ) = \sum_{\mu,
  \nu} c^{\lambda}_{\mu \nu} s_{\mu} ( x_1, \cdots, x_k ) s_{\nu'} ( \, y_1,
  \cdots, y_l ) . \]

Since $\iota$ is an automorphism of $\Lambda$ we have $c_{\mu
\nu}^{\lambda} = c_{\mu' \nu'}^{\lambda'}$ and so
\begin{equation}
\label{interchange} \tmop{LS}_{\lambda} ( x_1, \cdots, x_k ; \,
y_1, \cdots, y_l ) = \tmop{LS}_{\lambda'} ( y_1, \cdots, y_l ;
x_1, \cdots, x_k ) .
\end{equation}

With possible exception of Proposition \ref{lem:min2} the results
in this section are not new, though the proofs may be. We start
with the following generalization of Cauchy identity, due to
Berele and Remmel \cite{BR85}, who also gave a bijective proof.

\begin{proposition}
\label{generalizedcauchy}Let $\alpha_1, \cdots, \alpha_m$,
$\beta_1, \cdots, \beta_n$, $\gamma_1, \cdots, \gamma_s$ and
$\delta_1, \cdots, \delta_t$ be four sets of variables. We have
\begin{eqnarray}
   \sum \tmop{LS}_{\lambda} ( \alpha_1, \cdots, \alpha_m ; \beta_1, \cdots,
   \beta_n ) \, \tmop{LS}_{\lambda} ( \gamma_1, \cdots, \gamma_s ; \delta_1,
   \cdots, \delta_t ) & = &  \nonumber\\\qquad\qquad
   \prod_{i, k} ( 1 - \alpha_i \gamma_k )^{- 1} \prod_{i, l} ( 1 + \alpha_i
   \delta_l ) \prod_{j, k} ( 1 + \beta_j \gamma_k ) \prod_{j, l} ( 1 -
   \beta_j \delta_l )^{- 1} . &  &  \label{gencauchyformula}
\end{eqnarray}
\end{proposition}

\begin{proof}
By (\ref{explicitbranching}) and the Cauchy identity we have
\begin{eqnarray*}
   \sum_{\lambda} \sum_{\mu, \nu} c_{\mu \nu}^{\lambda} s_{\mu} ( \alpha_1,
   \cdots, \alpha_m ) s_{\nu} ( \beta_1, \cdots, \beta_n ) \sum_{\sigma,
   \tau} c_{\sigma \tau}^{\lambda} s_{\sigma} ( \gamma_1, \cdots, \gamma_s )
   s_{\tau} ( \delta_1, \cdots, \delta_t ) & = \\
   \sum_{\lambda} s_{\lambda} ( \alpha_1, \cdots, \alpha_m, \beta_1, \cdots,
   \beta_n ) \, s_{\lambda} ( \gamma_1, \cdots, \gamma_s, \delta_1, \cdots,
   \delta_t ) & = & \\
   \prod_{i, k} ( 1 - \alpha_i \gamma_k )^{- 1} \prod_{i, l} ( 1 - \alpha_i
   \delta_l )^{- 1} \prod_{j, k} ( 1 - \beta_j \gamma_k )^{- 1} \prod_{j, l}
   ( 1 - \beta_j \delta_l )^{- 1} . & &
\end{eqnarray*}
Formally, the identity follows on applying the involution $\iota$
in the variables $\beta$ and $\delta$. To make a rigorous proof
from this idea, we may proceed as follows. Rewrite the last
identity
\begin{eqnarray*}
   \sum_{\lambda} \sum_{\mu, \nu} c_{\mu \nu}^{\lambda} s_{\mu} ( \alpha_1,
   \cdots, \alpha_m ) s_{\nu} ( \beta_1, \cdots, \beta_n ) \sum_{\sigma,
   \tau} c_{\sigma \tau}^{\lambda} s_{\sigma} ( \gamma_1, \cdots, \gamma_s )
   s_{\tau} ( \delta_1, \cdots, \delta_t ) & = & \\
   \prod_{i, k} ( 1 - \alpha_i \gamma_k )^{- 1} \prod_{i = 1}^m \left(
   \sum_{N = 0}^{\infty} \alpha_i^N h_N ( \delta_1, \cdots, \delta_t )
   \right) &  & \\
   \times \prod_{j, k} ( 1 - \beta_i \gamma_k )^{- 1} \prod_{j = 1}^n \left(
   \sum_{N = 0}^{\infty} \beta_j^N h_N ( \delta_1, \cdots, \delta_t ) \right)
   . &  &
\end{eqnarray*}
Since this is true for all $t$ we may write
\begin{eqnarray*}
   \sum_{\lambda} \sum_{\mu, \nu} c_{\mu \nu}^{\lambda} s_{\mu} ( \alpha_1,
   \cdots, \alpha_m ) s_{\nu} ( \beta_1, \cdots, \beta_n ) \sum_{\sigma,
   \tau} c_{\sigma \tau}^{\lambda} s_{\sigma} ( \gamma_1, \cdots, \gamma_s )
   s_{\tau} & = & \\
   \prod_{i, k} ( 1 - \alpha_i \gamma_k )^{- 1} \prod_{i = 1}^m \left(
   \sum_{N = 0}^{\infty} \alpha_i^N h_N \right) &  & \\
   \times \prod_{j, k} ( 1 - \beta_i \gamma_k )^{- 1} \prod_{j = 1}^n \left(
   \sum_{N = 0}^{\infty} \beta_j^N h_N \right), &  &
\end{eqnarray*}
where now $s_{\tau}$ and $h_N$ are regarded as elements of the
ring $\Lambda$ which, we recall, is the inverse limit of the
$\Lambda^{( N )}$. Applying the involution, which replaces
$s_{\tau}$ by $s_{\tau'}$ and $h_N$ by $e_N$, then specializing
$s_{\tau} \longrightarrow s_{\tau} ( \delta_1, \cdots, \delta_t )$
gives
\begin{eqnarray*}
   \sum_{\lambda} \sum_{\mu, \nu} c_{\mu \nu}^{\lambda} s_{\mu} ( \alpha_1,
   \cdots, \alpha_m ) s_{\nu} ( \beta_1, \cdots, \beta_n )
   \tmop{LS}_{\lambda} ( \gamma_1, \cdots, \gamma_s ; \delta_1, \cdots,
   \delta_t ) & = & \\
   \prod_{i, k} ( 1 - \alpha_i \gamma_k )^{- 1} \prod_{i, l} ( 1 + \alpha_i
   \delta_l ) \prod_{j, k} ( 1 - \beta_i \gamma_k )^{- 1} \prod_{j, l} ( 1 +
   \beta_j \delta_l ) . &  &
\end{eqnarray*}
Now applying the same process again in the $\beta_i$ gives the
result.
\end{proof}

\begin{proposition}
\label{genlwr}We have
\[ \tmop{LS}_{\lambda} ( x_1, \cdots, x_{p + q} ; y_1, \cdots, y_l ) =
    \sum_{\mu, \nu} c^{\lambda}_{\mu \nu} \tmop{LS}_{\mu} ( x_1, \cdots, x_p
    ; y_1, \cdots, y_l ) s_{\nu} ( x_{p + 1}, \cdots, x_{p + q} ) . \]
\end{proposition}

\begin{proof}
By the definition of $\tmop{LS}_{\lambda}$ and
(\ref{explicitbranching}), we have
\begin{eqnarray*}
   &  & \tmop{LS}_{\lambda} ( x_1, \cdots, x_{p + q} ; y_1, \cdots, y_l )\\
   & = & \sum_{\theta, \phi} c^{\lambda}_{\theta \phi} s_{\theta} ( x_1,
   \cdots, x_{p + q} ) s_{\phi'} ( y_1, \cdots, y_l )\\
   & = & \sum_{\theta, \phi, \psi, \nu} c^{\lambda}_{\theta \phi}
   c^{\theta}_{\psi \nu} s_{\psi} ( x_1, \cdots, x_p ) s_{\phi'} ( y_1,
   \cdots, y_l ) s_{\nu} ( x_{p + 1}, \cdots, x_{p + q} )\\
   & = & \sum_{\mu, \phi, \psi, \nu} c^{\lambda}_{\mu \nu} c^{\mu}_{\psi
   \phi} s_{\psi} ( x_1, \cdots, x_p ) s_{\phi'} ( y_1, \cdots, y_l ) s_{\nu}
   ( x_{p + 1}, \cdots, x_{p + q} ),
\end{eqnarray*}
where we have used the identity
\[ \sum_{\theta} c^{\lambda}_{\theta \phi} c^{\theta}_{\psi \nu} =
    \sum_{\mu} c^{\lambda}_{\mu \nu} c^{\mu}_{\psi \phi}, \]
which is a reflection of the fact that $\Lambda$ is a commutative
ring. (Both sides equal $\left\langle \chi_{\lambda}, \chi_{\phi}
\chi_{\psi} \chi_{\nu} \right\rangle$ where the product is taken
over $U ( n )$ for any sufficiently large $n$.) The statement
follows.
\end{proof}

A special case of this is a generalization of Pieri's formula.
\begin{proposition}
\label{genpieri}We have
\begin{eqnarray*}
   & & \tmop{LS}_{\lambda} ( x_1, \cdots, x_k ; y_1, \cdots, y_l )
   \\& = &
   \sum_{\tmscript{\begin{array}{c}
     \mu \subseteq \lambda\\
     \text{$\lambda - \mu$ a horizontal strip}
   \end{array}}} \tmop{LS}_{\lambda} ( x_1, \cdots, x_{k - 1} ; y_1, \cdots,
   y_l ) \, x_k^{| \lambda | - | \mu |}\\
   &  & \\
   & = & \sum_{\tmscript{\begin{array}{c}
     \mu \subseteq \lambda\\
     \text{$\lambda - \mu$ a vertical strip}
   \end{array}}} \tmop{LS}_{\lambda} ( x_1, \cdots, x_k ; y_1, \cdots, y_{l -
   1} ) \, y_k^{| \lambda | - | \mu |} .
\end{eqnarray*}
\end{proposition}

\begin{proof}
The second formula follows from the first on applying
(\ref{interchange}). We prove the first. If $l = 0$, this is
Pieri's formula \eqref{e:pieri}. Applying this fact in
Proposition~\ref{genlwr} (with $p = k - 1$ and $q = 1$) gives the
result.
\end{proof}

We also have the following generalization of Lemma \ref{lem:min1}:

\begin{proposition}
\label{lem:min2}Suppose $\lambda$ of length $\leqslant K$ such
that $\lambda_L \geqslant \lambda_{L + 1} + Q$, let $\lambda =
\tau \cup \rho$ with
\[ \tau = ( \lambda_1, \ldots, \lambda_L ), \hspace{2em} \rho = (
    \lambda_{L + 1}, \cdots, \lambda_{L + K} ) . \]
Then
\begin{eqnarray}
   &&\tmop{LS}_{\lambda} ( \alpha_1, \cdots, \alpha_{L + K} ; \gamma_1, \ldots,
   \gamma_Q ) \nonumber \\ & = &\sum_{\sigma \in \Xi_{L, K}}
   \prod_{\tmscript{\begin{array}{c}
     1 \leqslant l \leqslant L\\
     1 \leqslant k \leqslant K
   \end{array}}} ( \alpha_{\sigma ( l )} - \alpha_{\sigma ( L + k )} )^{- 1}
   \nonumber\\&
   \times &\tmop{LS}_{\tau + \left\langle K^L \right\rangle} ( \alpha_{\sigma
   ( 1 )}, \cdots, \alpha_{\sigma ( L )} ; \gamma_1, \ldots, \gamma_Q )
   \nonumber\\&\times&
   \tmop{LS}_{\rho} ( \alpha_{\sigma ( L + 1 )}, \cdots, \alpha_{\sigma ( L +
   K )} ; \gamma_1, \ldots, \gamma_Q ) \label{lsminors}
\end{eqnarray}
\end{proposition}

\begin{proof}
We prove this by induction on $Q$. If $Q = 0$ this is
Lemma~\ref{lem:min1}. We assume that $Q > 0$ and that the
statement is true for $Q - 1$. We enumerate the partitions $\mu
\subseteq \lambda$ such that $\lambda - \mu$ is a vertical strip
as follows. Let $\tau_1 \subseteq \tau$ be a partition such that
$\tau - \tau_1$ is a vertical strip, and let $\subseteq$ be a
partition such that $\rho - \rho_1$ is a vertical strip. Then we
let $\mu = \tau_1 \cup \rho_1$ where
\[ \tau_1 = ( \mu_1, \cdots, \mu_L ) \hspace{2em} \rho_1 = ( \mu_{L + 1},
    \cdots \mu_{L + K} ) . \]
Since $\lambda_L - \lambda_{L + 1} \geqslant Q$ we have $\mu_L -
\mu_{L + 1} \geqslant Q - 1$, so $\mu$ is a partition, and the
induction hypothesis is also satisfied. By
Proposition~\ref{genpieri} we have
\begin{eqnarray*}
   &  & \tmop{LS}_{\lambda} ( \alpha_1, \cdots, \alpha_{L + K} ; \gamma_1,
   \ldots, \gamma_Q )\\
   & = & \sum_{\tmscript{\begin{array}{c}
     \tau_1 \subseteq \tau\\
     \rho_1 \subseteq \rho\\
     \tau - \tau_1, \rho - \rho_1\\
     \text{vertical strips}
   \end{array}}} \tmop{LS}_{\mu} ( \alpha_1, \cdots, \alpha_{L + K} ;
   \gamma_1, \ldots, \gamma_{Q - 1} ) \, \gamma_Q^{| \lambda | - | \mu |}\\
   & = & \sum_{\tmscript{\begin{array}{c}
     \tau_1 \subseteq \tau\\
     \rho_1 \subseteq \rho\\
     \tau - \tau_1, \rho - \rho_1\\
     \text{vertical strips}
   \end{array}}} \sum_{\sigma \in \Xi_{L, K}}
   \prod_{\tmscript{\begin{array}{c}
     1 \leqslant l \leqslant L\\
     1 \leqslant k \leqslant K
   \end{array}}} ( \alpha_{\sigma ( l )} - \alpha_{\sigma ( L + k )} )^{-
   1}\\
   & \times & \tmop{LS}_{\tau_1 + \left\langle K^L \right\rangle} (
   \alpha_{\sigma ( 1 )}, \cdots, \alpha_{\sigma ( L )} ; \gamma_1, \ldots,
   \gamma_{Q - 1} ) \gamma_Q^{| \tau | - | \tau_1 |}\\
   &  & \tmop{LS}_{\rho_1} ( \alpha_{\sigma ( L + 1 )}, \cdots,
   \alpha_{\sigma ( L + K )} ; \gamma_1, \ldots, \gamma_{Q - 1} ) \gamma_Q^{|
   \rho | - | \rho_1 |} .
\end{eqnarray*}
Applying Proposition~\ref{genpieri} again, the statement follows.
\end{proof}

\begin{proposition}
\label{littlewoodsquarecase} {\dueto{Littlewood}}We have
\[ \tmop{LS}_{\left\langle ( l + m )^k \right\rangle} ( x_1, \cdots, x_k ;
    y_1, \cdots, y_l ) = \left( \prod_{i = 1}^k x_i \right)^m
    \prod_{\tmscript{\begin{array}{c}
      1 \leqslant i \leqslant k\\
      1 \leqslant j \leqslant l
    \end{array}}} ( x_i + y_j ) . \]
\end{proposition}

\begin{proof}
We claim that if $N > k$ and $\mu \subseteq N^k$ then
\[ c^{\left\langle N^k \right\rangle}_{\mu \nu} = \left\{\begin{array}{ll}
      1 & \text{if $\nu = ( N - \mu_k, N - \mu_{k - 1}, \cdots, N - \mu_1 )$}
      ;\\
      0 & \text{otherwise.}
    \end{array}\right. \]
Indeed, let $\chi_{\mu}$, $\chi_{\nu}$ and $\chi_{\left\langle N^k
\right\rangle}$ be the corresponding associated characters of $U (
k )$. We have $\chi_{\left\langle N^k \right\rangle} = {\det}^N$.
The Littlewood-Richardson coefficient
\[ c^{\left\langle N^k \right\rangle}_{\mu \nu} = \left\langle \chi_{\mu}
    \chi_{\nu}, {\det}^N \right\rangle = \left\langle \chi_{\nu}, {\det}^N
    \otimes \overline{\chi_{\mu}} \right\rangle . \]
Now ${\det}^N \otimes \overline{\chi_{\mu}}$ is the character of
an irreducible representation; if $x_1, \cdots, x_k$ are the
eigenvalues of $g$, its value at $g$ is $s_{\nu_0} ( x_1, \cdots,
x_k )$, where
\[ \nu_0 = ( N - \mu_k, N - \mu_{k - 1}, \cdots, N - \mu_1 ) . \]
This may be seen directly from the definition of the Schur
polynomial $s_{\mu}$. The statement follows.

Now taking $N = l + m$,
\[ \tmop{LS}_{\left\langle ( l + m )^k \right\rangle} ( x_1, \cdots, x_k ;
    y_1, \cdots, y_l ) = \sum_{\mu} s_{\mu} ( x_1, \cdots, x_k ) s_{\nu'} (
    y_1, \cdots, y_l ), \]
where we sum over all $\mu \in N^k$ and $\nu = ( N - \mu_k, N -
\mu_{k - 1}, \cdots, N - \mu_1 )$. We may restrict ourselves to
$\mu$ for which $s_{\nu'} \neq 0$. This means that $\nu_1 = l (
\nu' ) \leqslant l$, which implies that $\mu_k \geqslant m$. Thus
$\left\langle m^k \right\rangle \subseteq \mu \subseteq
\left\langle N^k \right\rangle$, and we may write $\mu =
\left\langle m^k \right\rangle + \tilde{\mu}$ where now
$\tilde{\mu} \subseteq \left\langle l^k \right\rangle$,
$\tilde{\mu}' \subseteq \left\langle k^l \right\rangle$ and
\[ \nu = ( l - \tilde{\mu}_k, \cdots, l - \tilde{\mu}_1 ), \hspace{2em} \nu'
    = ( k - \tilde{\mu}'_l, \cdots, k - \tilde{\mu}'_1 ) . \]
We have
\[ s_{\nu'} ( y_1, \cdots, y_l ) = ( y_1 \cdots y_l )^k s_{\tilde{\mu}'} (
    y_1^{- 1}, \cdots, y_l^{- 1} ), \]
so by the dual Cauchy identity
\begin{eqnarray*}
   &  & \tmop{LS}_{\left\langle ( l + m )^k \right\rangle} ( x_1, \cdots,
   x_k ; y_1, \cdots, y_l )\\
   & = & \left( \prod_{i = 1}^k x_i \right)^m ( y_1 \cdots y_l )^k \sum
   s_{\mu} ( x_1, \cdots, x_k ) s_{\tilde{\mu}'} ( y_1^{- 1}, \cdots, y_l^{-
   1} )\\
   & = & \left( \prod_{i = 1}^k x_i \right)^m ( y_1 \cdots y_l )^k
   \prod_{\tmscript{\begin{array}{c}
     1 \leqslant i \leqslant k\\
     1 \leqslant j \leqslant l
   \end{array}}} ( 1 + x_i y_j^{- 1} ),
\end{eqnarray*}
and the statement follows.
\end{proof}

\begin{remark}
A substantial generalization of Proposition
\ref{littlewoodsquarecase} is furnished by the following result,
known as Sergeev-Pragacz Formula (see Macdonald \cite[p. 60 ex.
24]{Mac}). To describe it, let $\lambda$ be a partition with
$\lambda_{k+1}\leqslant l$.  Let $\mu$ be the part of $\lambda$
that falls within the $k\times l$ rectangle, let $\nu$ and $\eta$
be the remaining parts to the right and underneath this rectangle,
that is $\lambda=(\mu+\nu)\cup\eta$.  For example, if $\lambda =
(5, 5, 3, 2)$, $k=2$, $l=3$ we have $\mu=(3,3)$, $\nu=(2,2)$ and
$\eta=(2, 2,1)$.

\begin{figure}[!h]
\abovedisplayskip-.5\baselineskip \belowdisplayskip-\baselineskip
\begin{equation*}
\beginpicture \setcoordinatesystem units
<0.5cm,0.5cm>         \setplotarea x from 0 to 4, y from 1 to 3
\linethickness =0.5pt                          \putrule from 0 6
to 5 6
         \putrule from 0 5 to 5 5          \putrule from 0 4 to 5 4
     \putrule from 0 3 to 3 3          \putrule from 0 2 to 2 2
\putrule from 0 2 to 0 6        \putrule from 1 2 to 1 6 \putrule
from 2 2 to 2 6        \putrule from 3 3 to 3 6 \putrule from 4 4
to 4 6        \putrule from 5 4 to 5 6
\endpicture
\end{equation*}
\end{figure}

Then
\begin{multline} \label{e_spf}
\HS_{\lambda}(x_1, \ldots, x_k; y_1, \ldots,
y_l)=\frac{1}{\prod_{i<j}(x_i-x_j) \prod_{i<j}(y_i-y_j)}\times\\
\sum_{w \in S_k \times S_l}\varepsilon(w) w\left(x^{\nu+\delta_k}
y^{\eta' +\delta_l} \prod_{(i, j) \in \mu}(x_i  +y_j)\right),
\end{multline}

where
\[\delta_k =(k-1, k-1, \ldots, 1, 0).\]

In the special case when $\lambda$ contains $l^k$ (\ref{e_spf})
factorizes as follows:
\begin{equation} \label{e_br} \HS_{\lambda}(x_1, \ldots,
x_k; y_1, \ldots, y_l)= s_{\nu}(x_1, \ldots, x_k)s_{\eta'}(y_1,
\ldots, y_l) \prod_{i=1}^{k}\prod_{j=1}^{l}(x_i+y_j).
\end{equation}
Formula \eqref{e_br} is due to  Berele and
Regev~\cite[(6.20)]{BR87}; it will be used in Section~\ref{rect}.
\end{remark}

\begin{remark}
We conclude this section by remarking that
Proposition~\ref{generalizedcauchy} encodes an important property
of the ring $\Lambda$, namely, the fact that it is a {\tmem{Hopf
algebra}}. See Geissinger~{\cite{Geissinger}} and
Zelevinsky~{\cite{Zelevinsky}}.   We will not require this fact,
and the reader may skip it. It seems important enough to include.
The multiplication in $\Lambda$ induces a map $m : \Lambda \otimes
\Lambda \longrightarrow \Lambda$, whose adjoint with respect to
the basis for which the $s_{\lambda}$ are orthonormal is a map
$m^{\ast} : \Lambda \longrightarrow \Lambda \otimes \Lambda$.
Specifically we have
\[ m ( s_{\mu} \otimes s_{\nu} ) = \sum_{\lambda} c^{\lambda}_{\mu \nu}
  s_{\lambda}, \hspace{2em} m^{\ast} ( s_{\lambda} ) = \sum_{\mu, \nu}
  c^{\lambda}_{\mu \nu} \, s_{\mu} \otimes s_{\nu} . \]
The map $m^{\ast}$ is a comultiplication making $\Lambda$ a
coalgebra. The {\tmem{Hopf axiom}} is the commutativity of the
following diagram:
\[ \begin{array}{ccccc}
    \Lambda \otimes \Lambda & \longrightarrowlim^{m^{\ast} \otimes m^{\ast}}
    & \Lambda \otimes \Lambda \otimes \Lambda \otimes \Lambda &
    \longrightarrowlim^{1 \otimes \tau \otimes 1} & \Lambda \otimes \Lambda
    \otimes \Lambda \otimes \Lambda\\
    \downarrow {\scriptstyle m} &  &  &  & \downarrow
    {\scriptstyle m \otimes m}\\
    \Lambda &  & \longrightarrowlim^{m^{\ast}} &  & \Lambda \otimes \Lambda
  \end{array}, \]
where $\tau : R \otimes R \longrightarrow R \otimes R$ is the map
$\tau ( u \otimes v ) = v \otimes u$.

\begin{proposition}
{\dueto{Geissinger~{\cite{Geissinger}}}}The Hopf axiom is
satisfied.
\end{proposition}

\begin{proof}
The Hopf axiom reduces to the formula
\begin{equation}
   \label{hopfformula} \sum_{\lambda} c_{\mu \nu}^{\lambda}
   c^{\lambda}_{\sigma \tau} = \sum_{\varphi, \eta} c_{\varphi
   \eta}^{\sigma} c^{\tau}_{\psi \xi} c_{\varphi \xi}^{\mu} c_{\psi
   \eta}^{\nu},
\end{equation}
since if we apply $m^{\ast} \circ m$ to $s_{\mu} \otimes s_{\nu}$,
then extract the coefficient of $s_{\sigma} \otimes s_{\tau}$ we
obtain the left-hand side in (\ref{hopfformula}), while if we
perform the same computation with the map $( m \otimes m ) \circ (
1 \otimes \tau \otimes 1 ) \circ ( m^{\ast} \otimes m^{\ast} )$ we
obtain the right-hand side.

To deduce (\ref{hopfformula}) from the generalized Cauchy identity
we note that (in an obvious notation) the left-hand side of
(\ref{gencauchyformula}) equals
\[ \sum c_{\mu \nu}^{\lambda} s_{\mu} ( \alpha ) s_{\nu'} ( \beta )
    c^{\lambda}_{\sigma \tau} s_{\sigma} ( \gamma ) s_{\tau'} ( \delta ) \]
while the right-hand side equals
\begin{eqnarray*}
   &  & \sum s_{\varphi} ( \alpha ) s_{\varphi} ( \gamma ) s_{\psi'} ( \beta
   ) s_{\psi'} ( \delta ) s_{\xi} ( \alpha ) s_{\xi'} ( \delta ) s_{\eta'} (
   b ) s_{\eta} ( \gamma )_{}\\
   & = & \sum c_{\varphi \eta}^{\sigma} c^{\tau}_{\psi \xi} s_{\varphi} (
   \alpha ) s_{\xi} ( \alpha ) s_{\psi'} ( \beta ) s_{\eta'} ( \beta )
   s_{\sigma} ( \gamma ) s_{\tau'} ( \delta )\\
   & = & \sum c_{\varphi \eta}^{\sigma} c^{\tau}_{\psi \xi} c_{\varphi
   \xi}^{\mu} c_{\psi \eta}^{\nu} s_{\mu} ( \alpha ) s_{\nu'} ( \beta )
   c^{\lambda}_{\sigma \tau} s_{\sigma} ( \gamma ) s_{\tau'} ( \delta ) .
\end{eqnarray*}
Comparing, we obtain the result.
\end{proof}
\end{remark}

\subsection{Ratios}
The goal of this section is to give a simple proof of Theorem
\ref{t:1}, which was established by Conrey, Farmer and Zirnbauer
\cite{cfz} and by Conrey, Forrester and Snaith \cite{cfs}.

\begin{theorem} \label{t:1}
Let $\alpha_1, \cdots, \alpha_{K+L}$ be complex numbers; let
$\gamma_1, \dots, \gamma_Q$ and $\delta_1, \dots \delta_R$ be
complex numbers satisfying $|\gamma_q| < 1$ and $|\delta_r| < 1$.
Let $\Xi_{L, K}$ consist of all permutations $\sigma \in S_{K +
L}$ described in \eqref{perms}. If $N \geqslant Q, R$ we have
\begin{eqnarray}
   &  & \int_{U ( N )} \frac{\prod_{l = 1}^L \det ( I + \alpha_l^{- 1} \cdot
   g^{- 1} ) \cdot \prod_{k = 1}^K \det ( I + \alpha_{L + k} \cdot g
   )}{\prod_{q = 1}^Q \det ( I - \gamma_q \cdot g ) \prod_{r = 1}^R \det ( I
   - \delta_r \cdot g^{- 1} )} \hspace{0.25em} dg = \nonumber\\
   &  & \sum_{\sigma \in \Xi_{L, K}} \prod_{k = 1}^K ( \alpha_{\sigma ( L +
   k )}^{- 1} \alpha_{L + k} )^N \times \nonumber\\
   &  & \frac{\prod_{q = 1}^Q \prod_{l = 1}^L ( 1 + \gamma_q \alpha_{\sigma
   ( l )}^{- 1} ) \prod_{r = 1}^R \prod_{k = 1}^K ( 1 + \delta_r
   \alpha_{\sigma ( L + k )} )}{\prod_{k = 1}^K \prod_{l = 1}^L ( 1 -
   \alpha_{\sigma ( l )}^{- 1} \alpha_{\sigma ( L + k )} ) \prod_{r = 1}^R
   \prod_{q = 1}^Q ( 1 - \gamma_q \delta_r )} .  \label{unitaryratio}
\end{eqnarray}
\end{theorem}

\begin{proof}
By the dual Cauchy identity,
\begin{eqnarray}
   &  & \prod_{l = 1}^L \det ( I + \alpha_l^{- 1} \cdot g^{- 1} ) \cdot
   \prod_{k = 1}^K \det ( I + \alpha_{L + k} \cdot g ) \hspace{0.25em}
   \nonumber\\
   & = & \overline{\det ( g )^L} \hspace{0.75em} \prod_{l = 1}^L \alpha_l^{-
   N} \prod_{k = 1}^{K + L} \det ( I + \alpha_k g ) \hspace{0.25em}
   \nonumber\\
   & = & \overline{\det ( g )^L} \hspace{0.75em} \prod_{l = 1}^L \alpha_l^{-
   N} \sum_{\lambda} s_{\lambda} ( \alpha_1, \cdots, \alpha_{K + L} )
   \chi_{\lambda'} ( g )  \label{zirnalphaterms}
\end{eqnarray}
On the other hand by the Cauchy identity
\[ \prod_{q = 1}^Q \det ( I - \gamma_q g )^{- 1} = \sum_{\mu} s_{\mu} (
    \gamma_1, \cdots, \gamma_Q ) \hspace{0.25em} \chi_{\mu} ( g ) \]
and
\[ \prod_{r = 1}^R \det ( I - \delta_r \cdot g^{- 1} )^{- 1} = \sum_{\nu}
    s_{\nu} ( \delta_1, \cdots, \delta_R ) \hspace{0.25em}
    \overline{\chi_{\nu} ( g )} . \]
Since we are assuming that $N \geqslant Q, R$, the sums in these
identities is over all partitions $\mu$ of length $\leqslant Q$
and $\nu$ of length $\leqslant R$.

By Schur orthogonality the left hand side in (\ref{unitaryratio})
equals
\[ \sum_{\lambda, \mu, \nu} \left\langle \chi_{\lambda'} \chi_{\mu},
    {\det}^L \otimes \chi_{\nu} \right\rangle \prod_{l = 1}^L \alpha_l^{- N}
    s_{\lambda} ( \alpha_1, \cdots, \alpha_{L + K} ) s_{\mu} ( \gamma_1,
    \cdots, \gamma_Q ) s_{\nu} ( \delta_1, \cdots, \delta_R ) . \]
We rewrite this as
\begin{eqnarray*}
   \prod_{l = 1}^L \alpha_l^{- N} \sum_{\lambda, \mu, \nu}
   c^{\tilde{\nu}}_{\lambda' \mu} s_{\lambda} ( \alpha_1, \cdots, \alpha_{L +
   K} ) s_{\mu} ( \gamma_1, \cdots, \gamma_Q ) s_{\nu} ( \delta_1, \cdots,
   \delta_R ) & = & \\
   \prod_{l = 1}^L \alpha_l^{- N} \sum_{\nu} \tmop{LS}_{\tilde{\nu}} (
   \gamma_1, \ldots, \gamma_Q ; \alpha_1, \ldots, \alpha_{L + K} ) s_{\nu} (
   \delta_1, \cdots, \delta_R ) & = & \\
   \prod_{l = 1}^L \alpha_l^{- N} \sum_{\nu} \tmop{LS}_{\hat{\nu}} (
   \alpha_1, \ldots, \alpha_{L + K} ; \gamma_1, \ldots, \gamma_Q ) s_{\nu} (
   \delta_1, \cdots, \delta_R ), &  &
\end{eqnarray*}
where $\tilde{\nu} = \nu + \left\langle L^N \right\rangle$ and
$\hat{\nu} = \tilde{\nu}' = N^L \cup \nu'$. Now
Proposition~\ref{lem:min2} is applicable since $N \geqslant Q$. It
gives
\begin{equation*}
\begin{aligned}
&  \tmop{LS}_{\hat{\nu}} ( \alpha_1, \ldots, \alpha_{L + K} ;
\gamma_1,
   \ldots, \gamma_Q )  =
   \sum_{\sigma \in \Xi_{L, K}} \prod_{\tmscript{\begin{array}{c}
     1 \leqslant l \leqslant L\\
     1 \leqslant k \leqslant K
   \end{array}}} ( \alpha_{\sigma ( l )} - \alpha_{\sigma ( L + k )} )^{-
   1}\times
   \\
   & \tmop{LS}_{\left\langle ( N + K )^L \right\rangle} ( \alpha_{\sigma
   ( 1 )}, \cdots, \alpha_{\sigma ( L )} ; \gamma_1, \ldots, \gamma_Q )
   \tmop{LS}_{\nu'} ( \alpha_{\sigma ( L + 1 )}, \cdots, \alpha_{\sigma ( L +
   K )} ; \gamma_1, \ldots, \gamma_Q ).
   \end{aligned}
\end{equation*}
Substituting this expression, then using
Propositions~\ref{generalizedcauchy}
and~\ref{littlewoodsquarecase}, and the obvious fact that
\begin{equation}
   \label{zfcobvious} \prod_{l = 1}^L \alpha_l^{- N} \alpha_{\sigma ( l )}^N
   = \prod_{k = 1}^K ( \alpha_{\sigma ( L + k )}^{- 1} \alpha_{L + k} )^N,
\end{equation}
completes the proof.
\end{proof}

\begin{remark} We conclude this section by remarking that Theorem
\ref{t:1} in combination with Heine identity (see \cite{GS39} or
\cite{BD02}) easily implies a formula of Day \cite{day}. The
derivation is parallel to that in \cite[Section 3]{DG04} where the
formula due to Schmidt and Spitzer~\cite{SS60}, of which Day's
formula is a generalization, is deduced from Proposition
\ref{cpaverage}. A simple derivation of Day's formula was given by
Conrey, Forrester and Snaith \cite{cfs} using the method of Basor
and Forrester {\cite{bf94}}.
\end{remark}

\section{Common Features of the formulae} \label{comf}

This section contains a remark about the nature of the proofs of
the formulae for mean values of products and ratios of
characteristic polynomials. In each of the formulas for mean
values of products or ratios of characteristic polynomials, a sum
appears over certain Weyl group elements. For example, in
Theorem~\ref{t:1}, this is the sum over $\Xi_{L, K}$, while in
Theorem~\ref{ratsp} below, it is the sum over the $\varepsilon_i$.
In each case, the summation appears from the Laplace expansion of
a determinant, as in Lemma~\ref{lem:min1}.

Our method may be summarized in the following terms:
\begin{itemize}
  \item The quantity in question (a mean value of a product or ratio of
  characteristic polynomials) is expressed in terms of the character
  $\chi_{\lambda}$ of an irreducible representation of the group $G$,
  typically one whose highest weight vector is a partition of rectangular
  type;

  \item The shape of the highest weight vector suggests reduction with respect
  to a particular parabolic subgroup $P = M U$. If $W$ and $W_M$ denote the
  Weyl groups of $G$ and the Levi factor $M$, the irreducible character $\chi$
  is a sum over $W_M \backslash M$.
\end{itemize}
To make the paper as elementary as possible we always express
these reductions by an application of the Laplace expansion of a
determinant, but it seems worthwhile to note a more general reason
such reductions are possible.

Let $G$ be a reductive complex analytic Lie group, and let $P$ be
a parabolic subgroup with Levi decomposition $P = M U$, where $M$
is the Levi factor and $U$ the unipotent radical of $P$. Let $T$
be a maximal torus of $M$, which is therefore also a maximal torus
of $G$. We will denote by $\mathfrak{g}$ and $\mathfrak{t}$ the
Lie algebras of $G$ and $T$.

Let $\Phi \subset \mathfrak{t}^{\ast}$ be the root system of $G$,
and let $\Phi_M \subset \Phi$ be the root system of $M$. We choose
an ordering of the roots so that the roots in $U$ are positive.
Let $\rho, \rho_M$ be half the sum of the positive root in $\Phi$
and $\Phi_M$, respectively, and let $\rho_U$ be half the sum of
the positive roots in $U$, so that $\rho = \rho_M + \rho_U$.

Let $W$ and $W_M$ be the Weyl groups of $G$ and $M$ with respect
to~$\mathfrak{t}$. Let $\mathcal{C}$ and $\mathcal{C}_M$ be the
positive Weyl chambers, so
\begin{eqnarray*}
  \mathcal{C} & = & \{ x \in \mathfrak{t}^{\ast} \, | \, \text{$\left\langle
  \alpha, x \right\rangle \geqslant 0$ for all $\alpha \in \Phi^+$ \},}\\
  \mathcal{C}_M & = & \{ x \in \mathfrak{t}^{\ast} \, | \, \text{$\left\langle
  \alpha, x \right\rangle \geqslant 0$ for all $\alpha \in \Phi^+_M$ \} .}
\end{eqnarray*}
If $\Lambda \subset \mathfrak{t}^{\ast}$ is the lattice of
weights, that is, differentials of rational characters of $T$,
then it is known that
\begin{eqnarray}
  \Lambda \cap \mathcal{C}^{\circ} = \rho + ( \Lambda \cap \mathcal{C}), &  &
  \nonumber\\
  \Lambda \cap \mathcal{C}^{\circ}_M = \rho_M + ( \Lambda \cap \mathcal{C}_M )
  . &  &  \label{interiorpwc}
\end{eqnarray}
Each coset in $W_M \backslash W$ has a unique representative $w$
such that $w\mathcal{C} \subset \mathcal{C}_M$. Let $\Xi$ be this
set of coset representatives.

Now let $\lambda \subset \Lambda \cap \mathcal{C}$ be a dominant
weight. If $w \in \Xi$ then $w ( \lambda + \rho ) = \lambda_w +
\rho_M$ where by (\ref{interiorpwc}) we have $\lambda_w \in
\Lambda \cap \mathcal{C}_M$.

The Weyl character formula expresses the character
$\chi_{\lambda}^G$ as
\[ \frac{\sum_{w \in W} ( - 1 )^{l ( w )} e^{w ( \lambda + \rho )}}{e^{- \rho}
   \prod_{\alpha \in \Phi^+} ( 1 - e^{\alpha} )} = \frac{\sum_{w \in \Xi}
   \sum_{\sigma \in W_M} ( - 1 )^{l ( \sigma w )} e^{\sigma w ( \lambda + \rho
   )}}{e^{- \rho} \prod_{\alpha \in \Phi^+} ( 1 - e^{\alpha} )} . \]
Here $l$ is the length function on the Weyl group.

We note that if $w \in W_M$ then $w ( \rho_U ) = \rho_U$. It
follows that
\[ \chi_{\lambda}^G = \frac{1}{e^{- \rho_U} \prod_{\alpha \in \Phi^+ -
   \Phi_U^+} ( 1 - e^{\alpha} )} \sum_{w \in \Xi} ( - 1 )^{l ( w )}
   \chi^M_{\lambda_w} . \]
This generalization of Lemma~\ref{lem:min1} could be used
everywhere in this paper that the Laplace expansion is invoked.

\section{Symplectic Group}\label{sec:sp}

A unitary matrix $g$ is said to be {\tmem{symplectic}} if $gJ {^t
g} = J$ where
\[ J = \left( \begin{array}{cc}
     0 & I_N\\
     - I_N & 0
   \end{array} \right) . \]
A symplectic matrix has determinant equal to 1. The symplectic
group $\tmop{Sp} ( 2 N )$ is the group of $2 N \times 2 N$
symplectic matrices. The eigenvalues of a symplectic matrix are
\[ e^{\pm i \theta_1}, \cdots, e^{\pm i \theta_N} \]
with
\[ 0 \leqslant \theta_1 \leqslant \theta_2 \leqslant \cdots \leqslant \theta_N \leqslant \pi . \]
The Weyl integration formula {\cite{weyl}} for integrating a
symmetric function $f ( A ) = f ( \theta_1, \cdots, \theta_N )$
over $\tmop{Sp} ( 2 N )$ with $\tmop{Sp} ( 2 N )$ respect to Haar
measure is
\begin{eqnarray*}
  \mathbb{E}_{\tmop{Sp} ( 2 N )} f = \int_{\tmop{Sp} ( 2 N )} f ( g ) \; d g
   =  \frac{2^{N^2}}{\pi^N N!} \times&&\\
  \int_{[ 0, \pi ]^N} f ( \theta_1, \cdots, \theta_N ) \prod_{1 \leqslant j < k \le
  N} ( \cos \; \theta_k - \cos \; \theta_j )^2 \prod_{n = 1}^N \sin^2 \theta_n
  \, d \theta_1 \cdots d \theta_N . &  &
\end{eqnarray*}

Denoting the irreducible representation of $U ( 2 n )$ with
highest weight $\lambda$ by $F^{\lambda}_{( 2 n )}$ and the
irreducible representation of $\tmop{Sp} ( 2 n )$ with highest
weight $\mu$ by $V^{\mu}_{( 2 n )}$ we have the following
classical branching rule, due to Littlewood {\cite{Li2}},
{\cite{Li}}. Denoting by $[ F^{\lambda}_{( 2 n )}, V^{\mu}_{( 2 n
)} ]$ the multiplicity of $V_{( 2 n )}^{\lambda}$ in the
restriction to $\tmop{Sp} ( 2 n )$ of $F^{\lambda}_{( 2 n )}$,
Littlewood~{\cite{Li}} p.~295 gave the branching rule
\begin{equation}
  \label{e:bsym} [ F^{\lambda}_{( 2 n )}, V^{\mu}_{( 2 n )} ] =
  \sum_{\text{$\beta'$ even}} c^{\lambda}_{\mu \beta} \hspace{2.6em} ( l (
  \lambda ), l ( \mu ) < n ) .
\end{equation}
Here a partition $\lambda$ is called {\it even} if all its parts
are even (and similarly, it is called {\it odd} if all its parts
are odd. See also Howe, Tan and
Willenbring~{\cite{HoweTanWillenbring}} and
King~{\cite{KingBranching}} for this branching rule.

Denoting the irreducible character of the symplectic group
$\tmop{Sp} ( 2 n )$ labelled by partition $\lambda$ by
$\chi_{\lambda}^{\sp_{2n}}$, the Weyl character formula
{\cite{weyl}} in the case of symplectic group can be written
\begin{equation}
  \label{charsp} \chi_{\lambda}^{\sp_{2n}}( x_1^{\pm 1}, \cdots,
  x_n^{\pm 1} ) = \frac{\det_{1 \leqslant i, j \leqslant n} ( x_j^{\lambda_i +
  n - i + 1} - x_j^{- ( \lambda_i + n - i + 1 )} )}{\det_{1 \leqslant i, j
  \leqslant n} ( x_j^{n - i + 1} - x_j^{- ( n - i + 1 )} )} .
\end{equation}
The determinant in the denominator can be evaluated as by Weyl
{\cite{weyl}}:
\begin{equation}
  \label{detsp} \det_{1 \leqslant i, j \leqslant n} ( x_j^{n - i + 1} - x_j^{-
  ( n - i + 1 )} ) = \frac{\prod_{i < j} ( x_i - x_j ) ( x_i x_j - 1 ) \prod_i
  ( x_i^2 - 1 )}{( x_1 \cdots x_n )^n} .
\end{equation}

The Weyl dimension formula for the the dimension of the
irreducible
  representation of the symplectic group $\tmop{Sp} ( 2 k )$ labelled by
  partition $\lambda$ is
  \[\dim(\chi_{\lambda}^{\sp_{2k}}) = \frac{\prod_{i < j} ( \mu_i - \mu_j
     ) ( \mu_i + \mu_j + 2 ) \prod_i ( \mu_i + 1 )}{( 2 k - 1 ) ! ( 2 k - 3 )
     ! \cdots 1!}, \]
  where $\mu_i = \lambda_i + k - i$.

  An alternative expression is given by El-Samra and King {\cite{EK}}, in the
  following analogue of the hook formula \eqref{e_hhl}:
  \begin{equation}
    \label{elsamra} \dim ( \chi_{\lambda}^{\tmop{Sp}_{2 k}} ) = \prod_{u \in
    \lambda} \frac{2 k + c^{\tmop{Sp}} ( u )}{h ( u )},
  \end{equation}
  where
  \[ c^{\tmop{Sp}} ( i, j ) = \left\{\begin{array}{ll}
       i + j - \lambda_i' - \lambda_j' & \text{if $i \leqslant j$;}\\
       \lambda_i + \lambda_j + 2 - i - j & \tmop{if} i > j.
     \end{array}\right. \]

We will make crucial use of the following analog of Lemma
\ref{unidecomp}:

\begin{lemma}\label{symplecticdecomposition}
  For $\lambda \subseteq
  \left\langle N^k
  \right\rangle$ let $\tilde{\lambda} = ( k - \lambda_N', \cdots, k- \lambda_1' )$.
   Then we have
  \begin{eqnarray}
    \prod_{i = 1}^k \prod_{n = 1}^N ( x_i + x_i^{- 1} - t_n - t_n^{- 1} ) &  &
    = \nonumber\\
    \sum_{\lambda \subseteq N^k} ( - 1 )^{| \tilde{\lambda} |}
    \chi_{\lambda}^{\tmop{Sp} ( 2 k )} ( x_1^{\pm 1}, \cdots, x_k^{\pm 1} )
    \chi_{\tilde{\lambda}}^{\tmop{Sp} ( 2 N )} ( t_1^{\pm 1}, \cdots, t_N^{\pm
    1} ) . &  &  \label{ecfspone}
  \end{eqnarray}
\end{lemma}

\begin{proof}
Using Weyl character formula \eqref{charsp} and the Laplace
expansion we can rewrite the right-hand side of \eqref{ecfspone}
as follows:
\begin{equation} \label{e:pl1}
\begin{aligned}
&\det \left|
  \begin{array}{ccccc}
  x_1^{N+k}-x_1^{-(N+k)} &   x_1^{N+k-1}-x_1^{-(N+k-1)}&
\dots &  x_1^{1}-x_1^{-1}& \\
    \vdots & \vdots &\ddots & \vdots & \\
x_k^{N+k}-x_k^{-(N+k)} &   x_k^{N+k-1}-x_k^{-(N+k-1)}&
\dots &  x_k^{1}-x_k^{-1}& \\
t_1^{N+k}-t_1^{-(N+k)} &   t_1^{N+k-1}-t_1^{-(N+k-1)}&
\dots &  t_1^{1}-t_1^{-1}& \\
    \vdots & \vdots &\ddots & \vdots & \\
t_N^{N+k}-t_{N}^{-(N+k)} &   t_N^{N+k-1}-t_N^{-(N+k-1)}& \dots &
t_N^{1}-t_N^{-1}
\end{array} \right| \\
&\times  \frac {(x_1 \dots x_k)^{k}}{\prod_{1 \leqslant i<j \le
k}(x_i-x_j)(x_i x_j-1) \prod_{i=1}^{k}(x_i^2-1)} \\
&\times  \frac {(t_1 \dots t_N)^{N}}{\prod_{1 \leqslant i<j \le
N}(t_i-t_j)(t_i t_j-1) \prod_{i=1}^{N}(t_i^2-1)}.
\end{aligned}
\end{equation}

Now the determinant in the equation \eqref{e:pl1} can be evaluated
using the Weyl denominator formula \eqref{detsp} to be equal to
\begin{equation} \label{e:pl2}
\begin{aligned}&\prod_{1 \leqslant i<j \leqslant k}(x_i-x_j)(x_i x_j-1)
\prod_{i=1}^{k}(x_i^2-1) \prod_{1 \leqslant i<j \leqslant
N}(t_i-t_j)(t_i
t_j-1) \prod_{i=1}^{N}(t_i^2-1) \times\\
&\frac{ \prod_{i = 1}^k \prod_{n = 1}^N (x_i-t_n)(x_i t_n-1)}{(x_1
\dots x_k)^{N+k}  (t_1 \dots t_N)^{N+k}}.
\end{aligned}
\end{equation}

Finally combining \eqref{e:pl1} and \eqref{e:pl2} we get that the
expression on the right-hand side of \eqref{ecfspone} equals to
\begin{equation*}
\frac{ \prod_{i = 1}^k \prod_{n = 1}^N (x_i-t_n)(x_i t_n-1)}{(x_1
\dots x_k)^{N}  (t_1 \dots t_N)^{k}} =\prod_{i = 1}^k \prod_{n =
1}^N ( x_i + x_i^{- 1} - t_n - t_n^{- 1} ),
\end{equation*}
completing the proof.

\end{proof}

\begin{remark}
  Lemma~\ref{symplecticdecomposition} is stated in slightly different form in
  Jimbo and Miwa {\cite{JM}}. Another proof can be found in Howe
  {\cite{HowePerspectives}}, Theorem 3.8.9.3, which we digress to briefly
  discuss. We will describe an action of a group on a space of functions (or
  tensor fields) on some space as {\tmem{geometric}} if it is induced by an
  action of the group on the underlying space. As with our discussion of the
  Cauchy identity, we may consider the action of
  \[ ( g, h ) : X \longmapsto \det ( h )^{- k} \cdot g X {^t h} \]
  of $\tmop{Sp} ( 2 k ) \times \tmop{GL} ( N )$ on $\tmop{Mat}_{2 k \times N}
  (\mathbb{C}) =\mathbb{C}^{2 k N}$. This induces an action of $\tmop{Sp} (
  2 k ) \times \tmop{GL} ( N )$ on the symmetric and exterior algebras over
  $\mathbb{C}^{2 k N}$, which is geometric. The trace of $( g, h ) \in
  \tmop{Sp} ( 2 k ) \times \tmop{GL} ( N )$, where $t_i^{\pm 1}$ are the
  eigenvalues of $g$ and $x_n$ are the eigenvalues of $h$ is
  \[ \prod_{i = 1}^k \prod_{n = 1}^N x_n^{- 1} ( 1 + x_n t_i ) ( 1 + x_n
     t_i^{- 1} ) = \prod_{i = 1}^k \prod_{n = 1}^N ( x_i + x_i^{- 1} + t_n +
     t_n^{- 1} ) . \]
  Howe considers the centralizer of $\tmop{Sp} ( 2 k )$ in this geometric
  action on the exterior algebra $\Lambda (\mathbb{C}^{2 k N} )$ and observes
  that it is properly larger than $\tmop{GL} ( N )$; in fact, it is a group
  isomorphic to $\tmop{Sp} ( 2 N )$, though the second symplectic group does
  not act geometrically. He finds the complete isotypic decomposition of
  $\Lambda (\mathbb{C}^{2 k N} )$:
  \begin{equation}
    \label{hosp} \Lambda (\mathbb{C}^{2 k N} ) \cong \sum_{\lambda}
    V^{\lambda}_{( 2 k )} \otimes V^{\tilde{\lambda}}_{( 2 N )},
  \end{equation}
  where $V^{\lambda}_{( 2 k )}$ is an irreducible representation of $\tmop{Sp}
  ( 2 k )$ with highest weight $\lambda$, and $\lambda$ runs over all
  partitions fitting inside $k \times N$ rectangle. Taking traces gives
  \begin{equation}\label{good}
  \begin{aligned}
&    \prod_{i = 1}^k \prod_{n = 1}^N ( x_i + x_i^{- 1} + t_n +
t_n^{- 1} )
    =\\
&    \sum_{\lambda \subseteq N^k} \chi_{\lambda}^{\tmop{Sp} ( 2 k
)} ( x_1^{\pm
    1}, \cdots, x_k^{\pm 1} ) \chi_{\tilde{\lambda}}^{\tmop{Sp} ( 2 N )} (
    t_1^{\pm 1}, \cdots, t_N^{\pm 1} ),
    \end{aligned}
  \end{equation}
  and replacing the second symplectic matrix with its negative
  gives~(\ref{ecfspone}).
\end{remark}

\subsection{Products}

The goal of this section is to give simple proofs of the
Proposition~\ref{prop:symp}, first derived by  Conrey, Farmer,
Keating, Rubinstein, and  Snaith  in \cite{CFKRS2}, and of
Corollary~\ref{corsp}, first derived by Keating and
Snaith~{\cite{KS00}}.

\begin{proposition} \label{prop:symp}Notation being as above we
  have:
  \begin{eqnarray}
    &  & \int_{\tmop{Sp} ( 2 N )} \prod_{j = 1}^k \det ( I +x_j g ) \; d g =
    ( x_1 \ldots x_k )^N \chi_{\left\langle N^k \right\rangle}^{\tmop{Sp} ( 2
    k )} ( x_1^{\pm 1}, \cdots, x_k^{\pm 1} ) = \nonumber\\
    &  & \sum_{\varepsilon \in \{ \pm 1 \}} \prod_{j = 1}^k x_j^{N ( 1 -
    \varepsilon_j )} \prod_{i \leqslant j} ( 1 - x_i^{\varepsilon_i}
    x_j^{\varepsilon_j} )^{- 1} .  \label{esymp}
  \end{eqnarray}

\end{proposition}

\begin{proof}
  Denoting the eigenvalues of $g$ in $\tmop{Sp} ( 2 N )$ by $t_1^{\pm 1},
  \cdots, t_N^{\pm 1}$ we have:
  \[ \prod_{i = 1}^k \det ( I + x_i g ) = \prod_{i = 1}^k \prod_{n = 1}^N ( 1
     + x_i t_n ) ( 1 + x_i t_n^{- 1} ) . \]
  Using (\ref{good}) we have:
  \begin{equation*}
  \begin{aligned}
&    ( x_1 \cdots x_k )^{- N} \mathbb{E}_{\tmop{Sp} ( 2 N )}
\prod_{j = 1}^k
    \det ( I + x_j g )  = \\
   & \mathbb{E}_{\tmop{Sp} ( 2 N )} \prod_{n = 1}^N \prod_{i = 1}^k ( x_i +
    x_i^{- 1} + t_n + t_n^{- 1} )  =  \\
&    \mathbb{E}_{\tmop{Sp} ( 2 N )} \left( \sum_{\lambda \subseteq
N^k}  \chi_{\lambda}^{\tmop{Sp} ( 2 k )} ( x_1^{\pm 1},
    \cdots, x_k^{\pm 1} ) \chi_{\tilde{\lambda}}^{\tmop{Sp} ( 2 N )} (
    t_1^{\pm 1}, \cdots, t_N^{\pm 1} ) \right)  =  \\
&    \chi_{N^k}^{\tmop{Sp} ( 2 k )} ( x_1^{\pm 1}, \cdots,
x_k^{\pm 1} ) .
\end{aligned}
\end{equation*}

  In the last line we used the fact that
  \[ \mathbb{E}_{\tmop{Sp} ( 2 N )} \chi_{\lambda}^{\tmop{Sp} ( 2 N )} =
     \left\{\begin{array}{ll}
       1 & \text{if $\lambda = \varnothing$;}\\
       0 & \text{otherwise} .
     \end{array}\right. \]
  Consequently we obtain
  \begin{equation}
    \mathbb{E}_{\tmop{Sp} ( 2 N )} \prod_{j = 1}^k \det ( I + x_j g ) = ( x_1
    \ldots x_k )^N \chi_{N^k}^{\tmop{Sp} ( 2 k )} ( x_1^{\pm 1}, \cdots, x_k^{\pm 1}
    ),
  \end{equation}
  proving the first line of (\ref{esymp}).

  Now using the Weyl character formula (\ref{charsp}) for
  symplectic group and the evaluation (\ref{detsp}) of
  the denominator in (\ref{charsp}) we have:
  \begin{eqnarray}
    ( x_1 \ldots x_k )^N \chi_{N^k}^{\tmop{Sp} ( 2 k )} ( x_1^{\pm 1}, \cdots,
    x_k^{\pm 1} ) & = &  \nonumber\\
    \left| \begin{array}{ccccc}
      x_1^{N + k} - x_1^{- ( N + k )} & x_1^{N + k - 1} - x_1^{- ( N + k - 1
      )} & \cdots & x_1^{N + 1} - x_1^{- ( N + 1 )} & \\
      \vdots & \vdots & \ddots & \vdots & \\
      x_k^{N + k} - x_k^{- ( N + k )} & x_k^{N + k - 1} - x_k^{- ( N + k - 1
      )} & \cdots & x_k^{N + 1} - x_k^{- ( N + 1 )} &
    \end{array} \right| &  &  \nonumber\\
    \times \frac{( x_1 \cdots x_k )^{k + N}}{\prod_{1 \leqslant i < j
    \leqslant k} ( x_i - x_j ) ( x_i x_j - 1 ) \prod_{i = 1}^k ( x_i^2 - 1 )}
    . &  &  \label{edetone}
  \end{eqnarray}
  Now splitting the determinant in (\ref{edetone}) we can rewrite it as
  follows:
  \begin{eqnarray*}
    \sum_{\varepsilon \in \{ \pm 1 \}^k} \det \left| \begin{array}{ccccc}
      \varepsilon_1 x_1^{\varepsilon_1 ( N + k )} & \varepsilon_1
      x_1^{\varepsilon_1 ( N + k - 1 )} & \cdots & \varepsilon_1
      x_1^{\varepsilon_1 ( N + 1 )} & \\
      \vdots & \vdots & \ddots & \vdots & \\
      \varepsilon_k x_k^{\varepsilon_k ( N + k )} & \varepsilon_k
      x_k^{\varepsilon_k ( N + k - 1 )} & \cdots & \varepsilon_k
      x_k^{\varepsilon_k ( N + 1 )} &
    \end{array} \right| & = & \\
    \sum_{\varepsilon \in \{ \pm 1 \}^k} \prod_{i = 1}^k \varepsilon_i
    x_i^{\varepsilon_i ( N + 1 )} \det \left| \begin{array}{ccccc}
      x_1^{\varepsilon_1 ( k - 1 )} & x_1^{\varepsilon_1 ( k - 2 )} & \cdots &
      1 & \\
      \vdots & \vdots & \ddots & \vdots & \\
      x_k^{\varepsilon_k ( k - 1 )} & x_k^{\varepsilon_k ( k - 2 )} & \cdots &
      1 &
    \end{array} \right| & = & \\
    \sum_{\varepsilon \in \{ \pm 1 \}^k} \prod_{i = 1}^k \varepsilon_i
    x_i^{\varepsilon_i ( N + 1 )} \prod_{i < j} ( x_i^{\varepsilon_i} -
    x_j^{\varepsilon_j} ), &  &
  \end{eqnarray*}
  where in the last line we have used the Vandermonde determinant evaluation.

  Next, making use of the elementary identities
  \begin{equation}
    \label{el1} \frac{\prod_{i = 1}^k \varepsilon_i x_i^{( 1 + \varepsilon_i
    )}}{\prod_{i = 1}^k ( x_i^2 - 1 )} = \frac{1}{\prod_{i = 1}^k ( 1 - x^{- 2
    \varepsilon_i} )},
  \end{equation}
  \begin{equation}
    \label{el2} \frac{x_i^{\varepsilon_i} - x_j^{\varepsilon_j}}{( x_i - x_j )
    ( x_i x_j - 1 )} = \frac{x_i^{\varepsilon_i - 1} x_j^{\varepsilon_-
    1}}{x_i^{\varepsilon_i} x_j^{\varepsilon_j} - 1},
  \end{equation}
  and noting that
  \begin{equation}
    \label{el3} \prod_{1 \leqslant i < j \leqslant k} x_i^{\varepsilon_i - 1}
    x_j^{\varepsilon_- 1} = \prod_{i = 1}^k x_i^{( k - 1 ) ( \varepsilon_i - 1
    )},
  \end{equation}
  the right hand side of (\ref{edetone}) is easily seen to be equal to
  \begin{eqnarray*}
    \sum_{\varepsilon \in \{ \pm 1 \}^k} \prod_{j = 1}^k x_j^{N ( 1 +
    \varepsilon_j )} \prod_{i \leqslant j} ( 1 - x_i^{- \varepsilon_i} x_j^{-
    \varepsilon_j} )^{- 1} & = & \\
    \sum_{\varepsilon \in \{ \pm 1 \}^k} \prod_{j = 1}^k x_j^{N ( 1 -
    \varepsilon_j )} \prod_{i \leqslant j} ( 1 - x_i^{\varepsilon_i}
    x_j^{\varepsilon_j} )^{- 1}, &  &
  \end{eqnarray*}
  completing the proof.
\end{proof}

\begin{corollary}
\label{corsp}We have
  \begin{eqnarray}
    &  & \int_{\tmop{Sp} ( 2 N )} \det ( I - g )^k \, d g = \dim (
    \chi_{N^k}^{\tmop{Sp} ( 2 k )} ) = \nonumber\\
    &  & \frac{( N + k ) !}{N!k!} \prod_{i = 1}^k \frac{( k + 2 N + i )
    !i!}{( 2 i + 2 N ) ! ( 2 i - 1 ) !}  \label{bgsp}
  \end{eqnarray}
\end{corollary}

The mean value $\mathbb{E}_{\tmop{Sp} ( 2 N )} \det ( I - g )^k$
was computed by Keating and Snaith in {\cite{KS00}} using the
Selberg integral (without restriction that $k$ be an integer).
They found it to be
\begin{equation}
  \label{kssp} 2^{2 Nk} \prod_{j = 1}^N \frac{\Gamma ( 1 + N + j ) \Gamma (
  \frac{1}{2} + k + j )}{\Gamma ( 1 + N + k + j ) \Gamma ( \frac{1}{2} + j )}
  .
\end{equation}
For integer $k$ we can rewrite (\ref{kssp}) using the duplication
formula
\[ 2^{2 z - 1} \Gamma ( z ) \Gamma \left( z + \frac{1}{2} \right) = \sqrt{\pi}
   \Gamma ( 2 z ) \]
in the form
\[ \prod_{j = 1}^N \frac{( N + j ) ! ( 2 k + 2 j - 1 ) ! ( j - 1 ) !}{( N + k
   + j ) ! ( 2 j - 1 ) ! ( k + j - 1 ) !}, \]
and the latter expression is equal to the one appearing on the
right-hand side of (\ref{bgsp}) by an elementary computation.

\begin{proof}
  We have:
  \begin{equation}
    \int_{\tmop{Sp} ( 2 N )} \det ( I - g )^k \, d g = \dim (
    \chi_{N^k}^{\tmop{Sp} ( 2 k )} ) .
  \end{equation}

  Applying (\ref{elsamra}) to the partition $\lambda = N^k$ and recalling that
  the product of hook numbers for partition $N^k$ is given by
  \eqref{hook123}, we obtain
  \begin{eqnarray*}
    \dim ( \chi_{N^k}^{\tmop{Sp} ( 2 k )} ) & = & \prod_{j = 1}^N \frac{( j -
    1 ) !}{( j + k - 1 ) !} \prod_{j = 1}^k \frac{( j + k ) !}{j!} \prod_{i =
    1}^k \frac{( k + 2 N + i ) !}{( 2 i + 2 N ) !} \prod_{i = 1}^k \frac{i!}{(
    2 i - 1 ) !}\\
    & = & \frac{( N + k ) !}{N!k!} \prod_{i = 1}^k \frac{( k + 2 N + i )
    !i!}{( 2 i + 2 N ) ! ( 2 i - 1 ) !} .
  \end{eqnarray*}
  Thus we have established that
  \[ \mathbb{E}_{\tmop{Sp} ( 2 N )} \det ( I - g )^k = \frac{( N + k )
     !}{N!k!} \prod_{i = 1}^k \frac{( k + 2 N + i ) !i!}{( 2 i + 2 N ) ! ( 2 i
     - 1 ) !}, \]
  completing the proof of Corollary~\ref{corsp}.
\end{proof}

\subsection{Ratios}

The goal of this section is to give a simple proof of
Theorem~\ref{ratsp}, first established in Conrey, Farmer and
Zirnbauer~{\cite{cfz}} and in Conrey, Forrester and
Snaith~{\cite{cfs}}.

\begin{theorem}
  \label{ratsp}  Let $y_j$ be complex numbers with $|y_j| < 1$.
  Suppose $N \geqslant l$. Then we have:
  \begin{eqnarray*}
    \int_{\tmop{Sp} ( 2 N )} \frac{\prod_{j = 1}^k \det ( I + x_j g
    )}{\prod_{i = 1}^l \det ( I - y_i g )} \; d g & = & \\
    \sum_{\varepsilon \in \{ \pm 1 \}^k} \prod_{j = 1}^k x_j^{N ( 1 -
    \varepsilon_j )} \frac{\prod_{i = 1}^k \prod_{j = 1}^l ( 1 +
    x_i^{\varepsilon_i} y_j )}{\prod_{i \leqslant j} ( 1 - x_i^{\varepsilon_i}
    x_j^{\varepsilon_j} ) \prod_{1 \leqslant i < j \leqslant l} ( 1 - y_i y_j
    )} . &  &
  \end{eqnarray*}

\end{theorem}

\noindent {\bf Proof:}
  When $l ( \lambda ) \leqslant N$ we may write the branching rule
  (\ref{e:bsym}) in the form
  \begin{equation}
    s_{\lambda} ( t_1^{\pm 1}, \cdots, t_N^{\pm 1} ) = \sum_{\mu \subseteq
    \lambda} \chi_{\mu}^{\tmop{Sp} ( 2 N )} ( t_1^{\pm 1}, \cdots, t_N^{\pm 1}
    ) \left( \sum_{\text{$\beta'$ even}} c^{\lambda}_{\mu \beta} \right) .
  \end{equation}
  Also we have the following identity of Littlewood~{\cite{Li}}
  \[ \prod_{1 \leqslant i < j \leqslant l} ( 1 - y_i y_j )^{- 1} =
     \sum_{\text{$\beta'$ even}} s_{\beta} ( y_1, \cdots, y_l ) . \]
  These two formulas, together with the Cauchy identity \eqref{ratioscauchy} yields the
  following Cauchy identity for $\tmop{Sp} ( 2 N )$, known to Weyl
  {\cite{weyl}} and Littlewood {\cite{Li}}:
  \begin{eqnarray*}
    \frac{1}{\prod_{n = 1}^N \prod_{j = 1}^l ( 1 - y_j t_n ) ( 1 - y_j t_n^{-
    1} )} & = & \\
    \frac{1}{\prod_{i < j} ( 1 - y_i y_j )} \sum_{\mu} \chi_{\mu}^{\tmop{Sp} (
    2 N )} ( t_1^{\pm 1}, \cdots, t_N^{\pm 1} ) s_{\mu} ( y_1, \cdots, y_l
    ).
    &  &
  \end{eqnarray*}

  Combining this identity with (\ref{good}) we have, with $t_i^{\pm 1}$
  the eigenvalues of $g \in \tmop{Sp} ( 2 N )$
\begin{equation*}
\begin{aligned}
&    \frac{\prod_{j = 1}^k \det ( I + x_j g )}{\prod_{i = 1}^l
\det ( I - y_i g
    )} =\\
    & \frac{( x_1 \cdots x_k )^N}{\prod_{i < j} ( 1 - y_i y_j )}
\sum_{\lambda
    \subseteq N^k} \chi_{\lambda}^{\tmop{Sp} ( 2
    k
    )} (
    x_1^{\pm 1}, \cdots, x_k^{\pm 1} ) \chi_{\tilde{\lambda}}^{\tmop{Sp} ( 2 N
    )} ( t_1^{\pm 1}, \cdots, t_N^{\pm 1} )  \\
&    \sum_{\mu} \chi_{\mu}^{\tmop{Sp} ( 2 N )} ( t_1^{\pm 1},
\cdots, t_N^{\pm
    1} ) s_{\mu} ( y_1, \cdots, y_l ).
    \end{aligned}
\end{equation*}
  Since
  \[ \mathbb{E}_{\tmop{Sp} ( 2 N )} \chi_{\lambda}^{\tmop{Sp} ( 2 N
    )} ( g )
     \chi_{\mu}^{\tmop{Sp} ( 2 N
    )} ( g ) = \left\{\begin{array}{ll}
       1 & \text{if $\lambda = \mu$, $l ( \lambda ) \leqslant N$} ;\\
       0 & \text{otherwise} ,
     \end{array}\right. \]
  the theorem follows from the following Proposition.

\begin{proposition}
  \label{prop:alts}Notation being as in Lemma~\ref{symplecticdecomposition} we
  have
  \begin{eqnarray}
    \sum_{\lambda \subseteq \left\langle N^k \right\rangle}  \chi_{\lambda}^{\tmop{Sp} ( 2 k )} ( x_1^{\pm 1},
    \cdots, x_k^{\pm 1} ) s_{\tilde{\lambda}} ( y_1, \cdots, y_l ) & = &
    \nonumber\\
    \sum_{\varepsilon \in \{ \pm 1 \}^k} \prod_{j = 1}^k x_j^{- N
    \varepsilon_j} \frac{\prod_{i = 1}^k \prod_{j = 1}^l ( 1 +
    x_i^{\varepsilon_i} y_j )}{\prod_{i \leqslant j} ( 1 - x_i^{\varepsilon_i}
    x_j^{\varepsilon_j} )} . &  &  \label{qttt}
  \end{eqnarray}
\end{proposition}

\begin{proof}
  In order to yield a non-zero contribution to the sum on the left-hand side
  of (\ref{qttt}) $\lambda$ must be of the form $\lambda = ( N - l )^k + \mu$
  with $\mu \subseteq l^k$. Now keeping in mind the Weyl character formula for
  symplectic group (\ref{charsp}) and the numerator evaluation (\ref{detsp}),
  together with the definition (\ref{schurdefinition}) of the Schur function
  $s_{\tilde{\lambda}}$ and Vandermonde identity expressing the denominator in
  (\ref{schurdefinition}) as $\prod_{i < j} ( x_i - x_j )$, we can rewrite the
  expression on the left-hand side of (\ref{qttt}) as follows, using the
  Laplace expansion:
\begin{equation*}
\begin{aligned}
    & \det \left| \begin{array}{ccccc}
      x_1^{N+k} - x_1^{- ( N+k)} & x_1^{N+k-1} - x_1^{- (N+k-1)} & \cdots & x_1^{(N-l+1)}
       - x_1^{- ( N-l+1)} & \\
      \vdots & \vdots & \ddots & \vdots & \\
      x_k^{N+k} - x_k^{- ( N+k)} & x_k^{N+k-1} - x_k^{- (N+k-1)} & \cdots & x_k^{(N-l+1)}
       - x_k^{- ( N-l+1)} & \\
      ( - y_1 )^{l + k - 1} &  ( - y_1 )^{l + k - 2}& \cdots & 1 & \\
      \vdots & \vdots & \ddots & \vdots & \\
      ( - y_l )^{l + k - 1} &  ( - y_l )^{l + k - 2}& \cdots & 1 &
    \end{array} \right|  \\
    &\\
    & \times \frac{1}{\prod_{1 \leqslant i < j \leqslant l} ( y_j - y_i )}
    \frac{( x_1 \cdots x_k )^k}{\prod_{1 \leqslant i < j \leqslant k} ( x_i -
    x_j ) ( x_i x_j - 1 ) \prod_{i = 1}^k ( x_i^2 - 1 )} .
\end{aligned}
  \end{equation*}
  Now splitting the determinant in this expression we can rewrite it as
  follows:
  \begin{eqnarray*}
    \sum_{\varepsilon \in \{ \pm 1 \}^k} \det \left| \begin{array}{ccccc}
      \varepsilon_1 x_1^{\varepsilon_1 ( N+k )} & \varepsilon_1
      x_1^{\varepsilon_1 (N+k-1 )} & \cdots & \varepsilon_1
      x_1^{\varepsilon_1 ( N-l+1 )} & \\
      \vdots & \vdots & \ddots & \vdots & \\
      \varepsilon_k x_k^{\varepsilon_k ( N+k )} & \varepsilon_k
      x_k^{\varepsilon_k ( N+k-1 )} & \cdots & \varepsilon_k
      x_k^{\varepsilon_k ( N -l+1)} & \\
 ( - y_1 )^{l + k - 1} & ( - y_1 )^{l + k - 2}& \cdots &1 & \\
      \vdots & \vdots & \ddots & \vdots & \\
      ( - y_l )^{l + k - 1} & ( - y_l )^{l + k - 2}& \cdots &1 &
    \end{array} \right|;
\end{eqnarray*}

and this can be further expressed in the following way:
\begin{eqnarray*}
    \sum_{\varepsilon \in \{ \pm 1 \}^k} \prod_{i = 1}^k \varepsilon_i
    x_i^{\varepsilon_i ( N - l + 1 )} \det \left| \begin{array}{ccccc}
      x_1^{\varepsilon_1 ( l + k - 1 )} & x_1^{\varepsilon_1 ( l + k - 2 )} & \cdots &
      1 &
      \\
      \vdots & \vdots & \ddots & \vdots & \\
x_k^{\varepsilon_1 ( l + k - 1 )} & x_k^{\varepsilon_1 ( l + k - 2
)} & \cdots &
      1 &
      \\
       ( - y_1 )^{l + k - 1} & ( - y_1 )^{l + k - 2}& \cdots &1 & \\
      \vdots & \vdots & \ddots & \vdots & \\
      ( - y_l )^{l + k - 1} & ( - y_l )^{l + k - 2}& \cdots &1 &
    \end{array} \right| & = & \\
    \sum_{\varepsilon \in \{ \pm 1 \}^k} \prod_{i = 1}^k \varepsilon_i
    x_i^{\varepsilon_i ( N - l + 1 )} \prod_{i < j} ( x_i^{\varepsilon_i} -
    x_j^{\varepsilon_j} ) \prod_{1 \leqslant i < j \leqslant l} ( y_j - y_i )
    \prod_{i = 1}^k \prod_{j = 1}^l ( x_i^{\varepsilon_i} + y_j ), &  &
  \end{eqnarray*}
  where in the last line we have used the Vandermonde determinant evaluation.

  Next, making use of the elementary identities (\ref{el1}), (\ref{el2}),
  (\ref{el3}), the expression on the right-hand side is easily brought to the
  form
  \[ \sum_{\varepsilon \in \{ \pm 1 \}^k} \prod_{j = 1}^k x_j^{- N
     \varepsilon_j} \frac{\prod_{i = 1}^k \prod_{j = 1}^l ( 1 +
     x_i^{\varepsilon_i} y_j )}{\prod_{i \leqslant j} ( 1 -
     x_i^{\varepsilon_i} x_j^{\varepsilon_j} )} . \]
\end{proof}

\section{Orthogonal group} \label{sec:o}

A matrix $g$ is said to be {\tmem{orthogonal}} if it is real and
$g \, {^t g} = I$, where $^t g$ denotes the transpose of $g$. In
particular, it is unitary. We let $O ( N )$ denote the group of $N
\times N$ orthogonal matrices. We let $\tmop{SO} ( N )$ denote the
subgroup of $U ( N )$ consisting of $N \times N$ orthogonal
matrices with determinant equal to 1.

For any complex eigenvalue of an orthogonal matrix, its complex
conjugate is also an eigenvalue. The eigenvalues of $g \in
\tmop{SO} ( 2 N )$ can be written as
\[ e^{\pm i \theta_1}, \cdots, e^{\pm i \theta_N} \]
with
\[ 0 \leqslant \theta_1 \leqslant \theta_2 \leqslant \cdots \leqslant \theta_N
   \leqslant \pi . \]
The Weyl integration formula {\cite{weyl}} for integrating a
symmetric function $f ( g ) = f ( \theta_1, \cdots, \theta_N )$
over $\tmop{SO} ( 2 N )$ with respect to Haar measure is given by
\begin{eqnarray*}
  \mathbb{E}_{\tmop{SO} ( 2 N )} f ( g ) = \int_{\tmop{SO} ( 2 N )} f ( g ) ~
  dg & = & \\
  \frac{2^{( N - 1 )^2}}{\pi^N N!} \int_{[ 0, \pi ]^N} f ( \theta_1, \cdots,
  \theta_N ) \prod_{1 \leqslant j < k \leqslant N} ( \cos \theta_k - \cos
  \theta_j )^2 d \theta_1 \cdots d \theta_N . &  &
\end{eqnarray*}
The eigenvalues of $g \in \tmop{SO} ( 2 N + 1 )$ can be written as
\[ 1, e^{\pm i \theta_1}, \cdots, e^{\pm i \theta_N} \]
with
\[ 0 \leqslant \theta_1 \leqslant \theta_2 \leqslant \cdots \leqslant \theta_N
   \leqslant \pi . \]
The Weyl integration formula {\cite{weyl}} for integrating a
symmetric function $f ( g ) = f ( \theta_1, \cdots, \theta_N )$
over $\tmop{SO} ( 2 N + 1 )$ with respect to Haar measure is given
by
\begin{eqnarray*}
  \mathbb{E}_{\tmop{SO} ( 2 N + 1 )} f = \int_{\tmop{SO} ( 2 N + 1 )} f ( g )
  ~ dg & = & \\
  \frac{2^{1 + ( N - 1 )^2}}{\pi^{N - 1} ( N - 1 ) !} \int_{[ 0, \pi ]^{N -
  1}} f ( \theta_1, \cdots, \theta_{N - 1} ) &  & \\
  \prod_{1 \leqslant j < k \leqslant N} ( \cos \theta_k - \cos \theta_j )^2 d
  \theta_1 \cdots d \theta_{N - 1}, &  &
\end{eqnarray*}
where $\theta_N = 0$.

Denoting the irreducible representation of $U ( 2 n )$ with
highest weight $\lambda$ by $F^{\lambda}_{( 2 n )}$ and the
irreducible representation of $O ( 2 n )$ with highest weight
$\mu$ by $E^{\mu}_{( 2 n )}$ we have the following classical
branching rule, due to Littlewood {\cite{Li2,Li}}
\begin{equation}
  \label{e:borth} [ F^{\lambda}_{( 2 n )}, E^{\mu}_{( 2 n )} ] = \sum_{2
  \delta} c^{\lambda}_{\mu ( 2 \delta )} .
\end{equation}
Denoting the irreducible character of $\tmop{SO} ( 2 n + 1 )$
labelled by partition $\lambda$ by $\chi_{\lambda}^{\tmop{SO} ( 2
n + 1 )}$, the Weyl character formula {\cite{weyl}} in the case of
$\tmop{SO} ( 2 n + 1 )$ may be written
\begin{equation}
  \label{charsod} \chi_{\lambda}^{\tmop{SO} ( 2 n + 1 )} ( x_1^{\pm 1},
  \cdots, x_n^{\pm 1}, 1 ) = \frac{\det_{1 \leqslant i, j \leqslant n} (
  x_j^{\lambda_i + n - i + 1 / 2} - x_j^{- ( \lambda_i + n - i + 1 / 2 )}
  )}{\det_{1 \leqslant i, j \leqslant n} ( x_j^{n - i + 1 / 2} - x_j^{- ( n -
  i + 1 / 2 )} )} ;
\end{equation}
the determinant in the numerator can be evaluated as follows
{\cite{weyl}}:
\begin{equation}
  \label{detsod} \det_{1 \leqslant i, j \leqslant n} ( x_j^{n - i + 1} -
  x_j^{- ( n - i + 1 )} ) = \frac{\prod_{i < j} ( x_i - x_j ) ( x_i x_j - 1 )
  \prod_{i=1}^k ( 1 - x_i )}{( x_1 \cdots x_n )^{n - 1 / 2}} .
\end{equation}
In case of $\tmop{SO} ( 2 n )$ the situation is more subtle.
Denoting the irreducible character of $\tmop{SO} ( 2 n )$ labelled
by partition $\lambda$ by $\chi_{\lambda}^{\tmop{SO} ( 2 n )}$,
the Weyl character formula {\cite{weyl}} in the case of $\tmop{SO}
( 2 n )$ reads
\begin{eqnarray}
  \chi_{\lambda}^{\tmop{SO} ( 2 n )} ( x^{\pm 1} ) & = &\label{charso2}\\
  \frac{\det_{1 \leqslant i, j \leqslant n} ( x_j^{\lambda_i + n - i} + x_j^{-
  ( \lambda_i + n - i )} ) + \det_{1 \leqslant i, j \leqslant n} (
  x_j^{\lambda_i + n - i} - x_j^{- ( \lambda_i + n - i )} )}{\det_{1 \leqslant
  i, j \leqslant n} ( x_j^{n - i} + x_j^{- ( n - i )} )} .&  &  \nonumber
\end{eqnarray}
Here $\lambda_1 \geqslant \lambda_2 \geqslant \cdots \lambda_{n -
1} \geqslant | \lambda_n |$, so for a partition $\lambda$ we have
two characters: $\lambda_+$ associated with $( \lambda_1, \cdots,
\lambda_n )$ and $\lambda_-$ associated with $( \lambda_1, \cdots,
- \lambda_n )$ (this corresponds to the involution in the Dynkin
diagram of type $D_n$). The second term in the numerator in
(\ref{charso2}) changes sign when $\lambda_n$ is replaced by $-
\lambda_n$; in particular it vanished when $\lambda_n = 0$. When
$\lambda_n = 0$, the character $\chi_{\lambda}^{\tmop{SO} ( 2 n
)}$ also yields the character of the orthogonal group $O ( 2 N )$.
When $\lambda_n \ne 0$, the irreducible character of $O ( 2 N )$
is given by $\chi_{\lambda_+}^{\tmop{SO} ( 2 n )} +
\chi_{\lambda_-}^{\tmop{SO} ( 2 n )}$:
\begin{equation}
  \label{charo} \chi_{\lambda}^{O ( 2 n )} ( x^{\pm 1} ) = \frac{\det_{1
  \leqslant i, j \leqslant n} ( x_j^{\lambda_i + n - i} + x_j^{- ( \lambda_i +
  n - i )} )}{\det_{1 \leqslant i, j \leqslant n} ( x_j^{n - i} + x_j^{- ( n -
  i )} )} .
\end{equation}
The determinant in the numerator of (\ref{charso2}) can be
evaluated as follows {\cite{weyl}}:
\begin{equation}
  \label{detso2} \det_{1 \leqslant i, j \leqslant n} ( x_j^{n - i} + x_j^{- (
  n - i )} ) = \frac{\prod_{i < j} ( x_i - x_j ) ( x_i x_j - 1 )}{( x_1 \cdots
  x_n )^{n - 1}} .
\end{equation}

\subsection{Products and ratios for $\tmop{SO} ( 2 N )$}

\begin{lemma} \label{lem:soe}
  For $\lambda \subseteq k^N$ let $\tilde{\lambda} = ( k - \lambda_N', \cdots, k- \lambda_1' )$.
   Then we have
  \begin{eqnarray}
    \prod_{i = 1}^k \prod_{n = 1}^N ( x_i + x_i^{- 1} - t_n - t_n^{- 1} ) = &
    &  \nonumber\\
    \sum_{\lambda \subseteq N^k} ( - 1 )^{| \tilde{\lambda} |} (
    \chi_{\lambda_+}^{\tmop{SO} ( 2 k )} ( x_1^{\pm 1}, \cdots, x_k^{\pm 1} )
    \chi_{\tilde{\lambda}_+}^{\tmop{SO} ( 2 N )} ( t_1^{\pm 1}, \cdots,
    t_N^{\pm 1} ) + &  &  \nonumber\\
    \chi_{\lambda_-}^{\tmop{SO} ( 2 k )} ( x_1^{\pm 1}, \cdots, x_k^{\pm 1} )
    \chi_{\tilde{\lambda}_-}^{\tmop{SO} ( 2 N )} ( t_1^{\pm 1}, \cdots,
    t_N^{\pm 1} ) ) . &  &  \text{\label{e:cfso1}}
  \end{eqnarray}
\end{lemma}

\medbreak
\begin{proof}
  The lemma is stated in slightly different form in Jimbo and Miwa
  {\cite{JM}}. It may be proved by the method of  Lemma~\ref{unidecomp}
  and
  Lemma~\ref{symplecticdecomposition}.
\end{proof}

We begin by giving a simple proof of the following proposition,
first established in {\cite{CFKRS2}}.

\begin{proposition}
  \label{soprod}Notation being as above we have
  \begin{eqnarray}
    \mathbb{E}_{g \in SO ( 2 N )} \prod_{j = 1}^k \det ( I + x_j g ) = ( x_1
    \ldots x_k )^N \chi_{N^k}^{O_{2 k}} ( x_1^{\pm 1}, \cdots, x_k^{\pm 1} ) &
    = &  \nonumber\\
    \sum_{\varepsilon \in \{ \pm 1 \}} \prod_{j = 1}^k x_j^{N ( 1 -
    \varepsilon_j )} \prod_{i < j} ( 1 - x_i^{\varepsilon_i}
    x_j^{\varepsilon_j} )^{- 1} . &  &  \label{e:soprod}
  \end{eqnarray}
\end{proposition}

\begin{proof}
  Denoting the eigenvalues of $g$ in $\tmop{SO} ( 2 N )$ by $t_1^{\pm 1},
  \cdots, t_N^{\pm 1}$ we have:
  \[ \prod_{i = 1}^k \det ( I + x_i g ) = \prod_{i = 1}^k \prod_{n = 1}^N ( 1
     + x_i t_n ) ( 1 + x_i t_n^{- 1} ) . \]
  Using (\ref{e:cfso1}) we have (where $t_i$ are the eigenvalues of $g \in
  \tmop{SO} ( 2 N )$)
  \begin{eqnarray*}
    & ( x_1 \cdots x_k )^{- N} \mathbb{E}_{\tmop{SO} ( 2 N )} \prod_{j = 1}^k
    \det ( I +  x_j g ) = & \\
    & \mathbb{E}_{\tmop{SO} ( 2 N )} \prod_{n = 1}^N \prod_{i = 1}^k ( x_i +
    x_i^{- 1} + t_n  + t_n^{- 1} ) = & \\
    & \mathbb{E}_{\tmop{SO} ( 2 N )} ( \sum_{\lambda \subseteq N^k}  \chi_{\lambda_+}^{\tmop{SO} ( 2 k )} ( x_1^{\pm
    1}, \cdots, x_k^{\pm 1} ) \chi_{\widetilde{\lambda_+}}^{\tmop{SO} ( 2 N )}
    ( t_1^{\pm 1}, \cdots, t_N^{\pm 1} ) + & \\
    &  \chi_{\lambda_-}^{\tmop{SO} ( 2 k )} (
    x_1^{\pm 1}, \cdots, x_k^{\pm 1} ) \chi_{\widetilde{\lambda_-}}^{\tmop{SO}
    ( 2 N )} ( t_1^{\pm 1}, \cdots, t_N^{\pm 1} ) ) = & \\
    & \chi_{N^k}^{O ( 2 k )} ( x_1^{\pm 1}, \cdots, x_k^{\pm 1} ) . &
  \end{eqnarray*}
  In the last line we used the fact that
  \[ \mathbb{E}_{\tmop{SO} ( 2 N )} \chi_{\lambda}^{\tmop{SO} ( 2 N )} =
     \left\{\begin{array}{ll}
       1 & \text{if $\lambda = \varnothing$;}\\
       0 & \text{otherwise}
     \end{array}\right. \]
  and $\chi_{\lambda}^{O ( 2 n )} = \chi_{\lambda_+}^{\tmop{SO} ( 2 n )} +
  \chi_{\lambda_-}^{\tmop{SO} ( 2 n )}$ (see discussion preceding
  (\ref{charo})).

  Consequently we obtain
  \begin{equation}
    \label{e:so12} \mathbb{E}_{\tmop{SO} ( 2 N )} \prod_{j = 1}^k \det ( I
    +
    x_j g ) = ( x_1 \ldots x_k )^N \chi_{N^k}^{o_{2 k}} ( x_1^{\pm 1}, \cdots,
    x_k^{\pm 1} ),
  \end{equation}
  proving the first line of (\ref{e:soprod}).

  Now using the Weyl character formula for the orthogonal group (\ref{charo})
  and numerator evaluation (\ref{detso2}) we have:
  \begin{equation}\label{e:deto1}
  \begin{aligned}
&    ( x_1 \ldots x_k )^N \chi_{N^k}^{O ( 2 k )} ( x_1^{\pm 1},
\cdots,
    x_k^{\pm 1} ) = \\
    &\\
    &
    \det \left| \begin{array}{ccccc}
      x_1^{N + k - 1} + x_1^{- ( N + k - 1 )} & x_1^{N + k - 2} - x_1^{- ( N +
      k - 2 )} & \cdots & x_1^N - x_1^{- ( N )} & \\
      \vdots & \vdots & \ddots & \vdots & \\
      x_k^{N + k - 1} - x_k^{- ( N + k - 1 )} & x_k^{N + k - 2} - x_k^{- ( N +
      k - 2 )} & \cdots & x_k^N - x_k^{- ( N )} &
    \end{array} \right| \\
    &\\
    & \times \frac{( x_1 \cdots x_k )^{k + N - 1}}{\prod_{1 \leqslant i < j
    \leqslant k} ( x_i - x_j ) ( x_i x_j - 1 )} .
    \end{aligned}
  \end{equation}
  Next, splitting the determinant in this expression we can rewrite it as
  follows:
  \begin{eqnarray*}
    \sum_{\varepsilon \in \{ \pm 1 \}^k} \det \left| \begin{array}{ccccc}
      x_1^{\varepsilon_1 ( N + k - 1 )} & x_1^{\varepsilon_1 ( N + k - 2 )} &
      \cdots & x_1^{\varepsilon_1 ( N )} & \\
      \vdots & \vdots & \ddots & \vdots & \\
      x_k^{\varepsilon_k ( N + k - 1 )} & x_k^{\varepsilon_k ( N + k - 2 )} &
      \cdots & x_k^{\varepsilon_k ( N )} &
    \end{array} \right| & = & \\
    \sum_{\varepsilon \in \{ \pm 1 \}^k} \prod_{i = 1}^k x_i^{\varepsilon_i (
    N )} \det \left| \begin{array}{ccccc}
      x_1^{\varepsilon_1 ( k - 1 )} & x_1^{\varepsilon_1 ( k - 2 )} & \cdots &
      1 & \\
      \vdots & \vdots & \ddots & \vdots & \\
      x_k^{\varepsilon_k ( k - 1 )} & x_k^{\varepsilon_k ( k - 2 )} & \cdots &
      1 &
    \end{array} \right| & = & \\
    \sum_{\varepsilon \in \{ \pm 1 \}^k} \prod_{i = 1}^k x_i^{\varepsilon_i N}
    \prod_{i < j} ( x_i^{\varepsilon_i} - x_j^{\varepsilon_j} ), &  &
  \end{eqnarray*}
  where in the last line we have used the Vandermonde determinant evaluation.

  Now making use of the elementary identities (\ref{el2}) and (\ref{el3}) the
  right hand side of (\ref{e:deto1}) is easily seen to be equal to
  \[ \sum_{\varepsilon \in \{ \pm 1 \}} \prod_{j = 1}^k x_j^{- N
     \varepsilon_j} \prod_{i < j} ( 1 - x_i^{\varepsilon_i}
     x_j^{\varepsilon_j} )^{- 1}, \]
  completing the proof.
\end{proof}

We note that proceeding along exactly the same lines we easily
establish the following result:

\begin{proposition}\label{prodo} Notation being as above we have
\begin{equation} \begin{aligned}
&{\mathbb E}_{M \in O(2N)} \prod_{j=1}^k \det (I+x_j M)=(x_1 \dots
x_k)^{N} \chi_{(N^k)_{+}}^{\so_{2k}}(x_1^{\pm 1},\dots, x_k^{\pm 1})=\\
& \sum_{\substack{\varepsilon\in\{\pm 1\}\\
\sgn(\varepsilon)=1 }}\prod_{j=1}^{k}x_j^{N (1-\varepsilon_j) }
\prod_{i < j}(1-x_i^{\varepsilon_i}x_j^{\varepsilon_j})^{-1}.
\end{aligned}
\end{equation}
\end{proposition}

In particular, in analogy with the unitary and symplectic cases,
we have:
\begin{equation}
  \label{e:orth17} \mathbb{E}_{g \in O ( 2 N )} \; \det ( I - g )^k = \dim (
  \chi_{N^k}^{\tmop{SO} ( 2 k )} ) .
\end{equation}
The Weyl dimension formula for the the dimension of the
irreducible representation of the group $\tmop{SO} ( 2 k )$
labelled by partition $\lambda$ is given by
\[ \dim ( \chi_{\lambda}^{\tmop{SO} ( 2 k )} ) = 2^{k - 1} \frac{\prod_{i < j}
   ( \mu_i - \mu_j ) ( \mu_i + \mu_j )}{( 2 k - 2 ) ! ( 2 k - 4 ) ! \cdots
   2!}, \]
where $\mu_i = \lambda_i + k - i$.

An alternative expression, given by El-Samra and King {\cite{EK}},
is furnished by the following analogue of hook formula
(\ref{e_hhl}):
\begin{equation} \label{hookso}
\dim(\chi_{\lambda}^{\so_{2k}})=\prod_{u \in
\lambda}\frac{2k+c^{\so}(u)}{h(u)},
\end{equation}
where
\begin{equation*}
c^{\so}(i, j)=\begin{cases} i+j -\lambda'_i -\lambda'_j -2 \, \,
\text{if}\, \,  i < j \\
\lambda_i +\lambda_j -i-j \, \, \text{if} \, \, i \geqslant j.
\end{cases}
\end{equation*}
Applying (\ref{hookso}) to the partition $\lambda = N^k$ and
recalling that the product of hook numbers for partition $N^k$ is
given by (\ref{hook123}), we obtain, in view of (\ref{e:orth17}):
\begin{eqnarray*}
  \mathbb{E}_{g \in O ( 2 N )} \det ( I - g )^k = 2^k \frac{( N + k - 1 ) !}{(
  N - 1 ) ! ( k - 1 ) !} \times &  & \\
  \prod_{i = 1}^k \frac{( 2 i - 3 ) ! ( i + k + 2 N - 2 ) ! ( i + 2 N - 2 )
  !}{( i + N - 1 ) ! ( 2 i + 2 N - 2 ) ! ( i + k + N - 2 ) !} . &  &
\end{eqnarray*}
We remark that in {\cite{KS00}} Keating and Snaith derived the
following expression using Selberg's integral (without restriction
that $k$ be an integer):
\begin{equation}
  \label{kssotwo} \mathbb{E}_{\tmop{SO} ( 2 N )} \det ( I - g )^k = 2^{2 Nk}
  \prod_{j = 1}^N \frac{\Gamma ( N + j - 1 ) \Gamma ( k + j - \frac{1}{2}
  )}{\Gamma ( N + k + j - 1 ) \Gamma ( j - \frac{1}{2} )} ;
\end{equation}
for integer $k$ this is easily derived from (\ref{e:so12}) and the
Weyl dimension formula.

We now turn to ratios.

\begin{proposition}
  \label{prop:altso}Notation being as above we have
  \begin{equation} \label{e:altso1}
\begin{aligned}
&\sum_{\lambda \subseteq N^{k}}
\chi_{\lambda}^{\o_{2k}}(x_{1}^{\pm 1}, \dots, x_{k}^{\pm 1})
s_{\tilde{\lambda}}(y_1, \dots, y_l) = \\
& \sum_{\varepsilon\in\{\pm 1\}^k}
\prod_{j=1}^{k}x_j^{-N\varepsilon_j}\frac{\prod_{i=1}^{k}\prod_{j=1}^{l}(1+x_i^{\varepsilon_i}y_j)}
{\prod_{i < j}(1-x_i^{\varepsilon_i}x_j^{\varepsilon_j})}.
\end{aligned}
\end{equation}
\end{proposition}

\begin{proof}
  In order to yield a non-zero contribution to the sum on the left-hand side
  of (\ref{e:altso1}) $\lambda$ must be of the form $\lambda = ( N - l )^k +
  \mu$ with $\mu \subseteq l^k$. Now keeping in mind the Weyl character formula
  (\ref{charo}) and numerator evaluation (\ref{detso2}), together with the
  definition of the Schur polynomial, we can rewrite the expression on the
  left-hand side of (\ref{e:altso1}) as follows, using the Laplace expansion:
  \begin{equation}\label{e:detso2}
\begin{aligned}
& \det \left|
  \begin{array}{ccccc}
  x_1^{N+k-1}+x_1^{-(N+k-1)} &   x_1^{N+k-2}+x_1^{-(N+k-2)}&
\dots &  x_1^{N-l}+x_1^{-(N-l)}& \\
    \vdots & \vdots &\ddots & \vdots & \\
x_k^{N+k-1}+x_k^{-(N+k-1)} &   x_k^{N+k-2}+x_k^{-(N+k-2)}&
\dots &  x_k^{N-l}+x_k^{-(N-l)}& \\
(-y_1)^{l+k-1} &  (-y_1)^{l+k-2}& \dots & 1 &\\
\vdots & \vdots &\ddots & \vdots & \\
(-y_l)^{l+k-1} &  (-y_l)^{l+k-2}& \dots & 1 &
  \end{array} \right|\\
&\\
& \times \frac{1}{\prod_{1\leqslant i<j \leqslant l}(y_j-y_i)}
\frac {(x_1 \dots x_k)^{k-1}}{\prod_{1 \leqslant i<j \leqslant
k}(x_i-x_j)(x_i x_j-1) }.
\end{aligned}
\end{equation}
  Now splitting the determinant in this expression we can rewrite it as
  follows:
  \begin{eqnarray*}
    \sum_{\varepsilon \in \{ \pm 1 \}^k} \det \left| \begin{array}{ccccc}
      x_1^{\varepsilon_1 ( N+k-1 )} & x_1^{\varepsilon_1 ( N+k-2 )} &
      \cdots & x_1^{\varepsilon_1 ( N-l)} & \\
      \vdots & \vdots & \ddots & \vdots & \\
      x_k^{\varepsilon_k ( N+k-1 )} & x_k^{\varepsilon_k ( N+k-2)} &
      \cdots & x_k^{\varepsilon_k ( N-l )} & \\
     (-y_1)^{l+k-1} &  (-y_1)^{l+k-2}& \dots & 1 &\\
\vdots & \vdots &\ddots & \vdots & \\
(-y_l)^{l+k-1} &  (-y_l)^{l+k-2}& \dots & 1 &
    \end{array} \right| & = & \\
    \sum_{\varepsilon \in \{ \pm 1 \}^k} \prod_{i = 1}^k x_i^{\varepsilon_i (
    N - l )} \det \left| \begin{array}{ccccc}
      x_1^{\varepsilon_1 ( l + k - 1 )} & x_1^{\varepsilon_1 ( l + k - 2 )}& \cdots
      & 1 &
      \\
      \vdots & \vdots & \ddots & \vdots & \\
      x_k^{\varepsilon_k ( l + k - 1 )} & x_k^{\varepsilon_k ( l + k - 2 )} & \cdots & 1 &
      \\
      (-y_1)^{l+k-1} &  (-y_1)^{l+k-2}& \dots & 1 &\\
\vdots & \vdots &\ddots & \vdots & \\
(-y_l)^{l+k-1} &  (-y_l)^{l+k-2}& \dots & 1 &
    \end{array} \right| & = & \\
    \sum_{\varepsilon \in \{ \pm 1 \}^k} \prod_{i = 1}^k x_i^{\varepsilon_i (
    N - l )} \prod_{i < j} ( x_i^{\varepsilon_i} - x_j^{\varepsilon_j} )
    \prod_{1 \leqslant i < j \leqslant l} ( y_j - y_i ) \prod_{i = 1}^k
    \prod_{j = 1}^l ( x_i^{\varepsilon_i} + y_j ), &  &
  \end{eqnarray*}
  where in the last line we have used the Vandermonde determinant evaluation.

  Next, making use of the elementary identities (\ref{el2}), (\ref{el3}), the
  expression on the right-hand side of (\ref{e:detso2}) is easily brought to
  the form
  \[ \sum_{\varepsilon \in \{ \pm 1 \}^k} \prod_{j = 1}^k x_j^{- N
     \varepsilon_j} \frac{\prod_{i = 1}^k \prod_{j = 1}^l ( 1 +
     x_i^{\varepsilon_i} y_j )}{\prod_{i < j} ( 1 - x_i^{\varepsilon_i}
     x_j^{\varepsilon_j} )}, \]
  proving the proposition.
\end{proof}

\begin{theorem}
  \label{ratso}Assume $N \geqslant l$ and $|y_i| <1$. Then we
  have:
  \begin{eqnarray*}
    \mathbb{E}_{g \in SO ( 2 N )} \frac{\prod_{j = 1}^k \det ( I + x_j g
    )}{\prod_{i = 1}^l \det ( I - y_i g )} = & &\\
    \sum_{\varepsilon \in \{ \pm 1 \}} \prod_{j = 1}^k x_j^{N ( 1 -
    \varepsilon_j )} \frac{\prod_{i = 1}^k \prod_{j = 1}^l ( 1 +
    x_i^{\varepsilon_i} y_j )}{\prod_{i < j} ( 1 - x_i^{\varepsilon_i}
    x_j^{\varepsilon_j} ) \prod_{1 \leqslant i \leqslant j \leqslant l} ( 1 -
    y_i y_j )} . & &
  \end{eqnarray*}

\end{theorem}

\begin{proof}
  The Cauchy identity for $\tmop{SO} ( 2 N )$ has the following form:
  \begin{eqnarray}
    \frac{1}{\prod_{n = 1}^N \prod_{j = 1}^l ( 1 - y_j t_n ) ( 1 - y_j t_n^{-
    1} )} & = & \nonumber\\
    \frac{1}{\prod_{i \leqslant j} ( 1 - y_i y_j )} \sum_{\mu} \chi_{\mu}^{O (
    2 N )} ( t_1^{\pm 1}, \cdots, t_N^{\pm 1} ) s_{\mu} ( y_1, \cdots, y_l ) .
    &  &\label{cauchyo}
  \end{eqnarray}
  Combining (\ref{cauchyo}) and (\ref{e:cfso1}) we have:
  \begin{equation*}
\begin{aligned}
&\frac{\prod_{j=1}^k \det (I+x_j M)}{\prod_{i=1}^l \det (I-y_i
M)}= \frac{(x_1 \dots x_k)^{N}}{\prod_{i\leqslant j}(1-y_i y_j)}
\times
\\&\sum_{\mu}\chi_{\mu}^{\o_{2N}}(t_1^{\pm 1}, \dots, t_N^{\pm 1})
s_{\mu}(y_1, \dots, y_l) \times\\ &\sum_{\lambda \subseteq N^{k}}(
\chi_{\lambda_{+}}^{\so_{2k}}(x_{1}^{\pm 1}, \dots, x_{k}^{\pm 1})
\chi_{\tilde{\lambda}_{+}}^{\so_{2N}}(t_1^{\pm 1},
\dots, t_N^{\pm 1})+\\
& \chi_{\lambda_{-}}^{\so_{2k}}(x_{1}^{\pm 1}, \dots, x_{k}^{\pm
1}) \chi_{\tilde{\lambda}_{-}}^{\so_{2N}}(t_1^{\pm 1}, \dots,
t_N^{\pm 1}))
\end{aligned}
\end{equation*}
  Since
\begin{equation}
\mathbb{E}_{\SO(2N)} \chi_{\lambda_{\pm}}^{\so_{2N}}(M)
\chi_{\mu}^{\o_{2N}}(M) =
\begin{cases} 1, &\text{if $\lambda = \mu,\, l(\lambda) \leqslant N$;}\\
0, &\text{otherwise},
\end{cases}
\end{equation}
  the result now follows from Proposition~\ref{prop:altso}.
\end{proof}

We note that proceeding along the same lines we easily establish
the following result.

\begin{theorem}
  \label{thm:o1}Assume $N \geqslant l$ and $|y_i| <1$. Then we
  have:
  \begin{eqnarray*}
     \mathbb{E}_{O ( 2 n )} \frac{\prod_{j = 1}^k \det ( I + x_j g
    )}{\prod_{i = 1}^l \det ( I - y_i g )} = & &\\
    \sum_{\tmscript{\begin{array}{c}
      \varepsilon \in \{ \pm 1 \}\\
      \tmop{sgn} ( \varepsilon ) = 1
    \end{array}}} \prod_{j = 1}^k x_j^{N ( 1 - \varepsilon_j )} \frac{\prod_{i
    = 1}^k \prod_{j = 1}^l ( 1 + x_i^{\varepsilon_i} y_j )}{\prod_{i < j} ( 1
    - x_i^{\varepsilon_i} x_j^{\varepsilon_j} ) \prod_{1 \leqslant i \leqslant
    j \leqslant l} ( 1 - y_i y_j )} . & &
  \end{eqnarray*}
\end{theorem}

\subsection{Products and ratios for $\tmop{SO} ( 2 N + 1 )$.}

\begin{lemma} \label{lem:sod}
  For $\lambda \subseteq N^k$ let
  $\tilde{\lambda} = ( k - \lambda_N', \cdots, k- \lambda_1' )$. Then we have
  \begin{equation}\label{e:cfsod1}
  \begin{aligned}
&    \prod_{i = 1}^k \prod_{n = 1}^N ( x_i + x_i^{- 1} - t_n -
t_n^{- 1} )  = \\
&    \sum_{\lambda \subseteq N^k} ( - 1 )^{| \tilde{\lambda} |}
    \chi_{\lambda}^{\tmop{SO} ( 2 k + 1 )} ( x_1^{\pm 1}, \cdots, x_k^{\pm 1},
    1 ) \chi_{\tilde{\lambda}}^{\tmop{SO} ( 2 N + 1 )} ( t_1^{\pm 1}, \cdots,
    t_N^{\pm 1}, 1 ) .
\end{aligned}
\end{equation}
\end{lemma}

\begin{proof}
  The lemma is stated in slightly different form in Jimbo and
  Miwa~{\cite{JM}}. It may be proved by the method of  Lemma~\ref{unidecomp}
  and
  Lemma~\ref{symplecticdecomposition}.
\end{proof}

\begin{proposition}
  \label{prop:sod}Notation being as above we have
  \begin{eqnarray}
    \mathbb{E}_{\tmop{SO} ( 2 N + 1 )} \prod_{j = 1}^k \det ( I - x_j g ) & =
    &  \nonumber\\
    ( x_1 \ldots x_k )^N \prod_{i = 1}^k ( 1 - x_i ) \chi_{N^k}^{\tmop{SO} ( 2
    k + 1 )} ( x_1^{\pm 1}, \cdots, x_k^{\pm 1}, 1 ) & = &  \nonumber\\
    \sum_{\varepsilon \in \{ \pm 1 \}} \tmop{sgn} ( \varepsilon ) \prod_{j =
    1}^k x_j^{( N + 1 / 2 ) ( 1 - \varepsilon_j )} \prod_{i < j} ( 1 -
    x_i^{\varepsilon_i} x_j^{\varepsilon_j} )^{- 1} . &  &  \label{e:sod}
  \end{eqnarray}
\end{proposition}

\begin{proof}
  Using (\ref{e:cfsod1}) we have:
  \begin{eqnarray*}
    ( x_1 \cdots x_k )^{- N} \mathbb{E}_{\tmop{SO} ( 2 N + 1 )} \prod_{j =
    1}^k \det ( I - x_j g ) = &  & \\
    \prod_{i = 1}^k ( 1 - x_i ) \mathbb{E}_{\tmop{SO} ( 2 N + 1 )} \prod_{n =
    1}^N \prod_{i = 1}^k ( x_i + x_i^{- 1} - t_n - t_n^{- 1} ) = &  & \\
    \mathbb{E}_{\tmop{SO} ( 2 N + 1 )} \sum_{\lambda \subseteq N^k} ( - 1 )^{|
    \tilde{\lambda} |} \chi_{\lambda}^{\tmop{SO} ( 2 k + 1 )} ( x_1^{\pm 1},
    \cdots, x_k^{\pm 1}, 1 ) \chi_{\tilde{\lambda}}^{\tmop{SO} ( 2 N + 1 )} (
    t_1^{\pm 1}, \cdots, t_N^{\pm 1}, 1 ) = &  & \\
    \chi_{N^k}^{\tmop{SO} ( 2 k + 1 )} ( x_1^{\pm 1}, \cdots, x_k^{\pm 1}, 1 )
    . &  &
  \end{eqnarray*}
  In the last line we used the fact that
  \[ \mathbb{E}_{\tmop{SO} ( 2 N + 1 )} \chi_{\lambda}^{\tmop{SO} ( 2 N + 1 )}
     = \left\{\begin{array}{ll}
       1 & \text{if $\lambda =$} \varnothing ;\\
       0 & \text{otherwise} .
     \end{array}\right. \]
  Consequently we obtain
  \begin{equation}
    \label{e:sod12} \mathbb{E}_{\tmop{SO} ( 2 N + 1 )} \prod_{i = 1}^k \det (
    I - x_i g ) = ( x_1 \ldots x_k )^N \prod_{i = 1}^k ( 1 - x_i )
    \chi_{N^k}^{\tmop{SO} ( 2 k + 1 )} ( x_1^{\pm 1}, \cdots, x_k^{\pm 1}, 1
    ),
  \end{equation}
  proving the first line of (\ref{e:sod}).

  Next, using the Weyl character formula for $\tmop{SO}(2n + 1 )$ together with
  the Weyl denominator formula for $\tmop{SO} ( 2 n + 1 )$ we can rewrite the
  right-hand side of (\ref{e:sod12}) as follows:
  \begin{eqnarray}
    ( x_1 \ldots x_k )^N \prod_{i = 1}^k ( 1 - x_i ) \chi_{N^k}^{\tmop{SO} ( 2
    k + 1 )} ( x_1^{\pm 1}, \cdots, x_k^{\pm 1}, 1 ) & = & \label{e:detsod1}\\
    \det ( x_i^{N + k + \frac{1}{2} - j} - x_i^{- ( N + k + \frac{1}{2} - j )}
    )_{1 \leqslant i, j \leqslant k} \frac{( x_1 \cdots x_k )^{k + N - 1 /
    2}}{\prod_{1 \leqslant i < j \leqslant k} ( x_i - x_j ) ( x_i x_j - 1 )} .
    &  &  \nonumber
  \end{eqnarray}
  Splitting the determinant in this expression we can rewrite it as follows:
  \begin{eqnarray*}
    \sum_{\varepsilon \in \{ \pm 1 \}^k} \det \left| \begin{array}{ccccc}
      \varepsilon_1 x_1^{\varepsilon_1 ( N + k - 1 / 2 )} & \varepsilon_1
      x_1^{\varepsilon_1 ( N + k - 3 / 2 )} & \cdots & \varepsilon_1
      x_1^{\varepsilon_1 ( N + 1 / 2 )} & \\
      \vdots & \vdots & \ddots & \vdots & \\
      \varepsilon_k x_k^{\varepsilon_k ( N + k - 1 / 2 )} & \varepsilon_k
      x_k^{\varepsilon_k ( N + k - 3 / 2 )} & \cdots & \varepsilon_k
      x_k^{\varepsilon_k ( N + 1 / 2 )} &
    \end{array} \right| & = & \\
    \sum_{\varepsilon \in \{ \pm 1 \}^k} \prod_{i = 1}^k \varepsilon_i
    x_i^{\varepsilon_i ( N + 1 / 2 )} \det \left| \begin{array}{ccccc}
      x_1^{\varepsilon_1 ( k - 1 )} & x_1^{\varepsilon_1 ( k - 2 )} & \cdots &
      1 & \\
      \vdots & \vdots & \ddots & \vdots & \\
      x_k^{\varepsilon_k ( k - 1 )} & x_k^{\varepsilon_k ( k - 2 )} & \cdots &
      1 &
    \end{array} \right| & = & \\
    \sum_{\varepsilon \in \{ \pm 1 \}^k} \prod_{i = 1}^k \varepsilon_i
    x_i^{\varepsilon_i ( N + 1 / 2 )} \prod_{i < j} ( x_i^{\varepsilon_i} -
    x_j^{\varepsilon_j} ), &  &
  \end{eqnarray*}
  where in the last line we have used the Vandermonde determinant evaluation.

  Now making use of the elementary identities (\ref{el2}) and (\ref{el3}) the
  right hand side of (\ref{e:detsod1}) is easily seen to be equal to
  \begin{eqnarray*}
    \sum_{\varepsilon \in \{ \pm 1 \}^k} \tmop{sgn} ( \varepsilon ) \prod_{j =
    1}^k x_j^{( N + 1 / 2 ) ( 1 + \varepsilon_j )} \prod_{i < j} ( 1 - x_i^{-
    \varepsilon_i} x_j^{- \varepsilon_j} )^{- 1} & = & \\
    \sum_{\varepsilon \in \{ \pm 1 \}^k} \tmop{sgn} ( \varepsilon ) \prod_{j =
    1}^k x_j^{( N + 1 / 2 ) ( 1 - \varepsilon_j )} \prod_{i < j} ( 1 -
    x_i^{\varepsilon_i} x_j^{\varepsilon_j} )^{- 1}, &  & \\
    &  &
  \end{eqnarray*}
  completing the proof.
\end{proof}

We remark that proceeding along similar lines we easily establish
the following result.

\begin{proposition}
  \label{prosmin}Notation being as above we have
  \begin{eqnarray*}
    \mathbb{E}_{g \in O^- ( 2 N )} \prod_{j = 1}^k \det ( I + x_j g ) = ( x_1
    \cdots x_k )^N \chi_{( N^k )_-}^{\tmop{SO} ( 2 k )} ( x_1^{\pm 1}, \cdots,
    x_k^{\pm 1} ) & = & \\
    \sum_{\varepsilon \in \{ \pm 1 \}} \tmop{sgn} ( \varepsilon ) \prod_{j =
    1}^k x_j^{N ( 1 - \varepsilon_j )} \prod_{i < j} ( 1 - x_i^{\varepsilon_i}
    x_j^{\varepsilon_j} )^{- 1} . &  &
  \end{eqnarray*}
\end{proposition}

We now turn to ratios.

\begin{proposition}
  \label{prop:altsod}Notation being as above we have
  \begin{eqnarray}
    \sum_{\lambda \subseteq N^k} ( - 1 )^{| \tilde{\lambda} |}
    \chi_{\lambda}^{\tmop{SO} ( 2 k + 1 )} ( x_1^{\pm 1}, \cdots, x_k^{\pm 1},
    1 ) s_{\tilde{\lambda}} ( y_1, \cdots, y_l ) & = &  \label{e:altsod1}\\
    \frac{( x_1 \cdots x_k )^{1 / 2}}{\prod_{i = 1}^k ( 1 - x_i )}
    \sum_{\varepsilon \in \{ \pm 1 \}^k} \tmop{sgn} ( \varepsilon ) \prod_{j =
    1}^k x_j^{- ( N + 1 / 2 ) \varepsilon_j} \frac{\prod_{i = 1}^k \prod_{j =
    1}^l ( 1 + x_i^{\varepsilon_i} y_j )}{\prod_{i < j} ( 1 -
    x_i^{\varepsilon_i} x_j^{\varepsilon_j} )} . &  &  \nonumber
  \end{eqnarray}
\end{proposition}

\begin{proof}
  In order to yield a non-zero contribution to the sum on the left-hand side
  of (\ref{e:altsod1}), $\lambda$ must be of the form $\lambda = ( N - l )^k +
  \mu$ with $\mu \subseteq l^k$. Now keeping in mind the Weyl character
  formula (\ref{charsod}) for $\tmop{SO} ( 2 n + 1 )$  and the numerator
  evaluation (\ref{detsod}), together with the definition
  (\ref{schurdefinition}) of the Schur polynomial, and the Vandermonde
  determinant evaluation of the denominator in that formula, we can rewrite
  the expression on the left-hand side of (\ref{e:altsod1}) as follows, using
  the Laplace expansion:
  \begin{eqnarray}
    \det \left| \begin{array}{cccc}
     X_{1, k + l} & X_{1, k+l-1} & \cdots & X_{11}\\
      \vdots & \vdots & \ddots & \vdots\\
      X_{k, k+l} & X_{k, k+l-1} & \cdots & X_{k1}\\
      ( - y_1 )^{l + k - 1} & ( - y_1 )^{l + k - 2}& \cdots & 1\\
      \vdots & \vdots & \ddots & \vdots\\
      ( - y_l )^{l + k - 1} & ( - y_l )^{l + k - 2}& \cdots & 1
    \end{array} \right| & \times &  \nonumber\\
    \prod_{1 \leqslant i < j \leqslant l} \frac{1}{y_j - y_i} \prod_{1
    \leqslant i < j \leqslant k} \frac{( x_1 \cdots x_k )^{k - 1 / 2}}{( x_i -
    x_j ) ( x_i x_j - 1 )} \prod_{i = 1}^k \frac{1}{1 - x_i}, &  &
    \label{e:detsod2}
  \end{eqnarray}
\medbreak\noindent
  where $X_{i j} = x_i^{N - l - \frac{1}{2} + j} - x_i^{- ( N - l - \frac{1}{2} +  j )}$.
  Splitting the determinant in this expression we can rewrite it as follows:
  \begin{eqnarray*}
    \sum_{\varepsilon \in \{ \pm 1 \}^k} \det \left| \begin{array}{ccccc}
      \varepsilon_1 x_1^{\varepsilon_1 ( N+k- 1 / 2 )} & \varepsilon_1
      x_1^{\varepsilon_1 ( N+k-3 / 2 )} & \cdots & \varepsilon_1
      x_1^{\varepsilon_1 ( N -l+ 1 / 2 )} & \\
      \vdots & \vdots & \ddots & \vdots & \\
      \varepsilon_k x_k^{\varepsilon_k ( N+k-1 / 2 )} & \varepsilon_k
      x_k^{\varepsilon_k ( N+k-3 / 2 )} & \cdots & \varepsilon_k
      x_k^{\varepsilon_k ( N-l+1 / 2 )} & \\
       ( - y_1 )^{l + k - 1} & ( - y_1 )^{l + k - 2}& \cdots & 1\\
      \vdots & \vdots & \ddots & \vdots\\
      ( - y_l )^{l + k - 1} & ( - y_l )^{l + k - 2}& \cdots & 1
    \end{array} \right| & = & \\
    \sum_{\varepsilon \in \{ \pm 1 \}^k} \prod_{i = 1}^k \varepsilon_i
    x_i^{\varepsilon_i ( N - l + 1 / 2 )} \det \left| \begin{array}{ccccc}
      1 & x_1^{\varepsilon_1} & \cdots & x_1^{\varepsilon_1 ( l + k - 1 )} &
      \\
      \vdots & \vdots & \ddots & \vdots & \\
      1 & x_k^{\varepsilon_k} & \cdots & x_k^{\varepsilon_k ( l + k - 1 )} &
      \\
     ( - y_1 )^{l + k - 1} & ( - y_1 )^{l + k - 2}& \cdots & 1\\
      \vdots & \vdots & \ddots & \vdots\\
      ( - y_l )^{l + k - 1} & ( - y_l )^{l + k - 2}& \cdots & 1
    \end{array} \right| &  &
  \end{eqnarray*}
  or, using the Vandermonde determinant evaluation,
  \[ \sum_{\varepsilon \in \{ \pm 1 \}^k} \prod_{i = 1}^k \varepsilon_i
     x_i^{\varepsilon_i ( N - l + 1 / 2 )} \prod_{i < j} ( x_i^{\varepsilon_i}
     - x_j^{\varepsilon_j} ) \prod_{1 \leqslant i < j \leqslant l} ( y_j - y_i
     ) \prod_{i = 1}^k \prod_{j = 1}^l ( x_i^{\varepsilon_i} + y_j ), \]
  Now making use of the elementary identities (\ref{el2}), (\ref{el3}), the
  expression on the right-hand side of (\ref{e:detsod2}) is easily brought to
  the form expressed on the right-hand of (\ref{e:altsod1}).
\end{proof}

\begin{theorem}
  \label{ratsod}Assume $N \geqslant l$ and $|y_i| <1$. Then we
  have:
  \begin{eqnarray*}
\mathbb{E}_{\tmop{SO} ( 2 N + 1 )} \frac{\prod_{j = 1}^k \det ( I-
x_j g )}{\prod_{i = 1}^l \det ( I - y_i g
    )} & = & \\
    \sum_{\varepsilon \in \{ \pm 1 \}} \tmop{sgn} ( \varepsilon ) \prod_{j =
    1}^k x_j^{( N + 1 / 2 ) ( 1 - \varepsilon_j )} \frac{\prod_{i = 1}^k
    \prod_{j = 1}^l ( 1 + x_i^{\varepsilon_i} y_j )}{\prod_{i < j} ( 1 -
    x_i^{\varepsilon_i} x_j^{\varepsilon_j} ) \prod_{1 \leqslant i \leqslant j
    \leqslant l} ( 1 - y_i y_j )} . &  &
  \end{eqnarray*}
\end{theorem}

\begin{proof}
  The Cauchy identity for $\tmop{SO} ( 2 N + 1 )$ has the following form:
  \begin{eqnarray}
    &&\prod_{n = 1}^N \prod_{j = 1}^l \frac{1}{( 1 - y_j t_n ) ( 1 - y_j t_n^{-
    1} )} \prod_{j = 1}^l \frac{1}{1 - y_j} =  \nonumber\\&&
    \left( \prod_{i \leqslant j} \frac{1}{1 - y_i y_j} \right) \sum_{\mu}
    \chi_{\mu}^{\tmop{SO} ( 2 N + 1 )} ( t_1^{\pm 1}, \cdots, t_N^{\pm 1}, 1 )
    s_{\mu} ( y_1, \cdots, y_l ) .  \label{e:cauchysod}
  \end{eqnarray}
  Consequently, using (\ref{e:cauchysod}) and (\ref{e:cfsod1}) we have
  \begin{eqnarray*}
    \frac{\prod_{j = 1}^k \det ( I - x_j g )}{\prod_{i = 1}^l \det ( I - y_i g
    )} = \frac{( x_1 \cdots x_k )^N \prod_{i = 1}^k ( 1 - x_i )}{\prod_{i
    \leqslant j} ( 1 - y_i y_j )} \times &  & \\
    \sum_{\lambda \subseteq N^k} ( - 1 )^{| \tilde{\lambda} |}
    \chi_{\lambda}^{\tmop{SO} ( 2 k + 1 )} ( x_1^{\pm 1}, \cdots, x_k^{\pm 1},
    1 ) \chi_{\tilde{\lambda}}^{\tmop{SO} ( 2 N + 1 )} ( t_1^{\pm 1}, \cdots,
    t_N^{\pm 1}, 1 ) \times &  & \\
    \sum_{\mu} \chi_{\mu}^{\tmop{SO} ( 2 N + 1 )} ( t_1^{\pm 1}, \cdots,
    t_N^{\pm 1}, 1 ) s_{\mu} ( y_1, \cdots, y_l ) . &  &
  \end{eqnarray*}
  Since
  \[ \mathbb{E}_{\tmop{SO} ( 2 N + 1 )} \chi_{\lambda}^{\tmop{SO} ( 2 N + 1 )}
     ( g ) \chi_{\mu}^{\tmop{SO} ( 2 N + 1 )} ( g ) = \left\{\begin{array}{ll}
       1 & \text{if $\lambda = \mu$,} l ( \lambda ) \leqslant N ;\\
       0 & \text{otherwise,}
     \end{array}\right. \]
  The theorem now follows from Proposition~\ref{prop:altsod}.
\end{proof}

We note that proceeding along similar lines we easily establish
the following result.

\begin{theorem}
  \label{thm:o3}Suppose $N \geqslant l$ and $|y_i| <1$. Then we
  have:
  \begin{eqnarray*}
    \mathbb{E}_{g \in O^- ( 2 N )} \frac{\prod_{j = 1}^k \det ( I + x_j g
    )}{\prod_{i = 1}^l \det ( I - y_i g )} & = & \\
    \sum_{\varepsilon \in \{ \pm 1 \}} \tmop{sgn} ( \varepsilon ) \prod_{j =
    1}^k x_j^{N ( 1 - \varepsilon_j )} \frac{\prod_{i = 1}^k \prod_{j = 1}^l (
    1 + x_i^{\varepsilon_i} y_j )}{\prod_{i < j} ( 1 - x_i^{\varepsilon_i}
    x_j^{\varepsilon_j} ) \prod_{1 \leqslant i \leqslant j \leqslant l} ( 1 -
    y_i y_j )} . &  &
  \end{eqnarray*}

\end{theorem}

\section{Classical group characters of rectangular shape}\label{rect}

In this section using results of Sections \ref{sec:sp} and
\ref{sec:o} we give simple proofs of identities of Okada
{\cite{ok}} and Krattenthaler {\cite{krat}} and derive
generalizations of their results for Littlewood-Schur functions.

\subsection{Symplectic Group}

The branching rule (\ref{e:bsym}) implies
\begin{equation}
  \label{eschursymp} \mathbb{E}_{\tmop{Sp} ( 2 N )} s_{\lambda} ( g ) =
  \left\{\begin{array}{ll}
    1 & \text{if $\lambda'$ is even} ;\\
    0 & \tmop{otherwise} .
  \end{array}\right.
\end{equation}
The importance of (\ref{eschursymp}) in random matrix theory
context was emphasized by Rains {\cite{Ra98}} and Baik and Rains
{\cite{BR02}}.

Using (\ref{eschursymp}) we obtain:
\begin{equation}
  \label{elaba} \mathbb{E}_{Sp ( 2 N )} \prod_{j = 1}^k \det ( I + x_j g ) =
  \sum_{\tmscript{\begin{array}{c}
    \lambda_1 \leqslant 2 N\\
    \text{$\lambda$ even}
  \end{array}}} s_{\lambda} ( x_1, \ldots x_k ),
\end{equation}
an identity first noted by Baik and Rains {\cite{BR02}}.

Combining this with Proposition \ref{prop:symp}, we have
\begin{eqnarray}
  \sum_{\tmscript{\begin{array}{c}
    \lambda_1 \leqslant 2 N\\
    \text{$\lambda$ even}
  \end{array}}} s_{\lambda} ( x_1, \ldots x_k ) = ( x_1 \ldots x_k )^N
\chi_{N^k}^{\sp_{2k}}( x_1^{\pm 1}, \cdots, x_k^{\pm 1} ) = &  &
  \nonumber\\
  \sum_{\varepsilon \in \{ \pm 1 \}} \prod_{j = 1}^k x_j^{N ( 1 -
  \varepsilon_j )} \prod_{i \leqslant j} ( 1 - x_i^{\varepsilon_i}
  x_j^{\varepsilon_j} )^{- 1}, &  &  \label{okrat1}
\end{eqnarray}
an identity first derived by Okada {\cite{ok}} and
Krattenthaler~{\cite{krat}}.

Using Theorem \ref{ratsp} we now prove the following
generalization of (\ref{okrat1}):

\begin{proposition} Assume $N \geqslant l$. Then we
  have:
  \label{prop:ok1}
  \begin{eqnarray*}
    \sum_{\tmscript{\begin{array}{c}
      \lambda_1 \leqslant 2 N\\
      \text{$\lambda$ even}
    \end{array}}} \tmop{LS}_{\nu} ( x_1, \ldots, x_k ; y_1, \ldots, y_l ) = &
    & \\
    \sum_{\varepsilon \in \{ \pm 1 \}^k} \prod_{j = 1}^k x_j^{N ( 1 -
    \varepsilon_j )} \frac{\prod_{i = 1}^k \prod_{j = 1}^l ( 1 +
    x_i^{\varepsilon_i} y_j )}{\prod_{i \leqslant j} ( 1 - x_i^{\varepsilon_i}
    x_j^{\varepsilon_j} ) \prod_{1 \leqslant i < j \leqslant l} ( 1 - y_i y_j
    )} . &  &
  \end{eqnarray*}
\end{proposition}

\begin{proof}
  By the dual Cauchy identity
  \[ \prod_{j = 1}^k \det ( I + x_j g ) = \sum_{\eta} s_{\eta} ( x_1, \ldots,
     x_k ) s_{\eta'} ( t_1^{\pm 1}, \ldots, t_N^{\pm 1} ) . \]
  On the other hand, by the Cauchy identity
  \[ \frac{1}{\prod_{i = 1}^l \det ( I - y_i g )} = \sum_{\mu} s_{\mu} ( y_1,
     \ldots, y_l ) s_{\mu} ( t_1^{\pm 1}, \ldots, t_N^{\pm 1} ) . \]
  Consequently, using (\ref{eschursymp}), we have
\begin{equation}\label{e:1971}
\begin{aligned}
& {\mathbb E}_{M \in Sp(2N)} \frac{\prod_{j=1}^k \det (I+ x_j
M)}{\prod_{i=1}^l \det (I-y_i M)} = \sum_{\substack{l(\nu) \le
2N\\\nu' \, \text{even}}}c^{\nu}_{\eta' \mu} s_{\eta}(x_1,
\ldots, x_k) s_{\mu}(y_1, \ldots, y_l) =\\
&\sum_{\substack{l(\nu) \leqslant 2N\\\nu' \,
\text{even}}}\HS_{\nu}(y_1, \ldots, y_l; x_1, \ldots,
x_k)=\sum_{\substack{\nu_1 \leqslant 2N\\\nu \text{even}}}
\HS_{\nu}(x_1, \ldots, x_k; y_1, \ldots, y_l),
\end{aligned}
\end{equation}
an identity first noted by Baik and Rains  \cite{BR02}. Combining
\eqref{e:1971} with Theorem \ref{ratsp} completes the proof.
\end{proof}

We remark that Krattenthaler {\cite{krat}} (3.6) proved the
following generalization of (\ref{okrat1}):
\begin{equation}
  \label{kratgen} \sum_{\tmscript{\begin{array}{c}
    \lambda \subseteq ( 2 N )^k\\
    r ( \lambda ) = r
  \end{array}}} s_{\lambda} ( x_1, \ldots, x_k ) = ( x_1 \cdots x_k )^N
  \chi_{N^{k - r} \cup ( N - 1 )^r}^{\tmop{Sp}(2 k)} ( x_1^{\pm 1}, \cdots,
  x_k^{\pm 1} )
\end{equation}
where $r ( \lambda )$ denotes the number of odd rows of $\lambda$.

We now show how (\ref{kratgen}) can be used to give an alternative
proof of Theorem \ref{ratsp}. In (\ref{e:1971}) write $\nu =
\lambda \cup \tau$ with $\lambda \in ( 2 N )^k$. By the
Berele-Regev factorization (\ref{e_br}) we have
\[ \tmop{LS}_{\nu} ( x_1, \ldots, x_k ; y_1, \ldots, y_l ) =
   \tmop{LS}_{\lambda} ( x_1, \ldots, x_k ; y_1, \ldots, y_l ) s_{\tau'} (
   y_1, \ldots, y_l ) . \]
Consequently,
\begin{eqnarray*}
  \sum_{\tmscript{\begin{array}{c}
    \nu_1 \leqslant 2 N\\
    \text{$\nu$ even}
  \end{array}}} \tmop{LS}_{\nu} ( x_1, \ldots, x_k ; y_1, \ldots, y_l ) =
  \sum_{\tau' \text{even}} s_{\tau} ( y_1, \ldots, y_l ) \times &  & \\
  \sum_{\tmscript{\begin{array}{c}
    \lambda \subseteq ( 2 N )^k\\
    \text{$\lambda$ even}
  \end{array}}} \tmop{LS}_{\lambda} ( x_1, \ldots, x_k ; y_1, \ldots, y_l ) .
  &  &
\end{eqnarray*}
Since
\begin{equation}
  \sum_{\text{$\tau'$ even}} s_{\tau} ( y_1, \ldots, y_l ) = \frac{1}{\prod_{1
  \leqslant i < j \leqslant l} ( 1 - y_i y_j )},
\end{equation}
Theorem \ref{ratsp} follows from the following Proposition
(applied with $u = 0$).

\begin{proposition}
  \label{prop:sp3}Let $r ( \lambda )$ denote the number of odd rows of
  $\lambda$. Then
  \begin{eqnarray*}
    \sum_{\lambda \subseteq ( 2 N )^k} u^{r ( \lambda )} \tmop{LS}_{\lambda} (
    x_1, \ldots, x_k ; y_1, \ldots, y_l ) & = & \\
    \sum_{\varepsilon \in \{ \pm 1 \}^k} \prod_{j = 1}^k x_j^{N ( 1 -
    \varepsilon_j )} \frac{\prod_{i = 1}^k \prod_{j = 1}^l ( 1 +
    x_i^{\varepsilon_i} y_j ) \prod_{i = 1}^k ( 1 + x_i^{\varepsilon_i} u
    )}{\prod_{i \leqslant j} ( 1 - x_i^{\varepsilon_i} x_j^{\varepsilon_j} )}
    &  &
  \end{eqnarray*}
\end{proposition}

Proposition \ref{prop:sp3} in turn follows from the following
lemma by induction on the number of variables $y$ using the
generalized Pieri formula given in Proposition \ref{genpieri}.

\begin{lemma}
  \label{lem:sp3}Let $r ( \lambda )$ denote the number of odd rows of
  $\lambda$. Then
  \begin{equation}
    \label{elabb} \sum_{\lambda \subseteq ( 2 N )^k} u^{r ( \lambda )}
    s_{\lambda} ( x_1, \ldots, x_k ) = \sum_{\varepsilon \in \{ \pm 1 \}^k}
    \prod_{j = 1}^k x_j^{N ( 1 - \varepsilon_j )} \frac{\prod_{i = 1}^k ( 1 +
    x_i^{\varepsilon_i} u )}{\prod_{i \leqslant j} ( 1 - x_i^{\varepsilon_i}
    x_j^{\varepsilon_j} )}.
  \end{equation}
\end{lemma}

We remark that (\ref{elabb}) generalizes the classical formula due
to Littlewood {\cite{Li}}
\begin{equation}
  \sum_{\lambda} u^{r ( \lambda )} s_{\lambda} ( x_1, \ldots, x_k ) = \prod_{i
  = 1}^k \frac{1 + ux_i}{1 - x_i^2} \prod_{i < j} \frac{1}{1 - x_i x_j} .
\end{equation}
We now give a proof of Lemma \ref{lem:sp3} using Krattenthaler's
formula (\ref{kratgen}).

Let
$$\varphi_{k, N}(r) = (x_1 \dots x_k)^N \chi_{N^{k-r}
\cup (N-1)^r}^{\sp_{2k}}(x_1^{\pm 1}, \dots, x_k^{\pm 1}).$$ We
have
\begin{equation}
  \label{e:phi0} \sum_{\lambda \subseteq ( 2 N )^k} u^{r ( \lambda )}
  s_{\lambda} ( x_1, \ldots, x_k ) = \sum_{r = 0}^k u^r \varphi_{k, N} ( r ) .
\end{equation}
Let
\[ a_{ij}^{( r )} = \left\{\begin{array}{ll}
     x_j^{i - 1} - x_j^{2 N + 2 k + 1 - i} & \text{if $i \leqslant k - r$},\\
     x_j^i - x_j^{2 N + 2 k - i} & \text{if $i > k - r$} .
   \end{array}\right. \]
In light of the Weyl character formula (\ref{charsp}) applied to
partition $\lambda = N^{k - r} \cup ( N - 1 )^r$, combined with
(\ref{detsp}) we have
\begin{equation}
  \label{e:phi} \varphi_{k, N} ( r ) = \frac{\det ( a_{ij}^{( r )} )}{\prod_{i
  = 1}^k ( 1 - x^2_i ) \prod_{i < j} ( x_i - x_j ) ( x_i x_j - 1 )} .
\end{equation}
Now using (\ref{okrat1}) we have
\begin{eqnarray}
  \label{e:phibis}
\varphi_{k, N} ( 0 ) = \frac{\det ( a_{ij}^{( 0 )} )}{\prod_{i =
1}^k ( 1 -
  x^2_i ) \prod_{i < j} ( x_i - x_j ) ( x_i x_j - 1 )} & = & \\
  \frac{\det ( x_i^{j - 1} - x_i^{2 N + 2 k + 1 - j} )}{\prod_{i = 1}^k ( 1 -
  x^2_i ) \prod_{i < j} ( x_i - x_j ) ( x_i x_j - 1 )} = \sum_{\varepsilon \in
  \{ \pm 1 \}} \frac{\prod_{j = 1}^k x_j^{N ( 1 - \varepsilon_j )}}{\prod_{i
  \leqslant j} ( 1 - x_i^{\varepsilon_i} x_j^{\varepsilon_j} )} &  &
   \nonumber
\end{eqnarray}
Next, by the definition of the determinant,
\begin{eqnarray}
  \det ( a_{ij}^{( 0 )} ) = \det ( x_i^{j - 1} - x_i^{2 N + 2 k + 1 - j} ) & =
  &  \nonumber\\
  \sum_{w \in S_k} \tmop{sgn} ( w ) \sum_{\varepsilon}
  \prod_{\varepsilon_{w_j} = 1} x_{w_j}^{j - 1} \prod_{\varepsilon_{w_j} = -
  1} ( - x_{w_j}^{2 N + 2 k - j + 1} ), &  &  \label{e:phi1}
\end{eqnarray}
and
\begin{equation}\label{e:phi2}
\begin{aligned}
&\det(a_{ij}^{(r)})= \sum_{w \in S_k} \sgn(w) \sum_{\varepsilon}
\prod_{j=1}^{k-r}\left(\prod_{\varepsilon_{w_{j}}=1} x_{w_j}^{j-1}
\prod_{\varepsilon_{w_{j}}=-1}
(-x_{w_j}^{2N+2k-j+1})\right)\times\\&
\prod_{j=k-r+1}^{k}\left(\prod_{\varepsilon_{w_{j}}=1} x_{w_j}^{j}
\prod_{\varepsilon_{w_{j}}=-1} (-x_{w_j}^{2N+2k-j})\right)=\\&
\sum_{w \in S_k} \sgn(w) \sum_{\varepsilon}
\prod_{j=1}^{k-r}\left(\prod_{\varepsilon_{w_{j}}=1} x_{w_j}^{j-1}
\prod_{\varepsilon_{w_{j}}=-1} (-x_{w_j}^{2N+2k-j+1})\right)\times
\\&
\prod_{j=k-r+1}^{k}\left(\prod_{\varepsilon_{w_{j}}=1}
x_{w_j}^{j-1} \prod_{\varepsilon_{w_{j}}=-1}
(-x_{w_j}^{2N+2k-j+1})\right) \prod_{j=k-r+1}^{k}
x_{w(j)}^{\varepsilon_{w(j)}}=\\& \sum_{\varepsilon}\left(\sum_{w
\in S_k} \sgn(w)\prod_{\varepsilon_{w_{j}}=1} x_{w_j}^{j-1}
\prod_{\varepsilon_{w_{j}}=-1} (-x_{w_j}^{2N+2k-j+1})\right)
\sum_{i_1 <  \dots < i_r}x_{i_1}^{\varepsilon_{i_1}} \dots
x_{i_r}^{\varepsilon_{i_r}}.
\end{aligned}
\end{equation}

Consequently, using (\ref{e:phi}), (\ref{e:phibis}),
(\ref{e:phi1}) and (\ref{e:phi2}) we have
\begin{equation}
  \varphi_{k, N} ( r ) = \sum_{\varepsilon \in \{ \pm 1 \}} \frac{\prod_{j =
  1}^k x_j^{N ( 1 - \varepsilon_j )}}{\prod_{i \leqslant j} ( 1 -
  x_i^{\varepsilon_i} x_j^{\varepsilon_j} )} e_r ( x_1^{\varepsilon_1} \cdots
  x_k^{\varepsilon_k} ),
\end{equation}
and this combined with (\ref{e:phi0}) and
\[ \sum_{j = 0}^k e_j ( x_1, \cdots, x_k ) y^j = \prod_{i = 1}^k ( 1 + x_i y )
\]
completes the proof of Lemma \ref{lem:sp3}, and consequently
Proposition \ref{prop:sp3} and Theorem \ref{ratsp}.

\subsection{Orthogonal group}

Branching rule (\ref{e:borth}) implies
\begin{equation}
  \label{eschurorth} \mathbb{E}_{O ( 2 N )} s_{\lambda} ( g ) =
  \left\{\begin{array}{ll}
    1 & \text{if $\lambda$ is even, $l ( \lambda ) \leqslant 2 N$} ;\\
    0 & \text{otherwise} .
  \end{array}\right.
\end{equation}
where $2 \nu$ represents the partition produced by doubling each
elements of $\nu$; note that $\lambda = 2 \nu$ means that
$\lambda$ is even. The importance of (\ref{eschurorth}) in random
matrix theory context was emphasized by Rains  {\cite{Ra98}} and
Baik and Rains {\cite{BR02}}.

Proceeding exactly as in the case of symplectic group, we have,
using (\ref{eschurorth})
\begin{equation}
  \label{elabc} \mathbb{E}_{O ( 2 N )} \prod_{j = 1}^k \det ( I + x_j g ) =
  \sum_{\tmscript{\begin{array}{c}
    \lambda_1 \leqslant 2 N\\
    \text{$\lambda'$ even}
  \end{array}}} s_{\lambda} ( x_1, \ldots x_k ),
\end{equation}
an identity first noted by Baik and Rains {\cite{BR02}}. Combining
this with Proposition \ref{prodo}, we recover the following
formula due to Okada~{\cite{ok}} and Krattenthaler~{\cite{krat}}.
\begin{equation} \label{oke1} \begin{aligned}
&\sum_{ \substack{\lambda_1 \leqslant 2N \\ \lambda' \,
\text{even}}}s_{\lambda}(x_1, \ldots x_k)=(x_1 \dots
x_k)^{N} \chi_{(N^k)_{+}}^{\so_{2k}}(x_1^{\pm 1},\dots, x_k^{\pm 1})=\\
& \sum_{\substack{\varepsilon\in\{\pm 1\}\\
\sgn(\varepsilon)=1 }}\prod_{j=1}^{k}x_j^{N (1-\varepsilon_j) }
\prod_{i < j}(1-x_i^{\varepsilon_i}x_j^{\varepsilon_j})^{-1}.
\end{aligned}
\end{equation}
Next, considering separately the subgroup $\tmop{SO} ( 2 N )$ and
its coset $O^- ( 2 N )$, we have
\begin{equation}
  \mathbb{E}_{SO ( 2 N )} f ( g ) = \mathbb{E}_{g \in O ( 2 N )} ( 1 + \det (
  g ) ) f ( g ),
\end{equation}
and
\begin{equation}
  \mathbb{E}_{O^- ( 2 N )} f ( g ) = \mathbb{E}_{O ( 2 N )} ( 1 - \det ( g ) )
  f ( g ) .
\end{equation}
Now, denoting the eigenvalues of $M$ by $(t_1, \dots, t_{2N})$, we
have
\begin{eqnarray*}
  ( 1 + \det ( g ) ) \prod_{j = 1}^k \det ( I + x_j g ) & = & \\
  ( 1 + e_{2 N} ( t_1, \ldots, t_{2 N} ) ) \sum_{\lambda} s_{\lambda} ( x_1,
  \ldots, x_k ) s_{\lambda'} ( t_1, \ldots, t_{2 N} ), &  &
\end{eqnarray*}
and, therefore using (\ref{eschurorth}) and Pieri's formula
(\ref{e_pf2}), we obtain
\begin{eqnarray*}
  \mathbb{E}_{SO ( 2 N )} \prod_{j = 1}^k \det ( I + x_j g ) & = & \\
  \sum_{\tmscript{\begin{array}{c}
    \lambda_1 \leqslant 2 N\\
    \text{$\lambda'$ even}
  \end{array}}} s_{\lambda} ( x_1, \ldots x_k ) +
  \sum_{\tmscript{\begin{array}{c}
    \lambda_1 \leqslant 2 N\\
    \text{$\lambda'$ odd}
  \end{array}}} s_{\lambda} ( x_1, \ldots x_k ) &  &
\end{eqnarray*}
and
\begin{eqnarray*}
  \mathbb{E}_{O^- ( 2 N )} \prod_{j = 1}^k \det ( I + x_j g ) & = & \\
  \sum_{\tmscript{\begin{array}{c}
    \lambda_1 \leqslant 2 N\\
    \text{$\lambda'$ even}
  \end{array}}} s_{\lambda} ( x_1, \ldots, x_k ) -
  \sum_{\tmscript{\begin{array}{c}
    \lambda_1 \leqslant 2 N\\
    \text{$\lambda'$ odd}
  \end{array}}} s_{\lambda} ( x_1, \ldots, x_k ).&  &
\end{eqnarray*}
Using Proposition \ref{prosmin} and Proposition \ref{soprod} we
therefore recover the following formula due to Okada {\cite{ok}}
and Krattenthaler~{\cite{krat}}:
\begin{equation} \label{oke2} \sum_{
\substack{\lambda_1 \leqslant 2N
\\ \lambda' \,
\text{odd}}}s_{\lambda}(x_1, \ldots x_k) = \sum_{\substack{\varepsilon\in\{\pm 1\}\\
\sgn(\varepsilon)= -1 }}\prod_{j=1}^{k}x_j^{N (1-\varepsilon_j) }
\prod_{i < j}(1-x_i^{\varepsilon_i}x_j^{\varepsilon_j})^{-1}.
\end{equation}
Finally, proceeding exactly as in the case of symplectic group we
obtain using Proposition \ref{generalizedcauchy} and
(\ref{eschurorth}):
\[ \label{elabd} \mathbb{E}_{O ( 2 N )} \frac{\prod_{j = 1}^k \det ( I + x_j g
   )}{\prod_{i = 1}^l \det ( I - y_i g )} = \sum_{\tmscript{\begin{array}{c}
     \lambda_1 \leqslant 2 N\\
     \text{$\lambda'$ even}
   \end{array}}} \tmop{LS}_{\lambda} ( x_1, \ldots x_k ; y_1, \ldots, y_l ) ;
\]
\begin{equation*} \begin{aligned}& {\mathbb E}_{M \in SO(2N)}
\frac{\prod_{j=1}^k \det (I+ x_j M)}{\prod_{i=1}^l \det (I-y_i M)}
= \sum_{ \substack{\lambda_1 \leqslant 2N \\ \lambda' \,
\text{even}}}\HS_{\lambda}(x_1, \ldots x_k; y_1, \ldots, y_l) +\\
&\sum_{ \substack{\lambda_1 \leqslant 2N \\ \lambda' \,
\text{odd}}}\HS_{\lambda}(x_1, \ldots x_k; y_1, \ldots, y_l);
\end{aligned}
\end{equation*}
and
\begin{eqnarray*}
  \mathbb{E}_{O^- ( 2 N )} \frac{\prod_{j = 1}^k \det ( I + x_j g )}{\prod_{i
  = 1}^l \det ( I - y_i g )} & = & \\
  \sum_{\tmscript{\begin{array}{c}
    \lambda_1 \leqslant 2 N\\
    \text{$\lambda'$ even}
  \end{array}}} \tmop{LS}_{\lambda} ( x_1, \ldots x_k ; y_1, \ldots, y_l ) &
  & \\
  - \sum_{\tmscript{\begin{array}{c}
    \lambda_1 \leqslant 2 N\\
    \text{$\lambda'$ odd}
  \end{array}}} \tmop{LS}_{\lambda} ( x_1, \ldots x_k ; y_1, \ldots, y_l ), &
  &
\end{eqnarray*}

Consequently, using Theorem \ref{thm:o1}, Theorem \ref{ratso}, and
Theorem \ref{thm:o3}, we obtain the following generalizations of
Okada-Krattenthaler formulae \eqref{oke1} and \eqref{oke2}:

\begin{proposition}  Suppose $N \geqslant l$.  Then we have:

\begin{eqnarray*}
  \sum_{\tmscript{\begin{array}{c}
    \lambda_1 \leqslant 2 N\\
    \text{$\lambda'$ even}
  \end{array}}} \tmop{LS}_{\lambda} ( x_1, \ldots x_k ; y_1, \ldots, y_l ) & =
  & \\
  \sum_{\tmscript{\begin{array}{c}
    \varepsilon \in \{ \pm 1 \}\\
    \tmop{sgn} ( \varepsilon ) = 1
  \end{array}}} \prod_{j = 1}^k x_j^{N ( 1 - \varepsilon_j )} \frac{\prod_{i =
  1}^k \prod_{j = 1}^l ( 1 + x_i^{\varepsilon_i} y_j )}{\prod_{i < j} ( 1 -
  x_i^{\varepsilon_i} x_j^{\varepsilon_j} ) \prod_{1 \leqslant i \leqslant j
  \leqslant l} ( 1 - y_i y_j )}, &  & \\
  &  &
\end{eqnarray*}
and
\begin{eqnarray*}
  \sum_{\tmscript{\begin{array}{c}
    \lambda_1 \leqslant 2 N\\
    \text{$\lambda'$ odd}
  \end{array}}} \tmop{LS}_{\lambda} ( x_1, \ldots x_k ; y_1, \ldots, y_l ) & =
  & \\
  \sum_{\tmscript{\begin{array}{c}
    \varepsilon \in \{ \pm 1 \}\\
    \tmop{sgn} ( \varepsilon ) = - 1
  \end{array}}} \prod_{j = 1}^k x_j^{N ( 1 - \varepsilon_j )} \frac{\prod_{i =
  1}^k \prod_{j = 1}^l ( 1 + x_i^{\varepsilon_i} y_j )}{\prod_{i < j} ( 1 -
  x_i^{\varepsilon_i} x_j^{\varepsilon_j} ) \prod_{1 \leqslant i \leqslant j
  \leqslant l} ( 1 - y_i y_j )}. &  & \\
  &  &
\end{eqnarray*}
\end{proposition}

\end{document}